\documentclass{elsart}
\usepackage{natbib}
\bibpunct{[}{]}{;}{n}{,}{,}
\usepackage{amsmath}
\usepackage{amssymb}
\usepackage{graphicx}
\usepackage{epsfig}
\newcommand{\vprime}{{}'\!v}
\newcommand{\wprime}{{}'\!w}

\newcommand{\n}{\noindent}

\newcommand{\G}{\mathcal{G}}

\def\smallcircledV{{\scriptscriptstyle{\bigcirc\kern-5.66pt\vee\kern 1.33pt}}}

\def\div{\operatorname{div}}

\newcommand{\E}{{\mathcal E}}
\newcommand{\R}{{\mathbb R}}
\newcommand{\Beta}{{B}}

\numberwithin{equation}{section}
\begin{document}
\begin{frontmatter}
\title{Spectral - Lagrangian methods for Collisional Models of Non - Equilibrium Statistical States}
\author{Irene M. Gamba}
\address{Dept. of Mathematics \& Institute of Computational Engineering and Sciences, University of Texas Austin}
\author{Sri Harsha Tharkabhushanam}
\address{Institute of Computational Engineering and Sciences, University of Texas Austin}
\begin{abstract}
We propose a new spectral Lagrangian based deterministic solver for the non-linear Boltzmann Transport Equation for Variable Hard Potential (VHP) collision kernels with conservative or non-conservative binary interactions. The method is based on symmetries of the Fourier transform of the collision integral, where the complexity in its computing is reduced to a separate integral over the unit sphere $S^2$. In addition, the conservation of moments is enforced by Lagrangian constraints. The resulting scheme, implemented in free space is very versatile and adjusts in a very simple manner, to several cases that involve energy dissipation due to local micro-reversibility (inelastic interactions) or elastic model of slowing down process. Our simulations are benchmarked with the available exact self-similar solutions, exact moment equations and analytical estimates for homogeneous Boltzmann equation for both elastic and inelastic VHP interactions. Benchmarking of the simulations involves the selection of a time self-similar rescaling of the numerical distribution function which is performed using the continuous spectrum of the equation for Maxwell molecules as studied first in \cite{asympGranular} and generalized to a wide range of related models in \cite{BCG}. The method also produces accurate results in the case of inelastic diffusive Boltzmann equations for hard-spheres (inelastic collisions under thermal bath), where overpopulated non-Gaussian exponential tails have been conjectured in computations by stochastic methods in \cite{vanNoi-Ernst, ernst-brito, MSS, GRW04} and rigourously proven in \cite{diffGran} and \cite{highEnergyTails}.
\end{abstract}
\end{frontmatter}
\section{Introduction}
In a microscopic description of a rarefied gas, all particles are assumed to be traveling in a straight line with a fixed velocity until they enter into a collision. In such dilute flows, binary collisions are often assumed to be the main mechanism of particle interactions. The statistical effect of such collisions can be modeled by collision terms of the Boltzmann or Enskog transport equation type, where the kinetic dynamics of the gas are subject to the molecular chaos assumption. The nature of these interactions could be elastic, inelastic or coalescing. They could either be isotropic or anisotropic, depending on their collision rates as a function of the scattering angle. In addition, collisions are described in terms of inter-particle potentials and the rate of collisions is usually modeled as product of power laws for the relative speed and the differential cross section, at the time of the interaction. When the rate of collisions is independent of the relative speed, the interaction is referred to as of Maxwell type. When it corresponds to relative speed to a positive power less than unity, they are referred to as Variable Hard Potentials (VHP) and when the rate of collisions is proportional to the relative speed, it is referred to as hard spheres.
\\
\\
The Boltzmann Transport Equation (an integro-differential transport equation) describes the evolution of a single point probability distribution function $f(x, v, t)$ which is defined as the probability of finding a particle at position $x$ with velocity (kinetic) $v$ at time $t$. The mathematical and computational difficulties associated to the Boltzmann equation are due to the non local - non linear nature of the collision operator, which is usually modeled as a multi linear integral form in $d$-dimensional velocity space and unit sphere $S^{d-1}$.
\\
\\
From the computational point of view, of the well-known and well-studied methods developed in order to solve this equation is an stochastic based method called "Direct Simulation Monte-Carlo" (DSMC) developed initially by Bird~\cite{bird} and Nanbu~\cite{nanbu} and more recently by \cite{RjaWa96, RjaWa05}. This method is usually employed as an alternative to hydrodynamic solvers to model the evolution of moments or hydrodynamic quantities. In particular, this method have been shown to converge to the solution of the classical Boltzmann equation in the case of mono atomic rarefied gases \cite{Wagner92}. One of the main drawbacks of such methods is the inherent statistical fluctuations in the numerical results, which becomes very expensive or unreliable in presence of non-stationary flows or non equilibrium statistical states, where more information is desired about the evolving probability distribution. Currently, there is extensive work from Rjasanow and Wagner \cite{RjaWa05} and references therein, to determine accurately the high-velocity tail behavior of the distribution functions from DSMC data. Implementations for micro irreversible interactions such as inelastic collisions have been carefully studied in \cite{GRW04}.
\\
\\
In contrast, a deterministic method computes approximations of the probability distribution function using the Boltzmann equation, as well as approximations to the observables like density, momentum, energy, etc.,. There are currently two deterministic approaches to the computations of non-linear Boltzmann, one is the well known discrete velocity models and the second a spectral based method, both implemented for simulations of elastic interactions i.e. energy conservative evolution. Discrete velocity models were developed by Broadwell \cite{broadwell} and mathematically studied by Illner, Cabannes, Kawashima among many authors \cite{illner78, kawa81, cabannes}. More recently these models have been studied for many other applications on kinetic elastic theory in \cite{bobyCerci99,cerci-corn00,mieuss00,zheng-struchtrup,HerParSea}. These models have not adapted to inelastic collisional problems up to this point according to our best knowledge.
\\
\\
Spectral based models, which are the ones of our choice in this work, have been developed by Pareschi, Gabetta and Toscani \cite{Gab-Par-Tos}, and later by Bobylev and Rjasanow~\cite{bobylevRjasanow0} and Pareschi and Russo~\cite{pareschiRusso}. These methods are supported by the ground breaking work of Bobylev~\cite{bobylevFT} using the Fourier Transformed Boltzmann Equation to analyze its solutions in the case of Maxwell type of interactions. After the introduction of the inelastic Boltzmann equation for Maxwell type interactions and the use of the Fourier transform for its analysis by Bobylev, Carrillo and one of the authors here \cite{BoCaGa00}, the spectral based approach is becoming the most suitable tool to deal with deterministic computations of kinetic models associated with Boltzmann non-linear binary collisional integral, both for elastic or inelastic interactions. More recent implementations of spectral methods for the non-linear Boltzmann are due to Bobylev and Rjasanow~\cite{bobylevRjasanow2} who developed a method using the Fast Fourier Transform (FFT) for Maxwell type of interactions and then for Hard-Sphere interactions~\cite{bobylevRjasanow3} using generalized Radon and X-ray transforms via FFT. Simultaneously, L. Pareschi and B. Perthame~\cite{pareschiPerthame} developed  similar scheme using FFT for Maxwell type of interactions. Later, I. Ibragimov and S. Rjasanow~\cite{ibragRjasanow} developed a numerical method to solve the space homogeneous Boltzmann Equation on a uniform grid for a Variable Hard Potential interactions with elastic collisions. This particular work has been a great inspiration for the current work and was one of the first initiating steps in the direction of a new numerical method.
\\
\\
We mention that, most recently, Filbet and Russo~\cite{filbetRusso1},~\cite{filbetRusso} implemented a method to solve the space inhomogeneous Boltzmann equation using the previously developed spectral methods in \cite{pareschiRusso, pareschiPerthame}. Afore mentioned work in developing deterministic solvers for non-linear BTE have been restricted to elastic, conservative interactions. Finally, Mouhout and Pareschi~\cite{mouhotPareschi} are currently studying the approximation properties of the schemes. Part of the difficulties in their strategy arises from the constraint that the numerical solution has to satisfy conservation of the initial mass. To this end, the authors propose the use of a periodic representation of the distribution function to avoid aliasing. There is no conservation of momentum and energy in ~\cite{filbetRusso}, \cite{filbetRusso1} and ~\cite{mouhotPareschi}. Both methods (~\cite{filbetRusso}, ~\cite{filbetRusso1}, ~\cite{mouhotPareschi}), which are developed in 2 and 3 dimensions, do not guarantee the positivity of the solution due to the fact that the truncation of the velocity domain combined with the Fourier method makes the distribution function negative at times. This last shortcoming of the spectral approach remains in our proposed technique; however we are able to handle conservation in a very natural way by means of Lagrange multipliers. We also want to credit an unpublished calculation of V. Panferov and S. Rjasanow ~\cite{panferovRjasanow} who wrote a method to calculate the particle distribution function for inelastic collisions in the case of hard spheres, but there were no numerical results to corroborate the efficiency of the method. Our proposed approach is slightly different and it takes a less number of operations to compute the collision integral.
\\
\\
Our current approach, based on a modified version of the work in
\cite{bobylevRjasanow0} and \cite{ibragRjasanow}, works for elastic
or inelastic collisions and energy dissipative non-linear Boltzmann
type models for variable hard potentials. We do not use periodic
representations for the distribution function. The only restriction
of the current method is that it requires that the distribution
function at any time step be Fourier transformable. The required
conservation properties of the distribution function are enforced
through a Lagrange multiplier constrained optimization problem with
the desired conservation quantities set as the constraints. Such
corrections to the distribution function to make it conservative are
very small but crucial for the evolution of the probability
distribution function according to the Boltzmann equation.
\\
\\
This Lagrange optimization problem gives the freedom of not
conserving the energy, independent of the collision mechanism, as
long momentum is conserved. Such a technique plays a major role as
it gives the option of computing energy dissipative solutions by
just eliminating one constraint in the corresponding optimization
problem. The current method can be easily implemented in any
dimension. A novel aspect of the presented approach here lays on a
new method that uses the Fourier Transform as a tool to simplify the
computation of the collision operator that works, both for elastic
and inelastic collisions. It is based on an integral representation
of the Fourier Transform of the collision kernel as used in
\cite{bobylevRjasanow0}. If $N$ is the number of discretizations in
one direction of the velocity domain in $d$-dimensions, the total
number of operations required to solve for the collision integral
are of the order of $N^{2d}log(N) + O(N^{2d})$. And this number of
operations remains the same for elastic/ inelastic, isotropic/
anisotropic VHP type of interactions. However, when the differential
cross section is independent of the scattering angle, the integral
representation kernel is further reduced by an exact closed
integrated form that is used to save in computational number of
operations to $O(N^d log(N))$. This reduction is possible when
computing hard spheres in d+2 dimensions or Maxwell type models in
2-dimensions. Nevertheless, the method can be employed without much
changes for the other case. In particular the method becomes
$O(P^{d-1}\, N^d log(N))$, where $P$, the number of each angular
discretizations is expected to be much smaller than $N$ used for
energy discretizations. Such reduction in number of operations was
also reported in ~\cite{filbetRusso} with $O(N log(N))$ number of
operations, where the authors are assuming N to be the total number
of discretizations in the $d$-dimensional space (i.e. our $N^d$ and
$P$ of order of unity).
\\
\\
Our numerical study is performed for several examples of well
establish behavior associated to solutions of energy dissipative
space homogeneous collisional models under heating sources that
secure existence of stationary states with positive and finite
energy. We shall consider heating sources corresponding to randomly
heated inelastic particles in a heat bath, with and without
friction; elastic or inelastic collisional forms with
anti-divergence terms due to dynamically (self-similar) energy
scaled solutions \cite{diffGran, highEnergyTails} and a particularly
interesting example of inelastic collisions added to a slow down
linear process that can be derived as a weakly coupled heavy/light
binary mixture. On this particular case, when Maxwell type
interactions are considered, it is shown that
\cite{asympGranular,powerLikeGB,BCG}, on one hand dynamically energy scaled
solutions exist, they have a close, explicit formula in Fourier
space for a particular choice of parameters and their corresponding
anti Fourier transform in probability space exhibit a singularity at
the origin and power law high energy tails, while remaining
integrable and with finite energy. On the other hand they are stable
within a large class of initial states. We used this particular
example to benchmark our computations by spectral methods by
comparing the dynamically scaled computed solutions to the explicit
one self similar one.

Convergence and error results of the Fourier Transform Lagrangian method, locally in time,  are currently being developed  \cite{GTConv08},  and it is expected that the proposed spectral approximation of the free space problem will have optimal algorithm complexity using the non-equispaced FFT as obtained by Greengard and Lin \cite{greengardLin} for spectral approximation of the free space heat kernel.

Implementation of the space inhomogeneous case are also currently being considered. The spectral-Lagrangian scheme methodology proposed here can be extended to cases of Pareto tails, opinion dynamics and  $N$ player games, where the evolution and asymptotic behavior of probabilities are studied in Fourier space as well. \cite{Par-Tos,BCG}.
\\
\\
The paper is organized as follows. In section 2, some preliminaries and description of the various approximated models associated with the elastic or inelastic Boltzmann equation are presented. In section 3, the actual numerical method is discussed
with a small discussion on its discretization. In section 4, the special case of spatially homogeneous collisional model for a slow down process derived from a weakly coupled binary problem with isotropic elastic Maxwell type interactions is considered wherein an
explicit solution is derived and shown to have power-like tails in some particular cases corresponding to a cold thermostat problem. Section 5 deals with the numerical results and examples. Finally in section 6, direction of future work is proposed along with a
summary of the proposed numerical method.
\section{Preliminaries}
The initial value problem associated to space homogeneous Boltzmann
Transport Equation modeling the statistical (kinetic) evolution of a
single point probability distribution function $f(v, t)$ for
Variable Hard Potential (VHP) interactions is given by
\begin{eqnarray}
\frac{\partial}{\partial t} f(v, t) &\ = &\ Q(f, f)(v, t) \nonumber\\
&\ = &\ \int_{_{w \in \mathbb{R}^d, \sigma \in S^{d-1}}}
[J_{\beta}f(\vprime, t)
f(\wprime, t) - f(v, t)f(w, t)]\, B(|u|, \mu)\, d\sigma dw \nonumber \\
f(v, 0)&\ =&\ f_0(v) \, , \label{singleEq}
\end{eqnarray}
where the initial probability distribution $f_0(v)$ is assumed
integrable and $J_{\beta} = \frac{\partial (v', w')}{\partial(v, w)}$ is Jacobian of post with respect to pre collisional velocities which depend the local energy dissipation \cite{carcer95}. 
The problem  may or may not have finite initial energy
$\E_0=\int_{_{\mathbb{R}^d}} f_0(v) |v|^2 dv$ and the velocity
interaction law, written in center of mass and relative velocity
coordinates is
\begin{eqnarray}
\begin{aligned}
u = v - w : \ \text{the relative velocity}  \\
v'= v + \frac{\beta}{2}(|u|\sigma - u),\qquad \ w' = w - \frac{\beta}{2}(|u|\sigma - u) \ , \\
\mu = \cos(\theta) \ =\ \frac{u\cdot \sigma}{|u|} : \ \text{the cosine of the scattering angle}\ , \\
B(|u|, \mu) = |u|^{\lambda}\, b(\cos \theta) \qquad \text{with}\ 0\leq \lambda\leq 1, \\
\omega_{d-2}\int_{_{{0}}}^{{\pi}} b(\cos\theta) \sin^{d-2}\theta d \theta < K: \text{\emph{Grad cut-off assumption}} \\
\beta = \frac{1+e}{2} : \ \text{the energy dissipation parameter} \, ,
\end{aligned}
\label{singleEq2}
\end{eqnarray}
where the parameter $e \in [0, 1]$ is the restitution coefficient
corresponding from sticky to elastic interactions, where $J_\beta=J_1=1$.
\\
\\
We denote by $\vprime$ and $\wprime$ the pre-collision velocities
corresponding to $v$ and $w$. In the case of micro-reversible
(elastic) collisions one can replace $\vprime$ and $\wprime$ with
$v'$ and $w'$ respectively in the integral part of \eqref{singleEq}.
We assume the differential cross section function $b(\frac{u\cdot
\sigma}{|u|})$ is integrable with respect to the post-collisional
specular reflection direction $\sigma$ in the $d-1$ dimensional
sphere, referred as the {\sl Grad cut-off assumption}, and that
$b(\cos\theta)$ is renormalized such that
\begin{eqnarray}\label{grad-cut-off}
\int_{{S^{d-1}}} b(\frac{u \cdot \sigma}{|u|})\, d\sigma \ &=& \ \omega_{d-2} \int_{_{0}}^{^{\pi}}
b(\cos\theta) \sin^{d-2}\theta \,d\theta \ \nonumber \\
&=& \omega_{d-2} \int_{_{-1}}^{^{1}} b(\mu) (1 - \mu^2)^{(d-3)/2} d\mu \ = 1 \, ,
\end{eqnarray}
where the constant $\omega_{d-2}$ is the measure of the $d-2$ dimensional
sphere and the corresponding scattering angle is $\theta$ is defined by $\cos\theta=
\frac{\sigma\cdot u}{|u|}$.
\\
\\
The parameter $\lambda$ regulates the collision frequency as a function of the relative speed $|u|$. It accounts for inter particle potentials defining the collisional kernel and they are referred to as Variable Hard Potentials (VHP) whenever $0< \lambda< 1$, Maxwell Molecules type interactions (MM) for $\lambda=0$ and Hard Spheres (HS) for $\lambda=1$. The Variable Hard Potential collision kernel then takes the following general form:
\begin{itemize}
\item[\ ]
\begin{equation}
B(|u|, \mu) = C_{\lambda}(\sigma)|u|^{\lambda}\, , \end{equation}
with $C_{\lambda}(\sigma) = \frac{1}{4\pi} b(\theta),  \lambda = 0$ for Maxwell type of interactions; $C_{\lambda}(\sigma) = \frac{a^2}{4}, \lambda = 1$ for Hard Spheres. In addition, if $C_{\lambda}(\sigma)$ is independent of the scattering angle we call the interactions isotropic. Otherwise we referred to them as anisotropic Variable Hard Potential interactions.
\end{itemize}

\n For classical case of elastic collisions, it has been established
that the Cauchy problem for the space homogeneous Boltzmann equation
has a unique solution in the class of integrable functions with
finite energy (i.e. $C^1(L^1_2(\R^d))$), it is regular if initially
so, and $f(.,t)$ converges in $L^1_2(\R^d)$ to the Maxwellian
distribution $M_{\rho,V,\E}(v)$ associated to the $d+2$-moments of
the initial state $f(v,0)=f_0(v) \in L^1_2(\R^d)$. In addition, if
the initial state has Maxwellian decay, this property will remain
with a Maxwellian decay globally bounded in time
(\cite{Ga-Pa-Vi07}), as well as all derivatives if initial so (see
\cite{AlaGa07}).
\\
\\
Depending on their nature, collisions either conserve density,
momentum and energy (elastic) or density and momentum (inelastic) or
density (elastic - linear Boltzmann operator), depending on the
number of collision invariants the operator $Q(f,f)(t,v)$ has. In
the case of the classical Boltzmann equation for rarefied (elastic)
mono-atomic gases, the collision invariants are exactly $d+2$, that
is, according to the Boltzmann theorem, the number of polynomials
in velocity space $v$ that generate $\phi(v)= A +\bf{B}\cdot v +
C|v|^2$, with $C\leq 0$. In particular, one obtains the following
\emph{conserved quantities}
\begin{eqnarray}
\text{density}\ \ \ \rho(t) &\ = \int_{_{v \in \mathbb{R}^d}} f(v, t) dv \nonumber \\
\text{momentum}\ \ \ m(t)&\ =\ \int_{_{v \in \mathbb{R}^d}} v f(v, t) dv\\
\text{energy}\ \ \ \E(t) &\ =\ \frac{1}{2\rho(t)} \int_{_{v \in
\mathbb{R}^d}} |v|^2 f(v, t) dv \, . \nonumber \label{conserv}
\end{eqnarray}
Of significant interest from the statistical view point are the
evolution of moments or observables, at all orders. They are defined
by the dynamics of the corresponding time evolution equation for
the velocity averages, given by
\begin{eqnarray}\begin{aligned}
\frac{\partial}{\partial t}\, M_j(t)\ =\ \int_{_{v \in \mathbb{R}^d}}
f(v, t) v^{\smallcircledV j} dv \ =\ \int_{_{v \in \mathbb{R}^d}}
Q(f,f)(v, t) v^{\smallcircledV j} dv \ ,
\end{aligned}\label{moments1}\end{eqnarray}
where, $v^{\smallcircledV j} = $ the standard symmetric tensor product of $v$ with itself, $j$ times.
Thus, according to \eqref{conserv}, for the classical elastic
Boltzmann equation, the first $d+2$ moments are conserved, meaning,
$M_j(t)= M_{0,j} = \int_{_{v \in \mathbb{R}^d}} f_0(v)
v^{\smallcircledV j} dv $ for $j=0,1$; and $\E(t)=\text{tr}
(M_2)(t)= \E_0=\int_{_{v \in \mathbb{R}^d}} f_0(v) |v|^2 dv$. Higher
order moments or observables of interest are
\begin{eqnarray}\begin{aligned}
\text{Momentum Flow}\ \ \ M_2(t) &\ =\ \int_{_{\mathbb{R}^d}} v v^T
f(v, t) dv
\\
\text{Energy Flow} \ \ \ r(t)&\ =\
\frac{1}{2\rho(t)}\int_{_{\mathbb{R}^d}} v |v|^2 f(v, t) dv
\\
\text{ Bulk Velocity}\ \ \ V(t) &\ =\ \frac{m(t)}{\rho(t)}
\\
\text{ Internal Energy}\ \ \ \E(t) &\ =\ \frac{1}{2\rho} (tr(M_2) - \rho|V|^2) \\
\text{Temperature} \ \ \ T(t) &\ =\ \frac{2\E(t)}{\,{\bf k} d}
\end{aligned}
\label{moments2}
\end{eqnarray}
with ${\bf k} - $ Boltzmann constant.
\\
\\
We finally point out that, in the case of Maxwell molecules
($\lambda=0$),
it is possible to write recursion formulas for higher order moments of all
orders (\cite{Bob88} for the elastic case, and \cite{BoCaGa00} in the
inelastic case) which, in the particular case of isotropic
solutions depending only on $|v|^2/2$, take the form
\begin{eqnarray}\begin{aligned}\label{moments3}
m_n(t) &\ =\ \int_{\mathbb{R}^d} |v|^{2n}\, f(v, t) dv =\ e^{-\lambda_{n} t}m_n(0) + \\
& \sum_{k=1}^{n-1} \frac1{2(n+1)}\binom{2n+2}{2k+1}\,
\Beta_{\beta}(k,n-k) \int_0^t
m_k(\tau)\, m_{n-k}(\tau)\, e^{-\lambda_{n} (t-\tau)} \,d\tau \, ; \\
\text{with}\ &\ \ \\
&\lambda_n=1-\frac1{n+1}[\beta^{2n} + \sum_{k=0}^{n} (1-\beta)^{2k}]
\, , \\
&\Beta_{\beta}(k,n-k)\ =\ \beta^{2k}\int_0^1 s^k
(1-\beta(2-\beta)s)^{n-k} ds \, ,
\end{aligned}\end{eqnarray}
for $n\ge 1,\ 0\le\beta\le1,$ where $\lambda_0=0,\ m_0(t)=1, $ and
$m_n(0)=\int_{_{\mathbb{R}^d}} |v|^{2n}\, f_0(v) dv$.
\subsection{Boltzmann collisional models with heating sources}
A collisional model associated to the space homogeneous Boltzmann transport equation \eqref{singleEq} with grad cutoff assumption \eqref{singleEq2}, can be modified in order to accommodate for an energy or `heat source' like term $\mathcal{G}(f(t, v))$, where $\mathcal{G}$ is a differential or integral operator. In these cases, it is possible to obtain stationary states with finite energy as for the case of inelastic interactions. In such general framework, the corresponding initial
value problem model is
\begin{eqnarray}\begin{aligned}
&\frac{\partial}{\partial t} f(v, t)
\ =\ \zeta(t)\, Q(f, f)(v, t) + \mathcal{G}(f(t, v))\, ,\\
&\ f(v, 0)\ =\ f_0(v) \, ,
\end{aligned}
\label{bte-source}
\end{eqnarray}
where the collision operator $Q(f, f)(v, t)$ is as in
\eqref{singleEq} and $\mathcal{G}(f(t, v))$ models a `heating
source' due to different phenomena. The term $\zeta(t)$ may
represent a mean field approximation that allows from proper time
rescaling. See \cite{BoCaGa00} and \cite{highEnergyTails} for
several examples for these type of models and additional references.
\\
\\
Following the work initiated in \cite{highEnergyTails} and
\cite{powerLikeGB} on Non-Equilibrium Stationary States (NESS), our
computational approach we shall present several computational
simulations of non-conservative models for either elastic or
inelastic collisions associated to \eqref{bte-source} of
the Boltzmann equation with `heating' sources. In all the cases we
have addressed one can see that stationary states with finite
energy are admissible, but they may not be Maxwellian distributions. Of this type of model we show computational output for three cases.
First one is the pure diffusion thermal bath due to a randomly
heated background \cite{WiMa,vanNoi-Ernst,diffGran}, in which case
\begin{equation}
\G_1(f) = \mu \,\Delta {f}, \label{force:Gauss}
\end{equation}
where \(\mu>0\) is a constant. The second example relates to
self-similar solutions of equation \eqref{bte-source} for $\G(f)=0$
\cite{MoSa,ErBr2}, but dynamically rescaled by
\begin{equation}
\label{ssrescale} f(v,t) = \frac{1}{v_0^d(t)}\,{\tilde f}
\big({\tilde v}(v,t), {\tilde t}(t)\big),\quad {\tilde v}=\frac
v{v_0(t)},
\end{equation}
where
\begin{equation}
\label{ssrescale1} v_0(t) = (a+\eta t)^{-1}, \quad \ \tilde t(t) =
\frac {1}{\eta} \ln(1 +\frac\eta{a} t), \quad a,\,\eta >0.
\end{equation}
Then, the equation for $\tilde f (\tilde v, \tilde t)$ coincides
(after omitting the tildes) with equation \eqref{bte-source},for
\begin{equation}
\label{force:selfsimi} \G_2(f) = -\eta \div (v f), \qquad \eta>0\, .
\end{equation}
Of particular interest of dynamical time-thermal speed rescaling is
the case of collisional kernels corresponding to Maxwell type of
interactions. Since the second moment of the collisional integral is
a linear function of the energy, the energy evolves exponentially
with a rate proportional to the energy production rate, that is
\begin{equation}\label{enery-maxwell}
\frac{d}{dt}\E(t)\ =\ \lambda_0 \,\E(t), \qquad \text{or
equivalently} \ \E(t)=\E(0)\, e^{\lambda_0\,t}\, ,
\end{equation}
with $\lambda_0$ the energy production rate. Therefore the
corresponding rescaled variables and equations \eqref{ssrescale} and
\eqref{bte-source},\eqref{force:selfsimi} for the study of long time
behavior of rescaled solutions are
\begin{equation}
\label{ssrescale2} f(v,t) = \E^{-\frac d2}(t)\,{\tilde f} \big(\frac
v{\E^{\frac 12}(t)}\big) \ = \ (\E(0) e^{\lambda_0\,t})^{-\frac
d2}\, {\tilde f} ( v \,(\E(0) e^{\lambda_0\,t})^{-\frac 12} ) \, ,
\end{equation}
and ${\tilde f}$ satisfies the self-similar equation
\eqref{bte-source}
\begin{equation}
\label{force:selfsimi2} \G_{2'}(f) = -\lambda_0 x f_x, \qquad
\text{where} \ \ x = v{\E^{-\frac 12}(t)} \ \ \text{is the
self-similar variable} \, .
\end{equation}
We note that it has been shown that these dynamically self-similar
states are stable under very specific scaling for a large class of initial states \cite{BCG}.
\\
\\
The last source type we consider is given by a model, related to
weakly coupled mixture modeling slowdown (cooling) process
\cite{powerLikeGB} given by an elastic model in the presence of a
thermostat given by Maxwell type interactions of particles of mass
$\bf{m}$ having the Maxwellian distribution
\begin{equation}\label{maxwellian}
M_{\mathcal{T}}(v) = \frac{\bf{m}}{(2\pi {\mathcal{T}})^{d/2}}
e^{\frac{-\bf{m}|v|^2}{2{\mathcal{T}}}}\ ,\nonumber
\end{equation}
with a constant reference background or thermostat temperature
${\mathcal{T}}$ (i.e. the average of $\left.\int M_{\mathcal{T}}\,
dv=1\right.$ and $\int |v|^2 M_{\mathcal{T}}\, dv ={\mathcal{T}}$).
Define
\begin{equation}
Q_L(f)(v, t) \doteq \int_{{w \in \mathbb{R}^d, \sigma \in S^{d-1} }}
B_L(|u|, \mu) f(\vprime, t)M_{\mathcal{T}}(\wprime) - f(v,
t)M_{\mathcal{T}}(w)] \ d\sigma dw \, . \label{q_L}
\end{equation}
Then the corresponding evolution equation for $f(v, t)$ is given by
\begin{eqnarray}
\frac{\partial}{\partial t} f(v, t) \ &=& \ Q(f, f)(v, t) + \Theta Q_L(f)(v, t) \nonumber \\
f(v, 0)\ &=& \ f_0(v) \, . \label{mixEq}
\end{eqnarray}
where $Q(f, f)$, defined as in \eqref{singleEq}, is the classical
collision integral for elastic interactions (i.e. $\beta=1$), so
it conserves density, momentum and energy. The second integral term
in \eqref{mixEq} is a linear collision integral which conserves
just the density (but not momentum or energy) since
\begin{eqnarray}
\begin{aligned}\label{mixEq2}
u &\ = v - w \qquad \ \text{the relative velocity}\\
v' &\ = v + \frac{\bf{m}}{\bf{m}+1}(|u|\sigma - u),\qquad \ w' = w -
\frac{1}{\bf{m}+1}(|u|\sigma - u) \, .
\end{aligned}
\end{eqnarray}
The coupling constant $\Theta$ depends on the initial density, the
coupling constants and on $\bf{m}$. The collision kernel $B_L$ of the
linear part may not be the same as the one for the non-linear part
of the collision integral, however we assume that the {\sl Grad
cut-off assumption} \eqref{grad-cut-off} is satisfied and that, in
order to secure mass preservation, the corresponding differential
cross section functions $b_N$ and $b_L$, the non-linear and linear
collision kernels respectively, satisfy the renormalized condition
\begin{equation}\label{grad-cut-off2}
\int_{_{S^{d-1}}} b_N(\frac{u\cdot
\sigma}{|u|}) + \Theta b_L(\frac{u\cdot \sigma}{|u|})\, d\sigma \
=\ 1 + \Theta\, .
\end{equation}
This last model describes the evolution of binary interactions of
two sets of particles, heavy and light, in a weakly coupled limit,
where the heavy particles have reached equilibrium. The heavy
particles set constitutes the background or thermostat for the
second set of particles. It is the light particle distribution that
is modeled by \eqref{mixEq}. Indeed, $Q(f, f)$ corresponds to all
the collisions which the light particles have with each other and
the second linear integral term corresponds to collisions between
light and heavy particles at equilibrium given by a classical
distribution $M_{\mathcal{T}}(v)$. In this binary $3$-dimensional,
mixture scenario, collisions are assumed to be isotropic, elastic
and the interactions kernels of Maxwell type.
\\
\\
In the particular case of equal mass (i.e. $\bf{m}=1$), the model is of
particular interest for the development of numerical schemes and
simulations benchmarks. Even though the local interactions are
reversible (elastic), it does not conserve the total energy. In such a
case, there exists a special set of explicit, in spectral space,
self-similar solutions which are attractors for a large class of
initial states. When considering the case of Maxwell type of
interactions in three dimensions i.e. $B(|u|, \mu)= b(\mu) $ with a
cooling background process corresponding to a time temperature
transformation, ${\mathcal{T}}={\mathcal{T}}(t)$ such that
${\mathcal{T}}(t)\to 0$ as $t \to 0, $ the models have self similar
asymptotics \cite{powerLikeGB,BCG} for a large class of initial
states. Such long time asymptotics corresponding to dynamically
scaled solutions of \eqref{mixEq}, in the form of
\eqref{force:selfsimi2}, yields interesting behavior in $f(v, t)$
for large time, converging to states with power like decay tails in
$v$. In particular, such solution $f(v, t)$ of \eqref{mixEq} will
lose moments as time grows, even if the initial state has all
moments bounded. (see \cite{powerLikeGB,BCG} for the analytical proofs).
\\
\subsection{Collision Integral Representation}
One of the pivotal points in the derivation of the spectral numerical method for the computation of the non-linear Boltzmann equation lays in the representation of the collision integral in Fourier space by means of the weak form. Since for a suitably regular test function $\psi(v)$, the weak form of the collision integral takes the form (suppressing the time dependence in f)
\begin{equation}
\int_{_{v \in \mathbb{R}^d}} Q(f, f) \psi(v) dv = \int_{_{{(w,v) \in
\mathbb{R}^d\times\mathbb{R}^d ,\, \sigma \in S^{d-1}}}} f(v) f(w)
B(|u|, \mu) [\psi(v') - \psi(v)] d\sigma dw dv \, ,
\label{weakQ}
\end{equation}
with $v' = v + \frac{\beta}{2}(|u|\sigma - u)$. In particular, taking $\psi(v) = e^{-i \zeta \cdot v}/(\sqrt{2\pi})^d$, where $\zeta$ is the Fourier
variable, we get the Fourier Transform of the collision integral through its weak form as
\begin{eqnarray}
\widehat{Q}(\zeta) &=& \frac{1}{(\sqrt{2\pi})^d} \int_{_{v \in \mathbb{R}^d}} Q(f, f) e^{-i \zeta \cdot v} dv \nonumber \\
&=& \int_{_{(w,v) \in \mathbb{R}^d\times\mathbb{R}^d ,\, \sigma \in S^{d-1}}} f(v) f(w)  \frac{B(|u|, \mu)}{(\sqrt{2\pi})^d} [e^{-i \zeta \cdot v'} - e^{-i \zeta \cdot v}] d\sigma dw dv \, .
\label{weakQHat}
\end{eqnarray}
We will use $\widehat{.} = \mathcal{F}(.)$ - the Fourier transform and $\mathcal{F}^{-1}$ for the classical inverse Fourier transform. Plugging in the definitions of collision kernel $B(|u|, \mu) =
C_{\lambda}(\sigma)|u|^{\lambda}$ (which in the case of isotropic
collisions would just be the Variable Hard Potential collision
kernel) and $v'$
\begin{eqnarray}
\widehat{Q}(\zeta) &=& \frac{1}{(\sqrt{2\pi})^d} \int_{u \in \mathbb{R}^d} G_{\lambda, \beta}(u, \zeta) \int_{v \in \mathbb{R}^d} f(v) f(v - u) e^{-i \zeta \cdot v} dv du \nonumber \\
&=& \int_{u \in \mathbb{R}^d} G_{\lambda, \beta}(u, \zeta) [f(v) f(v - u)]\,^{\widehat\ } du \, , \label{weakQHat5}
\end{eqnarray}
where
\begin{eqnarray} \label{gDef1}
G_{\lambda, \beta}(u, \zeta) &=& \int_{_{\sigma \in S^{d-1}}} C_{\lambda}(\sigma) |u|^{\lambda}
[e^{-i \frac{\beta}{2}\zeta \cdot (|u|\sigma - u))} - 1] d\sigma \nonumber \\
&=& |u|^{\lambda} \left[ e^{i \frac{\beta}{2}\zeta \cdot u} \int_{_{\sigma \in S^{d-1}}}
C_{\lambda}(\sigma) e^{-i \frac{\beta}{2}|u| \zeta \cdot \sigma} d\sigma - \omega_{2} \right] \, .
\end{eqnarray}
Note that \eqref{gDef1} is valid for both isotropic and anisotropic
interactions. For the former type, a simplification ensues due to
the fact the $C_{\lambda}(\sigma)$ is independent of $\sigma \in
S^{d-1}$:
\begin{equation}
G_{\lambda, \beta}(u, \zeta) = C_{\lambda} \omega_{d-2} \, |u|^{\lambda}
\left[ e^{i \frac{\beta}{2}\zeta.u} \text{sinc}(\frac{\beta |u| |\zeta|}{2}) - 1 \right] \, . \\
\label{gDef2}
\end{equation}
Thus, it is seen that the dependence on $\sigma$ i.e. the
integration over the unit sphere $S^{d-1}$ is completely done
independently and there is actually a closed form expression for
this integration, given by \eqref{gDef2} in the case of isotropic
collisions. In the case of anisotropic collisions, the dependence of
$C_{\lambda}$ on $\sigma$ is again isolated into a separate integral
over the unit sphere $S^{d-1}$ as given in \eqref{gDef1}. The above expression can be transformed for elastic collisions
$\beta = 1$ into a form suggested by Rjasanow and Ibragimov in their
paper \cite{ibragRjasanow}. The corresponding expression for
anisotropic collisions is given by \eqref{gDef1}.
\\
\newline
Further simplification of \eqref{weakQHat5} is possible by observing that the Fourier transform inside the integral can be written in terms of the Fourier transform of $f(v)$ since it can also be written as a convolution of the Fourier transforms. Let $h(v) = f(v-u)$
\begin{eqnarray}
\widehat{Q}(\zeta) &=& \int_{_{u \in \mathbb{R}^d}} G_{\lambda, \beta}(u, \zeta) [f(v) h(v)]\,^{\widehat\ } du \quad = \int_{{u \in \mathbb{R}^d}} G_{\lambda, \beta}(u, \zeta) \frac{1}{(\sqrt{2\pi})^d} (\hat{f} \ast \hat{h})(\zeta) du \nonumber \\
&=& \int_{_{u \in \mathbb{R}^d}} G_{\lambda, \beta}(u, \zeta) \frac{1}{(\sqrt{2\pi})^d} \int_{\xi \in \mathbb{R}^d} \hat{f}(\zeta - \xi) \hat{f}(\xi) e^{-i \xi \cdot u} d\xi du \nonumber \\
&=& \frac{1}{(\sqrt{2\pi})^d} \int_{_{\xi \in \mathbb{R}^d}} \hat{f}(\zeta - \xi) \hat{f}(\xi) \hat{G}_{\lambda, \beta}(\xi, \zeta) d\xi ,
\label{qHatN2}
\end{eqnarray}
where $\hat{G}_{\lambda, \beta}(\xi, \zeta) = \int_{{u \in \mathbb{R}^d}} G_{\lambda, \beta}(u, \zeta) e^{-i \xi \cdot u} du$. Let $u = r \textbf{e}$, $\textbf{e} \in S^{d-1}, r \in \mathbb{R}$
For $d = 3$,
\begin{eqnarray}
\hat{G}_{\lambda, \beta}(\xi, \zeta) &=& \int_r \int_{\textbf{e}} r^2 G(r \textbf{e}, \zeta) e^{-i r \xi \cdot \textbf{e}} d\textbf{e} dr \nonumber \\
&=& 16 \pi^2 C_{\lambda} \int_r r^{\lambda + 2} [ \text{sinc}(\frac{2 \beta |\zeta|}{2}) \text{sinc}(r |\xi - \frac{\beta}{2} \zeta|) - \text{sinc}(r |\xi|) ] dr \, . \nonumber
\end{eqnarray}
Since the domain of computation is restricted to $\Omega_v = [-L, L)^3$, $u \in [-2L,2 L)^3 \Rightarrow r \in [0, 2\sqrt{3} L]$
\begin{equation}
\hat{G}_{\lambda, \beta}(\xi, \zeta) = 16 \pi^2 C_{\lambda} \int_0^{2\sqrt{3}L} r^{\lambda + 2} [ \text{sinc}(\frac{2 \beta |\zeta|}{2}) \text{sinc}(r |\xi - \frac{\beta}{2} \zeta|) - \text{sinc}(r |\xi|) ] dr \, .
\label{gHatExpression}
\end{equation}
A point worth noting is that the above formulation \eqref{qHatN2} results in $O(N^{2d})$  number of operations, where $N$ is the number of discretizations in each velocity direction. Also, exploiting the symmetric nature in particular cases of the collision kernel one can reduce the number of operations to $O(N^d log N)$.

\section{Numerical Method}
\subsection{Discretization of the Collision Integral}
Coming to the discretization of the velocity space, it is assumed
that the two interacting velocities and the corresponding relative
velocity
\begin{eqnarray}
v \, , w, \text{and}\ w &\ \in [-L, L)^d \, ,\\ \qquad \text{while}\qquad \zeta &\ \in [-L_{\zeta}, L_{\zeta})^d \, ,
\label{disc}
\end{eqnarray}
where the velocity domain $L$ is chosen such that $u = v - w \in [-L, L)^d$ through an assumption that $supp(f) \in [-L, L)^d$. For a sufficiently large $L$, the computed distribution will not lose mass, since the initial momentum is conserved (there is no convection in space homogeneous problems), and is renormalized to zero mean velocity. We assume a uniform grid in the velocity space and in the fourier space with $h_v$ and $h_{\zeta}$ as the respective grid element sizes. $h_v$ and $h_{\zeta}$ are chosen such that $h_v h_{\zeta} = \frac{2\pi}{N}$, where $N$ = number of discretizations of $v$ and $\zeta$ in each direction.

\subsection{Time Discretization}
\noindent To compute the actual particle distribution function, one needs to use an approximation to the time derivative of $f$. For this, a second-order Runge-Kutta scheme or a Euler forward step method were used. Since a non-dimensional Boltzmann equation is computed, for numerical computations the value of time step $dt$ is chosen such that it corresponds in its dimensional form to $0.1$ times the time between consecutive collisions (which depends on the collision frequency). During the standard process of non-dimensionalization of the Boltzmann Equation, such a reference quantity (time between collisions) comes up. With time discretizations taken as $t^n = n dt$, the discrete version of the Runge-Kutta scheme is given by
\begin{eqnarray}
f^{0}(v^j) &=& f_0(v^j) \nonumber \\
\tilde{f}(v^j) &=& f^{t^n}(v^j) + \frac{dt}{2} Q_{\lambda, \beta}
[f^{t^n}(v^j), f^{t^n}(v^j)] \nonumber \\
f^{t^{n+1}}(v^j) &=& f^{t^n}(v^j) + dt Q_{\lambda, \beta}
[\tilde{f}(v^j), \tilde{f}(v^j)]\ \, . \nonumber \\
\label{shDt}
\end{eqnarray}
The corresponding Forward Euler scheme with smaller time step is
given by
\begin{equation}
\tilde{f}(v^j) = f^{t^n}(v^j) + dt Q(f^{t^n}, f^{t^n}) \qquad \, .
\label{euler}
\end{equation}
\subsection{Conservation Properties - Lagrange Multipliers}
Since the calculation of $Q_{\lambda, \beta}(f, f)(v)$ involves computing Fourier Transforms with respect to $v$, we extensively use Fast Fourier Transform. Note that the total number of operations in computing the collision integral reduces to the order of $3N^{2d}log(N) + O(N^{2d})$ for \eqref{weakQHat5} and $O(N^{2d})$ for \eqref{qHatN2}. Observe that, choosing $1/2 \le \beta \le 1$, the proposed scheme works for both elastic and inelastic collisions. As a note, the method proposed in the current work can also be extended to lower dimensions in velocity space.
\\
\\
In the current work, due to the discretizations and the use of
Fourier Transform, the accuracy of the proposed method relies
heavily on the size of the grid and the number of points taken in
each velocity/ Fourier space directions. Because of this it is
seen that the computed $Q_{\lambda, \beta}[f, f](v)$ does not really
conserve quantities it is supposed to i.e. $\rho, m, e$ for elastic
collisions, $\rho$ for Linear Boltzmann Integral and $\rho, m$ for
inelastic collisions. Even though the difference between the
computed (discretized) collision integral and the continuous one is
not great, it is nevertheless essential that this issue be resolved.
To remedy this, a simple constrained Lagrange multiplier method is
employed where the constraints are the required conservation
properties. Let $M = N^d$, the total number of discretizations of
the velocity space. Assume that the classical Boltzmann collision
operator is being computed. So $\rho, m=(m1, m2, m3)$ and $e$ are conserved. Let
$\omega_j$ be the integration weights where $j = 1, 2, ..., M$. Let
\[ \tilde{f} = \left( \begin{array}{ccccc}
\tilde{f}_1 & \tilde{f}_2 & . & . & \tilde{f}_M
\end{array} \right)^T \]
be the distribution vector at the computed time step and
\[ f = \left( \begin{array}{ccccc}
f_1 & f_2 & . & . f_M \\
\end{array} \right)^T \]
be the corrected distribution vector with the required moments conserved. Let
\[ C_{_{(d+2) \times M}} =  \left( \begin{array}{ccc}
 & \omega_j & \\
 & v_i \omega_j & \\
 & |v_j|^2\omega_j &  \\
\end{array} \right) \]
and
\[ a_{_{(d+2) \times 1}} =  \left( \begin{array}{ccccc}
\rho & m1 & m2 & m3 & e \\
\end{array} \right)^T \]
be the vector of conserved quantities. Using the above vectors, the
conservation can be written as a constrained optimization problem:
\[(*) \left\{ \begin{array}{ll}
\| \tilde{f} - f \|_2^2 & \rightarrow min\\
Cf = a; & C \in \mathbb{R}^{{d+2} \times M}, f \in \mathbb{R}^M, a \in \mathbb{R}^{d+2} \\
\end{array} \right. \, . \]
\\
To solve $(*)$, one can employ the Lagrange multiplier
method. Let $\lambda \in \mathbb{R}^{d+2}$ be the Lagrange
multiplier vector. Then the scalar objective function to be
optimized is given by
\begin{equation}
L(f, \lambda) = \sum_{j = 1}^M |\tilde{f}_j - f_j|^2 + \lambda^T(Cf
- a) \, .\label{lagrange}
\end{equation}
\\
Equation \eqref{lagrange} can actually be solved explicitly for the
corrected distribution value and the resulting equation of
correction be implemented numerically in the code. Taking the
derivative of $L(f, \lambda)$ with respect to $f_j, j = 1, . . ., M$
and $\lambda_i, i = 1, . . . , {d+2}$ i.e. gradients of $L$,
\begin{eqnarray}
\frac{\partial L}{\partial f_j} &=& 0; j = 1, . . ., M \nonumber \\
& \Rightarrow & \nonumber \\
f &=& \tilde{f} + \frac{1}{2} C^T \lambda \, . \label{lambEq}
\end{eqnarray}
And
\begin{eqnarray}
\frac{\partial L}{\partial \lambda_1} &=& 0; i = 1, . . ., d+2 \nonumber \\
& \Rightarrow & \nonumber \\
C f &=& a
\end{eqnarray}
i.e. retrieves the constraints.
\\
Solving for $\lambda$,
\begin{equation}
C C^T \lambda = 2 (a - C \tilde{f}) \, .
\end{equation}
\\
Now $C C^T$ is symmetric $(C C^T)^T = C C^T$ and because $C$ is the
integration matrix, $C C^T$ is positive definite. By linear algebra, the inverse of $C C^T$ exists.
In particular one can compute the value of $\lambda$ by
\begin{equation} \lambda = 2 (C C^T)^{-1} (a - C \tilde{f})\, .
\end{equation}
\\
Substituting $\lambda$ into \eqref{lambEq},
\begin{equation}
f = \tilde{f} + C^T (C C^T)^{-1} (a - C \tilde{f}) \, .
\label{fEqCons}
\end{equation}
\\
Using equation for forward Euler scheme \eqref{euler}, the complete
scheme is given by ($f^{t^n}(v^j) = f^n_j$) $\forall
j$:
\begin{eqnarray}
\begin{aligned}
\tilde{f}_j = f^{n}_j + dt Q(f^n_j, f^n_j) \\
f^{n+1}_j = \tilde{f}_j + C^T (C C^T)^{-1} (a - C \tilde{f}) \, .
\end{aligned}
\label{completeScheme}
\end{eqnarray}
\\
So,
\begin{eqnarray}
f^{n+1}_j &=& f^n_j + dt Q(f^n_j, f^n_j) + C^T (C C^T)^{-1} (a - C \tilde{f}) \nonumber \\
&=& f^n_j + dt Q(f^n_j, f^n_j) + C^T (C C^T)^{-1} (a - a - dt C Q(f^n_j, f^n_j)) \nonumber \\
&=& f^n_j + dt Q(f^n_j, f^n_j) - dt C^T (C C^T)^{-1} C Q(f^n_j, f^n_j) \nonumber \\
&=& f^n_j + dt[ \mathbb{I} - C^T (C C^T)^{-1} C ] Q(f^n_j, f^n_j) \, ,\nonumber \\
\label{schemeFinal}
\end{eqnarray}
with $\mathbb{I}$ - $N \times N$ identity matrix. Letting $\Lambda_N(C) = \mathbb{I} - C^T (C C^T)^{-1} C$ with
$\mathbb{I}$ - $N \times N$ identity matrix, one obtains
\begin{equation}
f^{n+1}_j = f^n_j + dt \Lambda_N(C) Q(f^n_j, f^n_j) \, ,
\label{completeScheme1}
\end{equation}
where we expect the required observables are conserved and the
solution approaches a stationary state, since $ { \lim_{n\to \infty}
\|\Lambda_N(C)\, Q(f^n_j, f^n_j)\|_{\infty} = 0 }$ .
\\
\\
Identity \eqref{completeScheme1} summarizes the whole conservation
process. As described previously, setting the conservation
properties as constraints to a Lagrange multiplier optimization
problem ensures that the required observables are conserved. Also, the optimization method
can be extended to have the distribution function satisfy more (higher order) moments from \eqref{moments3}. In this case,
$a(t)$ will include entries of $m_n(t)$ from \eqref{moments}.
\\
\\
We point out that for the linear Boltzmann collision operator used
in the mixture problem conserves density and not momentum(unless one
computes isotropic solutions) and energy. For this problem, the
constraint would just be the density equation. For inelastic
collisions, density and momentum are conserved and for this case the
constraint would be the energy and momentum equations. And for the
elastic Boltzmann operator, all three quantities (density, momentum
and energy) are conserved and thus they become the constraints for
the optimization problem. The behavior of the conservation correction for Pseudo-Maxwell
Potentials for Elastic collisions will be numerically studied in the
numerical results section. This approach of using Lagrangian constraints in order to secure moment preservation differs from the one proposed in ~\cite{filbetRusso1},~\cite{filbetRusso} for spectral solvers.

\section{Self-Similar asymptotics for a general elastic or inelastic BTE of Maxwell type or the cold thermostat problem - power law tails}
As mentioned in introduction, a new interesting benchmark problem for our scheme is that of a dynamically scaled solutions or self-similar asymptotics. More precisely, we present simulation where the computed solution in properly scaled time approaches a self similar solution. This is of interest because of the power tail behavior i.e. higher order moments of the computed solution are bounded. For the completeness of this presentation, the analytical description of such asymptotics is given in the following two sub sections.
\subsection{Self-Similar Solution for a non-negative Thermostat Temperature}
We consider the Maxwell type equation from \eqref{mixEq} related to a space homogeneous model for a weakly coupled mixture modeling slowdown process. The content of this section is dealt in detail in~\cite{powerLikeGB} for a particular choice of zero background temperature (cold thermostat). For the sake of brevity, we refer to \cite{powerLikeGB} for details. However, a slightly more general form of the self-similar solution for non zero background temperature is derived here from the zero background temperature solution. Without loss of generality for our numerical test, we assume the differential cross sections $b_L$ for collision kernel of the linear and $b_N$, the corresponding one for the nonlinear part, are the same, both denoted by $b(\frac{k.\sigma}{|k|})$, satisfying the {\sl Grad cut-off} conditions \eqref{grad-cut-off}. In particular, condition \eqref{grad-cut-off2} is automatically satisfied.
\\
\\
In \cite{powerLikeGB}, Fourier transform of the isotropic self-similar solution associated to problem in \eqref{mixEq} will take the form:
\begin{equation}\label{bteF02}
\phi(x, t) \ = \ \psi(xe^{-\mu t}) \ =\ 1 - a(xe^{-\mu t})^p, \ \ \
\text{as}\ \ \ xe^{-\mu t} \rightarrow 0 ,\ \ \ \ \text{with} \quad p\leq 1\, ,
\end{equation}
where $x = |\zeta|^2/2$ and $\mu$ and $\Theta$ are related by
\begin{eqnarray}\label{equz}
\mu &\ = &\ \frac{2}{3p^2}\ \ \ \ \text{and}\ \ \ \Theta \ =\
\frac{(3p + 1)(2 - p)}{3p^2} \, . \nonumber
\end{eqnarray}
Note that $p=1$ corresponds to initial states with finite energy. It was shown in \cite{powerLikeGB} for $\mathcal{T} = 0$ (i.e. cold
thermostat), the Fourier transform of the self-similar, isotropic solutions of \eqref{mixEq} is given by
\begin{eqnarray}\label{originalsolu}
\phi(x, t) &=& \frac{4}{\pi} \int_0^{\infty} \frac{1}{(1 + s^2)^2}
e^{-x e^{\frac{-2t}{3}} a s^2} ds \, ,
\end{eqnarray}
and its corresponding inverse Fourier transform, both for $p = 1$,
$\mu = \frac{2}{3}$ and $\Theta=\frac 43$ (as computed in
\cite{powerLikeGB}) is given by
\begin{equation}
f^{ss}_0(|v|, t)  = e^t F_0(|v|e^{t/3}) \quad \text{with} \quad F_0(|v|)  = \frac{4}{\pi} \int_0^{\infty} \frac{1}{(1 + s^2)^2} \frac{e^{-|v|^2/2s^2}}{(2\pi s^2)^{\frac{3}{2}}} ds .
\label{fTSolution}
\end{equation}
\\
{\bf Remark:} \textsl{It is interesting to observe that, as computed
originally in \cite{BCLT}, for $p=\frac{1}{3}$ or
$p = \frac{1}{2}$ in \eqref{equz} yields $\Theta=0$, and one can
construct explicit solutions to the elastic BTE with {\sl infinite
initial energy}. It is clear now that in order to have self-similar
explicit solutions with finite energy one needs to have this weakly
couple mixture model for slowdown processes, or bluntly speaking the
linear collisional term added to the elastic energy conservative
operator.}
\\
\\
Finally, in order to recover the self-similar solution for the
original equilibrium positive temperature ${\mathcal{T}}$ (i.e. hot thermostat
case) for the linear collisional term, we denote, including time
dependence for convenience,
\begin{eqnarray}
\phi_0(x, t) &\ =\ \phi(x, t)_{Thermostat = 0} \ \ \ \text{and}
\ \ \ \ \phi_{\mathcal{T}}(x, t) \ = \ \phi(x, t)_{Thermostat = {\mathcal{T}}} \nonumber \\
&\text{so that} \ \ \ \phi_{\mathcal{T}}(x, t) \ = \ \phi_0(x, t) e^{-{\mathcal{T}}x} \, .
\label{phi0T}
\end{eqnarray}
Note that the solution constructed in \eqref{originalsolu} is
actually $\phi_0(x, t)$. Then the self-similar solution for non zero background temperature, denoted by
$\phi_{\mathcal{T}}(x, t) $ satisfies
\begin{eqnarray}
\phi_{\mathcal{T}}(k, t) & = & \frac{4}{\pi} \int_0^{\infty} e^{ -|k|^2 e^{-2t/3}
a s^2/2} \frac{1}{(1 + s^2)^2} e^{-|k|^2 {\mathcal{T}}/2} ds \nonumber \\
& = & \frac{4}{\pi} \int_0^{\infty} e^{ -|k|^2 [e^{-2t/3} a s^2 + {\mathcal{T}}]/2 } \frac{1}{(1 + s^2)^2} ds .
\label{phiTEq}
\end{eqnarray}
In particular, let $\bar{T} = e^{-2t/3} a s^2 + {\mathcal{T}}$ then,
taking the inverse Fourier Transform, we obtain the corresponding
self-similar state, according to \eqref{ssrescale2}, in probability space 
\begin{eqnarray}
f^{ss}_{\mathcal{T}}(|v|, t)  = e^t F_{\mathcal{T}}(|v|e^{t/3})\ \text{with}\  F_{\mathcal{T}}(|v|) = \frac{4}{\pi} \int_{_{0}}^{^{\infty}} \hspace*{-.1cm}\frac{1}{(1 + s^2)^2} \frac{e^{-|v|^2/2\bar{T}}}{(2\pi \bar{T})^{\frac{3}{2}}} ds .
\label{fTSolution2}
\end{eqnarray}
Then, letting $t \rightarrow \infty$, since $\bar{T} = {\mathcal{T}} + a s^2
e^{\frac{-2t}{3}} \rightarrow {\mathcal{T}}$, yields
\begin{eqnarray}\label{fTSolution3}
F_{\mathcal{T}}(|v|) &{\to_{_{t\to\infty}}}& \ \frac{4}{\pi}\,\frac1{(2\pi
{{\mathcal{T}}})^{\frac{3}{2}}} {e^{-|v|^2/2{{\mathcal{T}}}}} \, \int_0^{\infty} \frac{1}{(1
+ s^2)^2} ds =\ M_{\mathcal{T}}(v) \, ,
\end{eqnarray}
\begin{equation}\label{fTSolution4}
\textrm{since}\qquad\qquad\ \ \frac{4}{\pi}\, \int_0^{\infty} \frac{1}{(1 + s^2)^2} ds \ =\ \frac{2}{\pi}\,\left( \frac{s}{1+s^2} + \arctan(s) \right) \arrowvert_0^\infty =\ 1 .
\end{equation}
So, the self-similar particle distribution $f^{ss}_{\mathcal{T}}(v,t) $ approaches a rescaled Maxwellian
distribution with the background temperature ${\mathcal{T}}$, that is according to \eqref{ssrescale2},
\begin{eqnarray}\label{fTSolution5}
f^{ss}_{\mathcal{T}}(|v|, t)&\ =\ e^t F_{\mathcal{T}}(|v|e^{t/3}) \approx \frac{e^t}{(2\pi
{{\mathcal{T}}})^{\frac{3}{2}}} \, e^{-(|v|^2\, e^{2t/3})/{2{\mathcal{T}}} + t} \, , \ \ \
\text{as} \ \ {t\to\infty} \ \, .
\end{eqnarray}
{\bf Remark:} \textsl{As pointed out in the previous remark, such  asymptotic behavior, for finite initial energy, is due to the balance of the binary term and the linear collisional term in \eqref{mixEq}.}
\\
\\
In addition, very interesting behavior is seen on $F_{\mathcal{T}}(|v|)$ as ${\mathcal{T}} \rightarrow
0$ (cold thermostat problem), where the particle distribution approaches a distribution with
power-like tails (i.e. a power law decay for large values of $|v|$)
and an integral singularity at the origin. Indeed, in \cite{powerLikeGB} an asymptotic behavior
is derived for $F_0(|v|)$ from \eqref{fTSolution}, for large and small values of $|v|$, leading to
\begin{eqnarray}
F(|v|) &=& 2 (\frac{2}{\pi})^{5/2} \frac{1}{|v|^6} [1 + O(\frac{1}{|v|})], \ \ \ \text{for} \ \ |v| \rightarrow \infty, \nonumber \\
F(|v|) &=& \frac{2^{1/2}}{\pi^{5/2}} \frac{1}{|v|^2} [1 + 2|v|^2 ln(|v|) + O(|v|^2)], \ \ \ \text{for} \ \ |v| \rightarrow 0 .
\label{ssAsym}
\end{eqnarray}
In particular the self-similar particle distribution function $F(|v|)$, $v\in\mathbb{R}^3$, behaves like $\frac{1}{|v|^6}$ as $|v|
\rightarrow \infty$, and as $\frac{1}{|v|^2}$ as $|v| \rightarrow 0$, which indicates a very anomalous, non-equilibrium behavior as
function of velocity; but, nevertheless, remains with finite mass and kinetic temperature. This asymptotic effect can be described as an overpopulated (with respect to Maxwellian), large energy tails and infinitely many particles at zero energy. This interesting, unusual behavior is observed in problems of soft condensed matter \cite{GreLe05}.
\\
\\
We shall see, then in the following section, that our solver captures these states with spectral accuracy and consequently the self similar solutions are attractors for a large class of initial states. These numerical tests are a crucial aspect of the spectral Lagrangian deterministic solver used to simulate this type of non-equilibrium phenomena, where all these explicit formulas for our probability distributions allow us to carefully benchmark the proposed numerical scheme.
\subsection{Self-Similar asymptotics for a general problem}
The self-similar nature of the solutions $F(|v|)$ for a general
class of problems, for a wide range of values for the parameters
$\beta$, $p$, $\mu$ and $\Theta$, was addressed in~\cite{BCG} with
much detail. Three different behaviors have been clearly explained.
Of particular interest for our present numerical study are the
mixture problem with a cold background and the inelastic Boltzmann
cases. Interested readers are referred to~\cite{BCG}.
\\
\\
For the purpose of our presentation, let $\phi = \mathcal{F}[f]$ be
the Fourier transform of the probability distribution function
satisfying the initial value problem
\eqref{singleEq}-\eqref{singleEq2} or \eqref{bte-source}. Let's
denote by $\Gamma(\phi) = \mathcal{F}[Q^+(f,f)] $ the Fourier
transform of the gain part of the collisional term associated with
the initial value problem. It was shown in that the
operator $\Gamma(\phi)$, defined over the Banach space of continuous
bounded functions with the $L^\infty$-norm (i.e. the space of
characteristic functions, that is the space of Fourier transforms of
probability distributions), satisfies the following three
properties~\cite{BCG}:
\begin{itemize}
\item[{\bf 1 - }]
$\Gamma(\phi)$ preserves the unit ball in the Banach space.
\item[{\bf 2 - }]
$\Gamma(\phi)$ is $L$-Lipschitz operator, i.e. there exists a bounded
linear operator $L$ in the Banach space, such that
\begin{equation}\label{llips}
|\Gamma(u_1) - \Gamma(u_2)|(x, t) \ \leq \ L(|u_1 - u_2|(x, t)),
\qquad \forall \qquad \| u_i \| \leq 1; i = 1, 2 \nonumber \,.
\end{equation}
\item[{\bf 3 - }] $\Gamma(\phi)$ is invariant under transformations
(dilations)
\begin{equation}\label{invdil}
e^{\tau\mathcal{D}} \Gamma (u) = \Gamma (e^{\tau\mathcal{D}} u)\
,\quad \mathcal{D} = x \frac{\partial}{\partial x}\ ,\quad
e^{\tau\mathcal{D}} u(x) = u(xe^\tau),\quad \tau\in\mathbb{R}^+\ .
\end{equation}
\end{itemize}
In the particular case of the initial value problem associated to
Boltzmann type of equations for Maxwell type of interactions, the bounded linear operator that
satisfies property {\bf 2}, is the one that linearizes the Fourier
transform of the gain operator about the state $u=1$.
\\
\\
Next, let $x^p$ be the eigenfunction corresponding to the eigenvalue
$\lambda(p)$ of the linear operator $L$ associated to $\Gamma$, i.e. $L(x^p) = \lambda(p) x^p$.
\\
\\
Define the {\sl spectral function associated to $\Gamma$}
given by $\mu(p) = \frac{\lambda(p) - 1}{p}$ defined for $p>0$. It
was shown in~\cite{BCG} that $\mu(0+) = +\infty$ (i.e. $p=0$ is a
vertical asymptote) and that for the problems associated to the
initial value problems \eqref{singleEq}-\eqref{singleEq2} or
\eqref{bte-source}, there exists a unique minimum for $\mu(p)$
localized at $p_0>1$, and that $\mu(p)\to 0^-$ as $p\to +\infty$.
\\
\\
Then, the existence of self-similar states and convergence of the solution to the initial value problem
to such self-similar distribution function was described in \cite{BCG} in the following four statements:
\\
\begin{itemize}
\item[{\bf (i)}] \textsl{\cite{BCG} - Lemma 5.1 (existence):} There exists a unique isotropic solution $f(|v|, t)$ to the initial
value problem \eqref{singleEq}-\eqref{singleEq2} or \eqref{bte-source} for Maxwell type interactions, in
the class of probability measures, satisfying $f(|v|, 0) = f_0(|v|)
\geq 0, \int_{\mathbb{R}^d} f_0(|v|) dv = 1$ such that for the
Fourier transform problem $x = \frac{|k|^2}{2}, u_0 =
\mathcal{F}[f_0(|v|)] = 1 + O(x)$, as $ x \rightarrow 0,$
\\
\\
\item[{\bf (ii)}] \textsl{Self similar states - Theorem 7.2:} $f(|v|, t)$ has self-similar asymptotics in the following sense:
\\
Taking the Fourier transform of the initial state to satisfy
\begin{equation}\label{init-ss}
u_0 + \mu(p)\ x^p \ u'_0 = \Gamma(u_0) + O(x^{p + \epsilon}),\ \ \text{such that} \ \ p + \epsilon <p_0 \, ,
\end{equation}
(i.e. $\mu(p); \mu'(p) < 0$). Then, there exists a unique,
non-negative, self-similar solution
\begin{equation} f^{ss}(|v|, t) =
e^{-\frac{d}{2}\mu(p)t}F_p(|v|e^{-\frac{1}{2}\mu(p)t}) \, , \nonumber
\end{equation}
with $\mathcal{F}(F_p(|v|)) = w(x), x = |k|^2/2$ s.t. $\mu(p) x^p w'(x) + w(x) = \Gamma(w)$.
\\
\\
\item[{\bf (iii)}] \textsl{Self similar asymptotics - Section 9 and Theorem 11.1 in \cite{BCG}:} There exists a unique (in the class of probability
measures) solution $f(|v|, t)$ satisfying $f(|v|,0) = f_0(|v|) \geq 0,$ with $\int_{_{\mathbb{R}^d}} f(|v|) dv = 1, $
such that, for $x = \frac{|k|^2}{2}$, and
\begin{equation} \mathcal{F}[f_0(|v|)]
= 1 - a\, x^p + O(x^{p + \epsilon}), x \rightarrow 0, 0 \leq p \leq
1 \, \ \ \text{with} \ p + \epsilon < p_0 \ \, . \nonumber
\end{equation}
\\
Then, for any given $0 \le p \leq 1$, there exists a unique non-negative
self-similar solution $f_{ss}^{(p)}(|v|, t) =
e^{-\frac{d}{2}\mu(p)t} F_p(|v|e^{-\frac{1}{2}\mu(p)t})$ such that
\begin{equation}\label{self-lim}
f(|v|, t) \rightarrow_{t\rightarrow \infty} e^{-\frac{d}{2}\mu(p)t} F_p(|v|e^{-\frac{1}{2}\mu(p)t}) \ \, .
\end{equation}
or equivalently
\begin{equation}\label{self-lim2}
e^{\frac{d}{2}\mu(p)t} f(|v|e^{\frac{1}{2}\mu(p)t}, t)
\rightarrow_{t \rightarrow \infty} F_p(|v|) \ \, ,
\end{equation}
where $\mu(p)$ is the value of spectral function associated to the
linear bounded operator $L$ as described above.
\\
\\
\item[{\bf (iv)}] \textsl{Power tail behavior of the asymptotic limit:} If $\mu(p) <0$, then the self-similar limiting
function $F_p(|v|)$ does not have finite moments of all orders.
\noindent In addition, if $0 \le p \le 1$ then all moments of order less than
$p$ are bounded; i.e. $m_q = \int_{\mathbb{R}^d} F_p(|v|)|v|^{2q} dv
\le \infty; 0 \le q \le p$. However, if $p = 1$ (finite energy case) then, the boundedness of
moments of any order larger than 1, depend on the conjugate
value of $\mu(1)$ by the spectral function $\mu(p)$. That means
$m_q \le \infty$ only for $0 \le q \le p_{*}$, where $p_{*} \ge p_0 >1$ is the unique maximal
root of the equation $\mu(p_{*}) = \mu(1)$.
\end{itemize}
\textbf{Remark 1:} \textsl{When $p = 1$, $\mu(1)$ is the energy dissipation rate, and $\mathcal{E}(t)= e^{\mu(1)t}$ the kinetic energy evolution function. So, $\mathcal{E}(t)^{d/2} f(v \mathcal{E}(t), t) \rightarrow F_1(|v|).$}
\\
\\
\textbf{Remark 2:} \textsl{We point out that condition \eqref{init-ss} on the
initial state is easily satisfied by taking a sufficiently
concentrated Maxwellian distribution as shown in~\cite{BCG}, and as
done for our simulations in the next section.}
\\
\\
However, rescaling with a different rate, it is not possible to pick
up the non-trivial limiting state $f^{ss}$, since
\begin{equation} f(|v|e^{\frac{1}{2}\eta t}, t) \rightarrow_{t\rightarrow \infty}
e^{-\frac{d}{2} \eta t}\delta_0(|v|); \hspace{0.15in} \eta > \mu(1) \, ,
\label{ssAsympIC_1}
\end{equation}
and
\begin{equation} f(|v|e^{\frac{1}{2}\eta t}, t) \rightarrow_{t\rightarrow \infty}
0; \hspace{0.15in} \mu(p_{min}) < \mu(1 + \delta) < \eta < \mu(1) \, .
\label{ssAsympIC_2}
\end{equation}
\begin{figure}
\centering
\includegraphics[height=3.5in, width=4.5in]{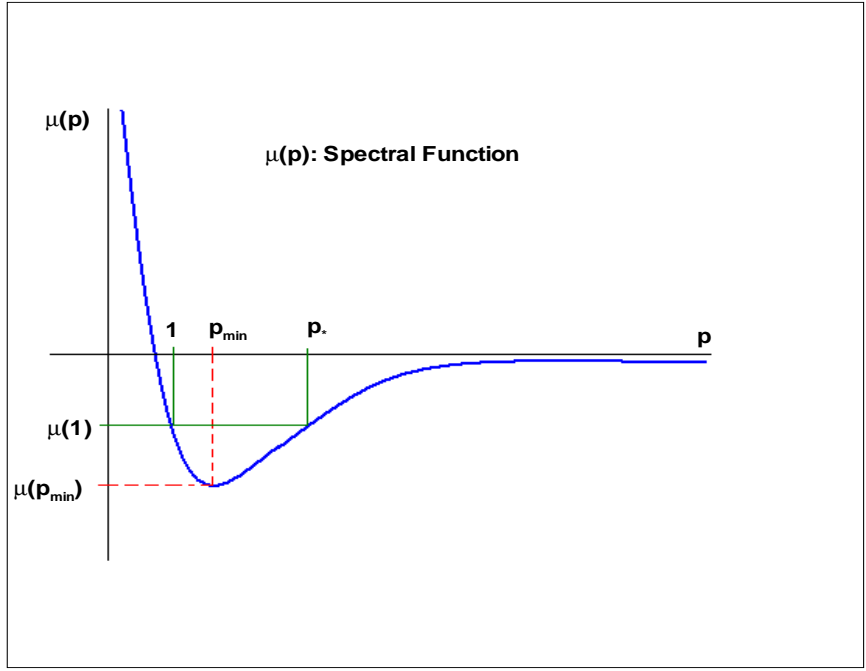}
\caption{Spectral Function $\mu(p)$ for a general homogeneous Boltzmann collisional problem of Maxwell type} \label{fig: mu_p}
\end{figure}
These results are also true for any $p \leq 1$. For the general space
homogeneous (elastic or inelastic) Boltzmann model of Maxwell type
or the corresponding mixture problem, the spectral function
$\mu(p)$ is given in Figure \ref{fig: mu_p}.

\section{Numerical Results}
We benchmark the new proposed numerical method to compute several
examples of $3-D$ in velocity and time for initial value problems associated with non-conservative models where some analysis is available, as are exact moment formulas for Maxwell type of
interactions as well as qualitative analysis for solutions of VHS
models. We shall plot our numerical results versus the exact
available solutions in several cases. Because all the computed problems converge to an isotropic long time state, we choose to plot the distribution function in only one direction, which is chosen to be the one with the initial anisotropies. All examples considered in this
manuscript are assumed to have isotropic, VHS collision kernels, i.e.
differential cross section independent from scattering angle.
We simulate the homogeneous problem associated to the following
problems for different choices of the parameters $\beta$ and $\lambda$,
and the Jacobian $J_{\beta}$ and heating force term $\mathcal{G}(f)$.
\subsection{Maxwell type of Elastic Collisions}
\begin{figure}
\centering
\begin{tabular}{cc}
\epsfig{file=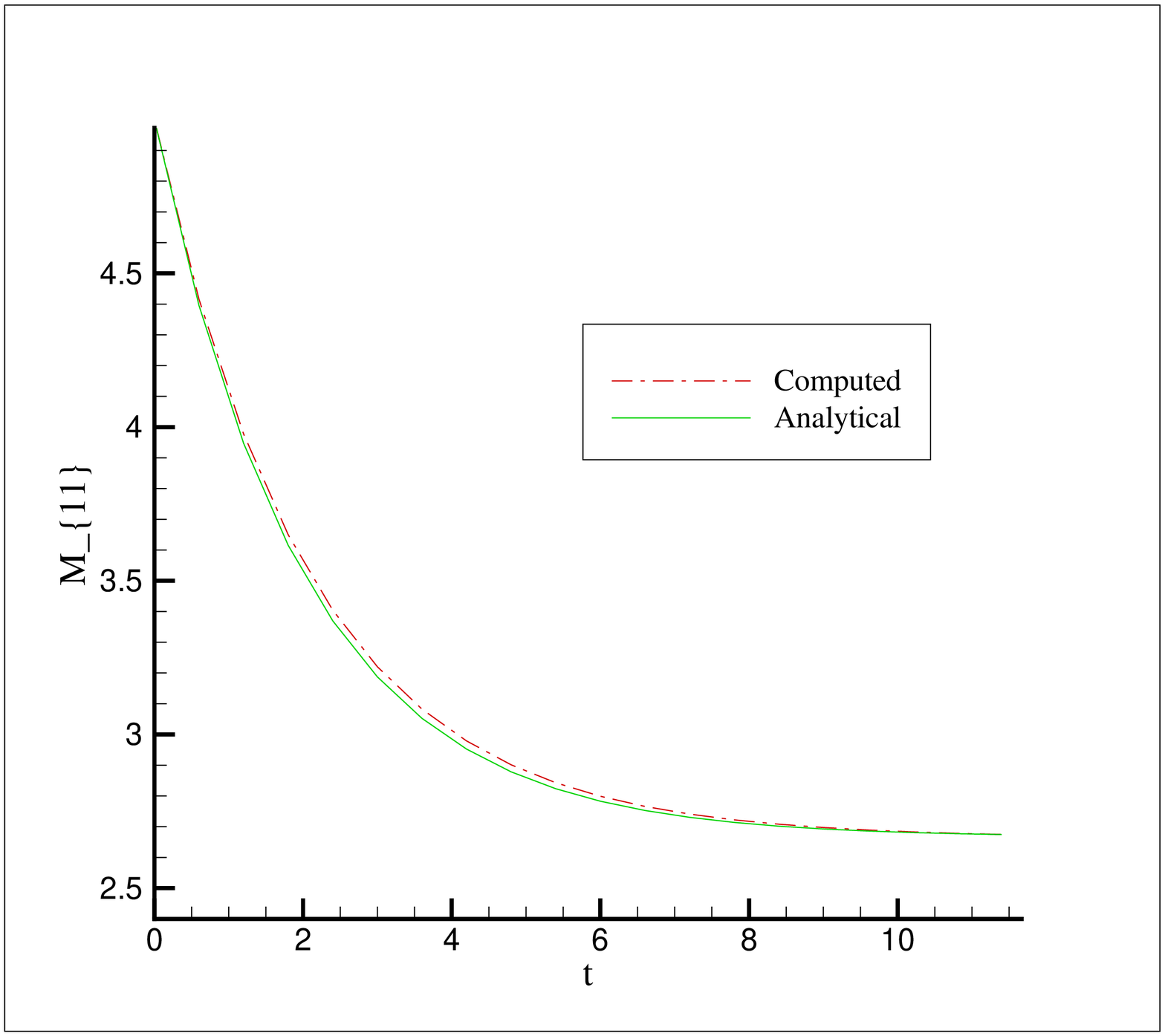,width=2.5in,height=2in} &
\epsfig{file=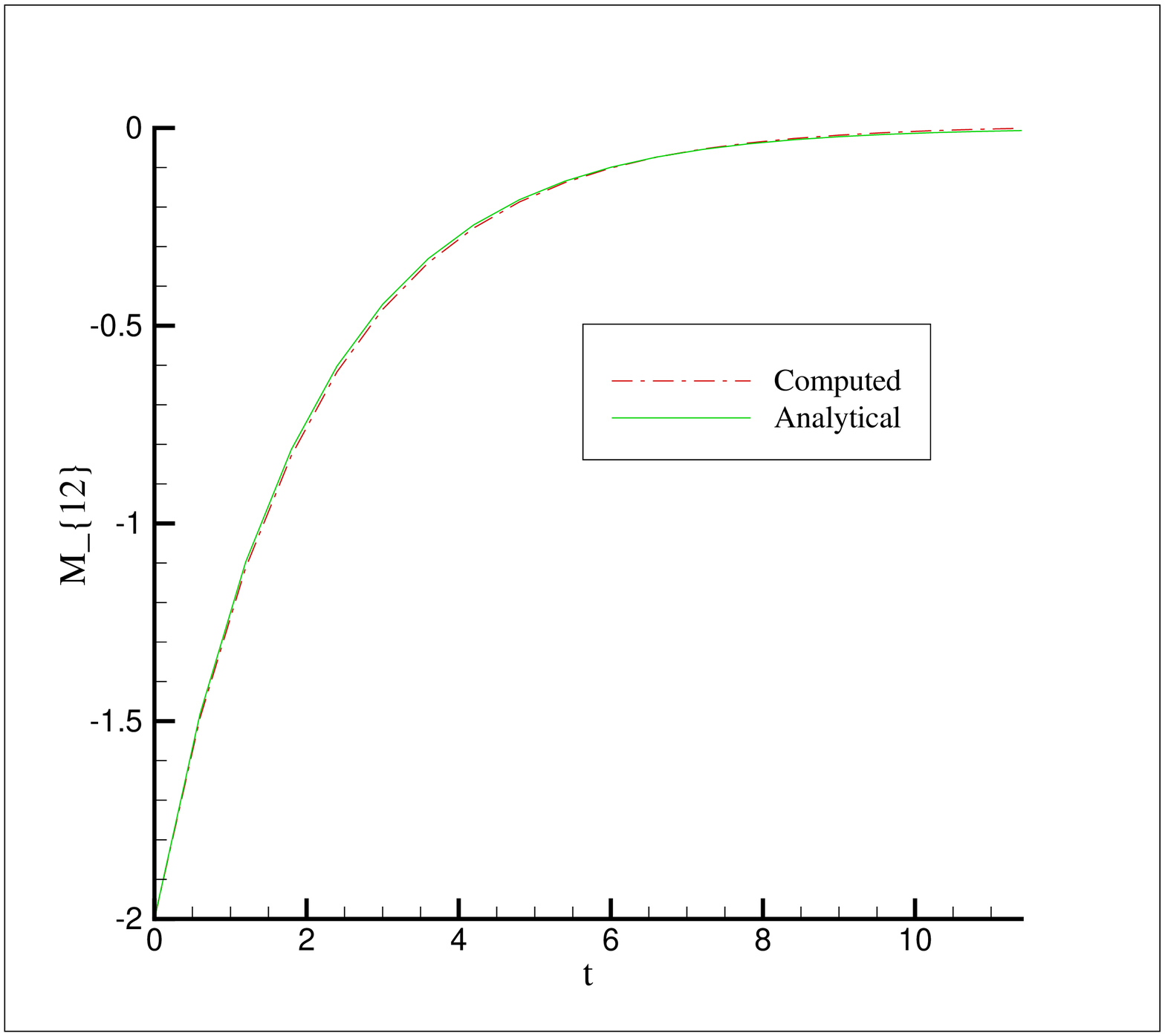,width=2.5in,height=2in} \\
($M_{11}$) & ($M_{12}$)\\
\epsfig{file=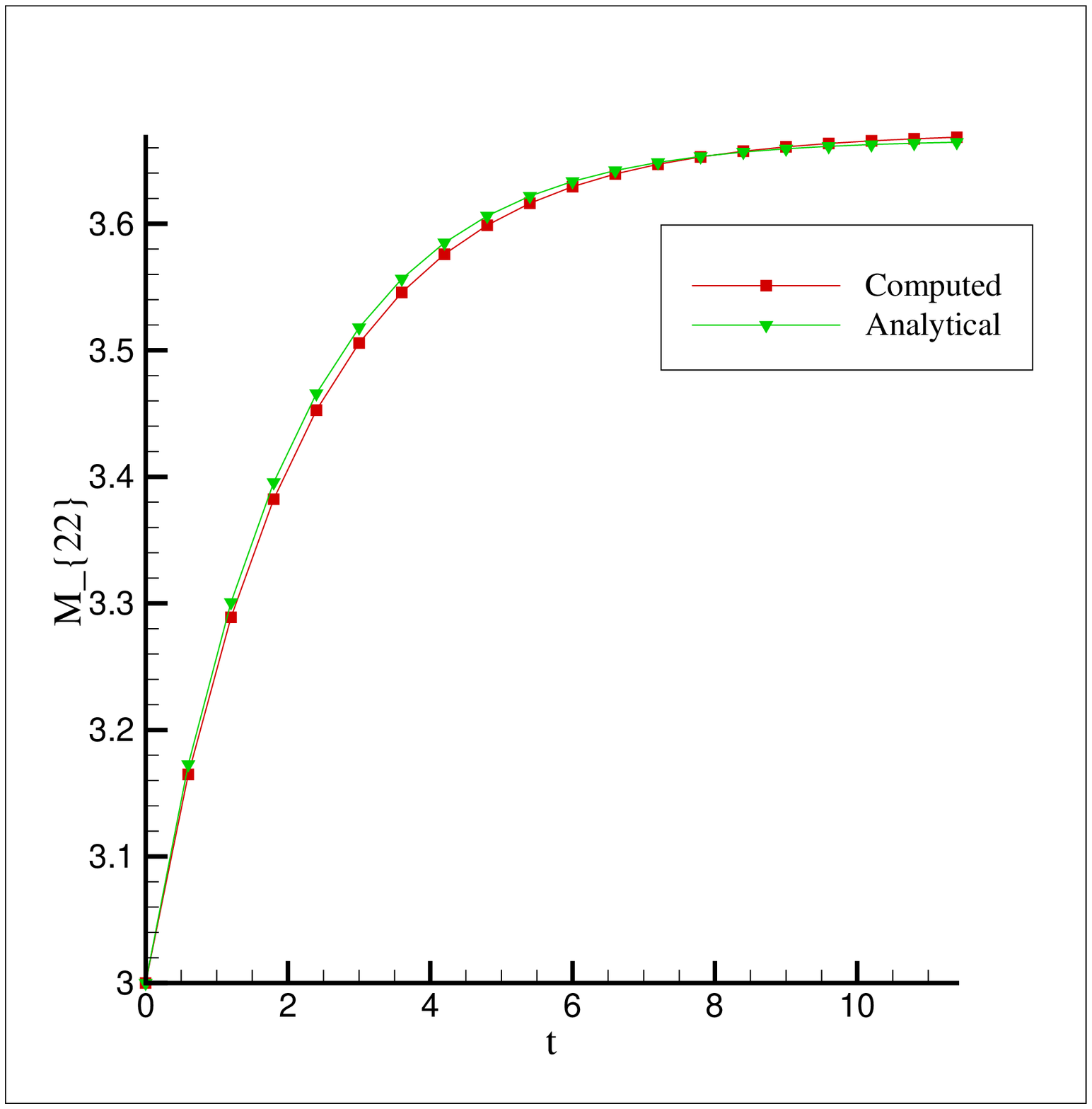,width=2.5in,height=2in} &
\epsfig{file=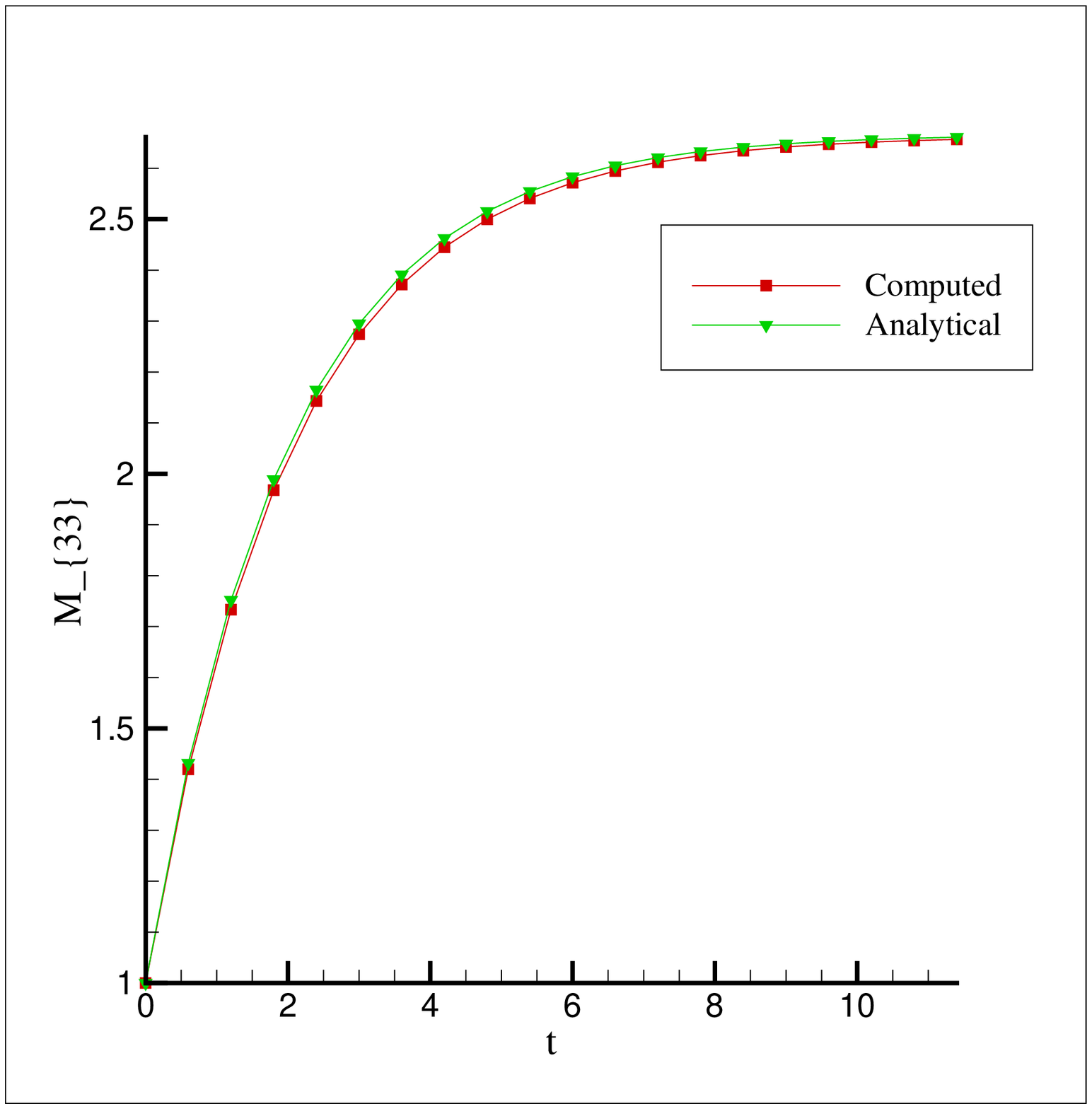,width=2.5in,height=2in} \\
($M_{22}$) & ($M_{33}$)\\
\epsfig{file=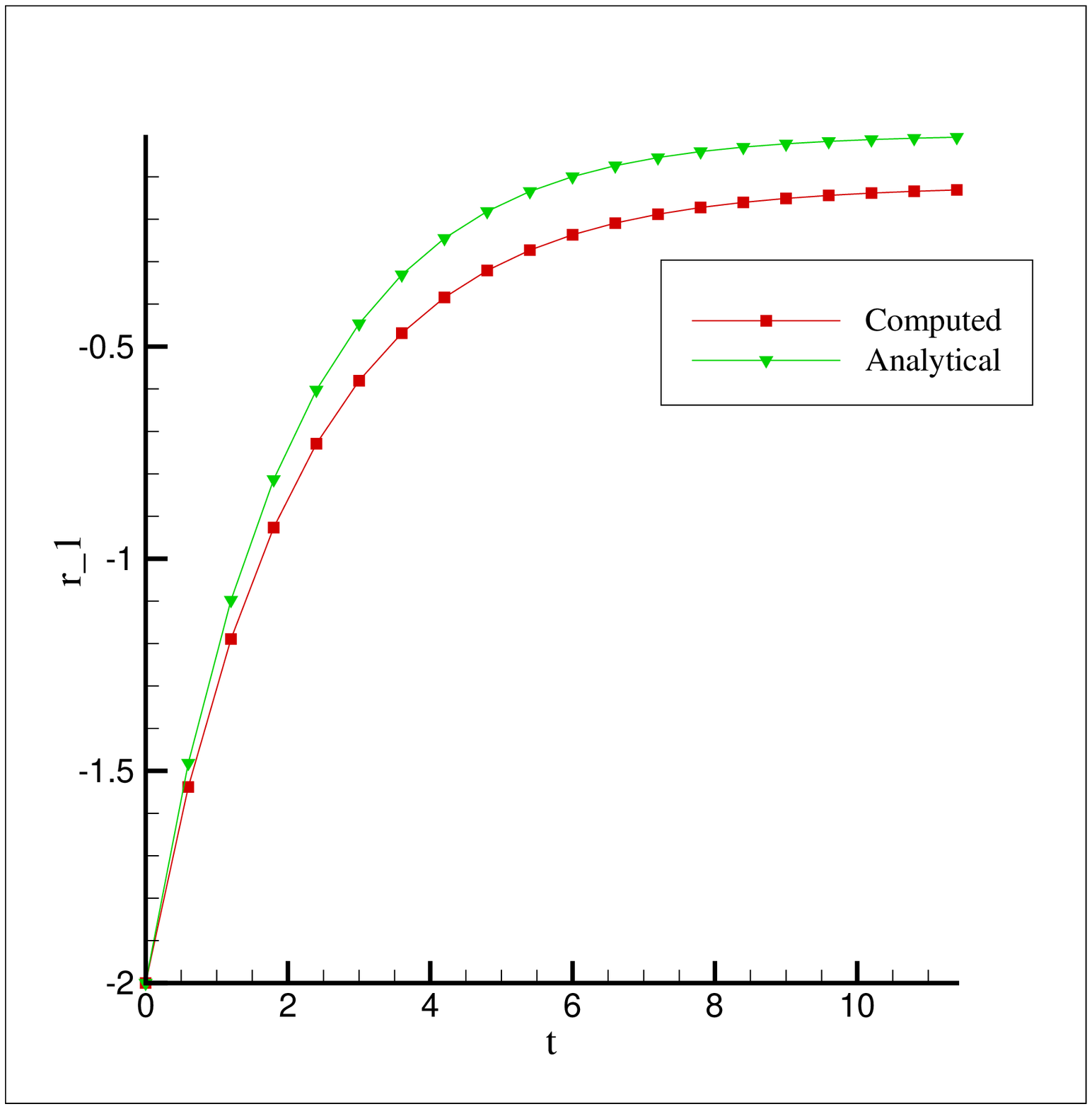,width=2.5in,height=2in} &
\epsfig{file=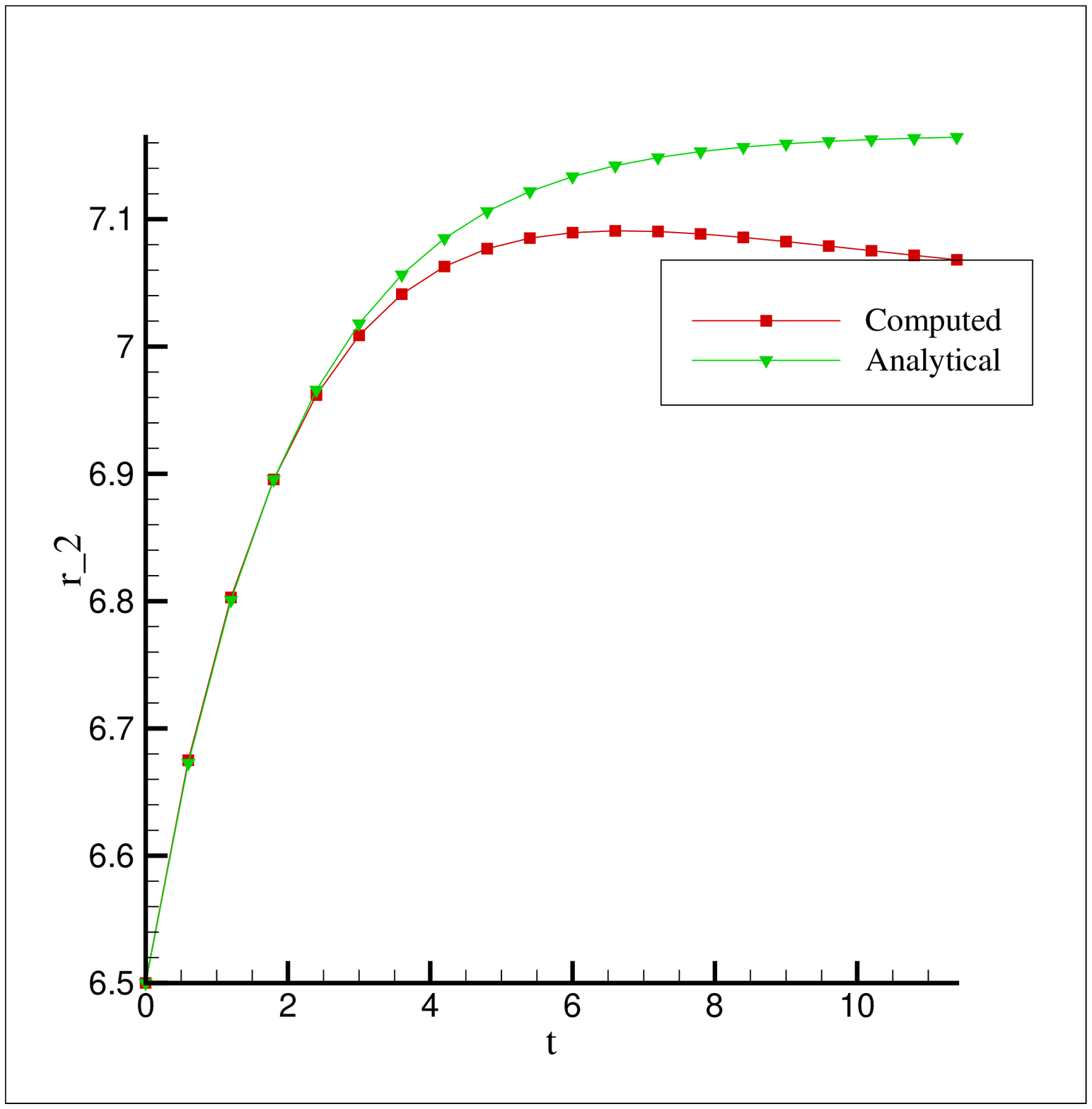,width=2.5in,height=2in} \\
($r_{1}$) & ($r_{2}$)\\
\label{fig:allMom}
\end{tabular}
\caption{Maxwell type of Elastic collisions: Momentum Flow $M_{11}, M_{12}, M_{22}, M_{33}$, Energy Flow $r_{1}, r_{2}$}
\end{figure}

\begin{figure}
\centering
\includegraphics[height=3in, width=3in]{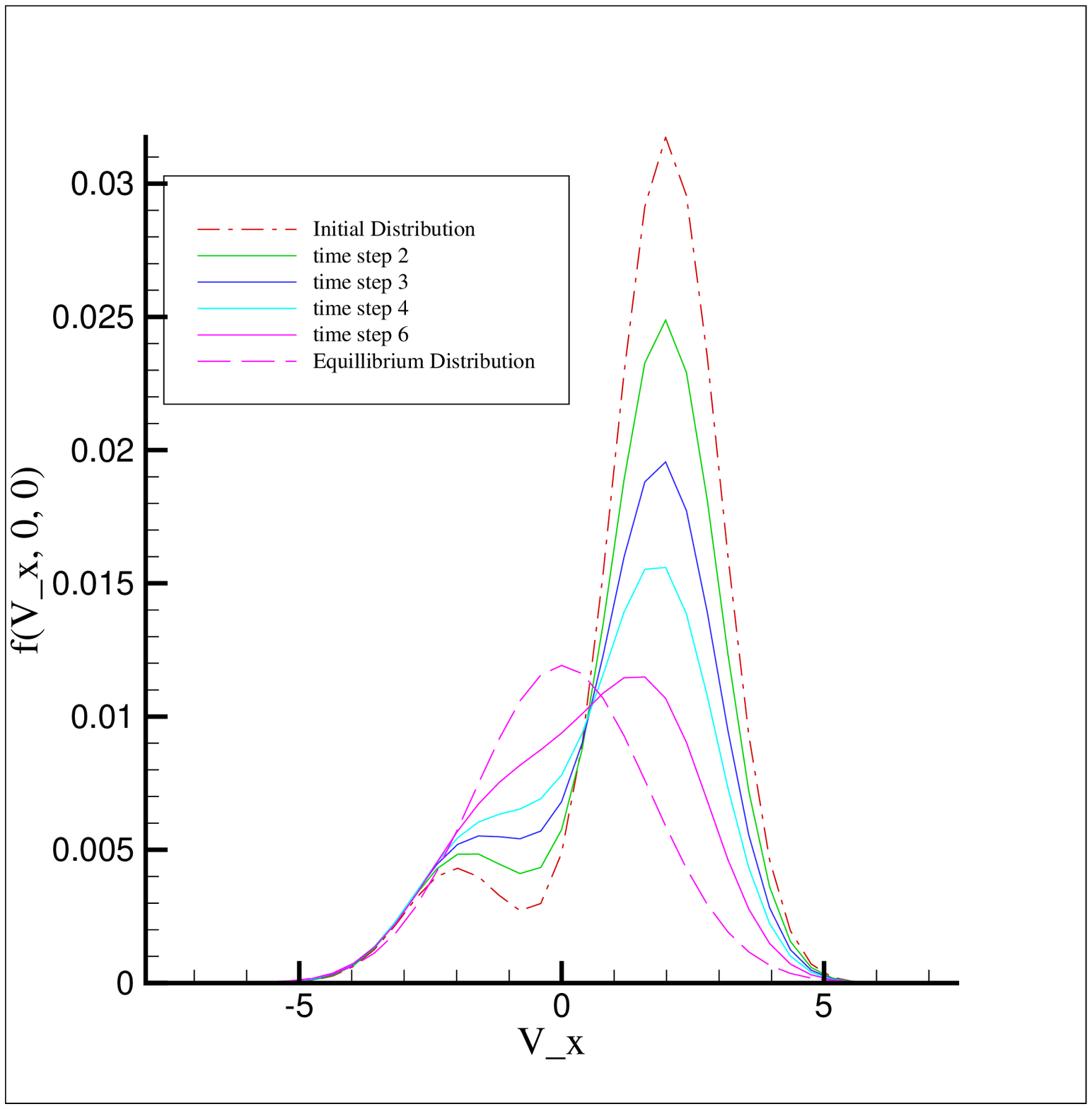}
\caption{Maxwell type of Elastic collisions: Evolution of the Distribution function}
\label{fig:evolFEl_40}
\end{figure}

Consider the initial value problem \eqref{singleEq}, \eqref{singleEq2}, with $B(|u|, \mu) = \frac{1}{4\pi}|u|^{\lambda}$. In \eqref{singleEq}, \eqref{singleEq2}, the value of the parameters are $\beta = 1, J_{\beta} = 1$ and $\lambda = 0$ with the pre-collision velocities defined from \eqref{singleEq2}.
In this case, for a general initial state with finite mass, mean and
kinetic energy, there is no exact expression for the evolving
distribution function. However there are exact expressions for all the
statistical moments (observables). Thus, the numerical method is
compared with the known analytical moments for different
discretizations in the velocity space.
\\
The initial states we take are convex combinations of two shifted
Maxwellian distributions. So consider the following case of initial
states with unit mass $\int_{\mathbb{R}^3} f_0(v) dv =1$ given by
convex combinations of shifted Maxwellians
\begin{eqnarray}
f(v, 0) = f_0(t) = \gamma M_{T_{1}}(v-V_1) + (1 - \gamma)M_{T_{2}}(v-V_2); \hspace{0.05in} \text{with} \hspace{0.05in} 0 \leqslant \gamma \leqslant 1 \nonumber
\end{eqnarray}
where $M_T(v-V) = \frac1{(2\pi T)^{3/2}}{e^{\frac{-|v - V|^2}{(2T)}}}$. Then, taking $\gamma = 0.5$ and mean fields for the initial state determined by
\begin{eqnarray*}
V_1  = [-2, 2, 0]^T \, , \ \ \ \ \ \ \ V_2 \ =\ [2, 0, 0]^T ; \ \ \ \ \ \ \ T_1  = 1 \ \ \ \ \ \ \, , T_2 \ =\ 1 \ \ \ \ \, ,
\end{eqnarray*}
enables the first five moment equations corresponding to the collision
invariants to be computed from those of the initial state. All higher
order moments are computed using the classical moments recursion
formulas for Maxwell type of interactions \eqref{moments3}. In particular, it is
possible to obtain the exact evolution of moments as functions of
time. Thus
\begin{equation*}
\rho(t) = \rho_0 = 1 \ \ \ \ \text {and} \qquad V(t) = V_0 = [0, 1, 0]^T \ \,.
\end{equation*}
By a corresponding moment calculation as in \eqref{moments3}, the complete evolution of the second moment tensor \eqref{moments2} is given by
\begin{equation*}
M(t) = \left( \begin{array}{ccc}
5 & -2 & 0 \\
-2 & 3 & 0 \\
0 & 0 & 1 \end{array}
\right)e^{-t/2} + \frac{1}{3}
\left( \begin{array}{ccc}
8 & 0 & 0 \\
0 & 11 & 0 \\
0 & 0 & 8 \end{array} \right)(1 - e^{-t/2}) \, ,
\end{equation*}
and the energy flow \eqref{moments2}
\begin{equation*} r(t) =
\frac{1}{2}\left( \begin{array}{c}
-4 \\
13 \\
0 \end{array} \right) e^{-t/3} + \frac{1}{6}\left( \begin{array}{c}
0 \\
43 \\
0 \end{array}\right) (1 - e^{-t/3}) -\frac{1}{6}\left( \begin{array}{c}
12 \\
4 \\
0 \end{array} \right) (e^{-t/2} - e^{-t/3}) \ \, ,
\end{equation*}
and the kinetic temperature is conserved, so
\begin{equation}
T(t) = T_0 = \frac{8}{3} \, .
\label{moments}
\end{equation}
The above moments along with their numerical approximations for
different discretizations in velocity space are plotted in Figures
\ref{fig:allMom}.
In Figure \ref{fig:evolFEl_40}, the evolution of the computed distribution function into a Maxwellian is plotted for $N = 40$.
\begin{figure}
\centering
\includegraphics[width=3in,height=3in]{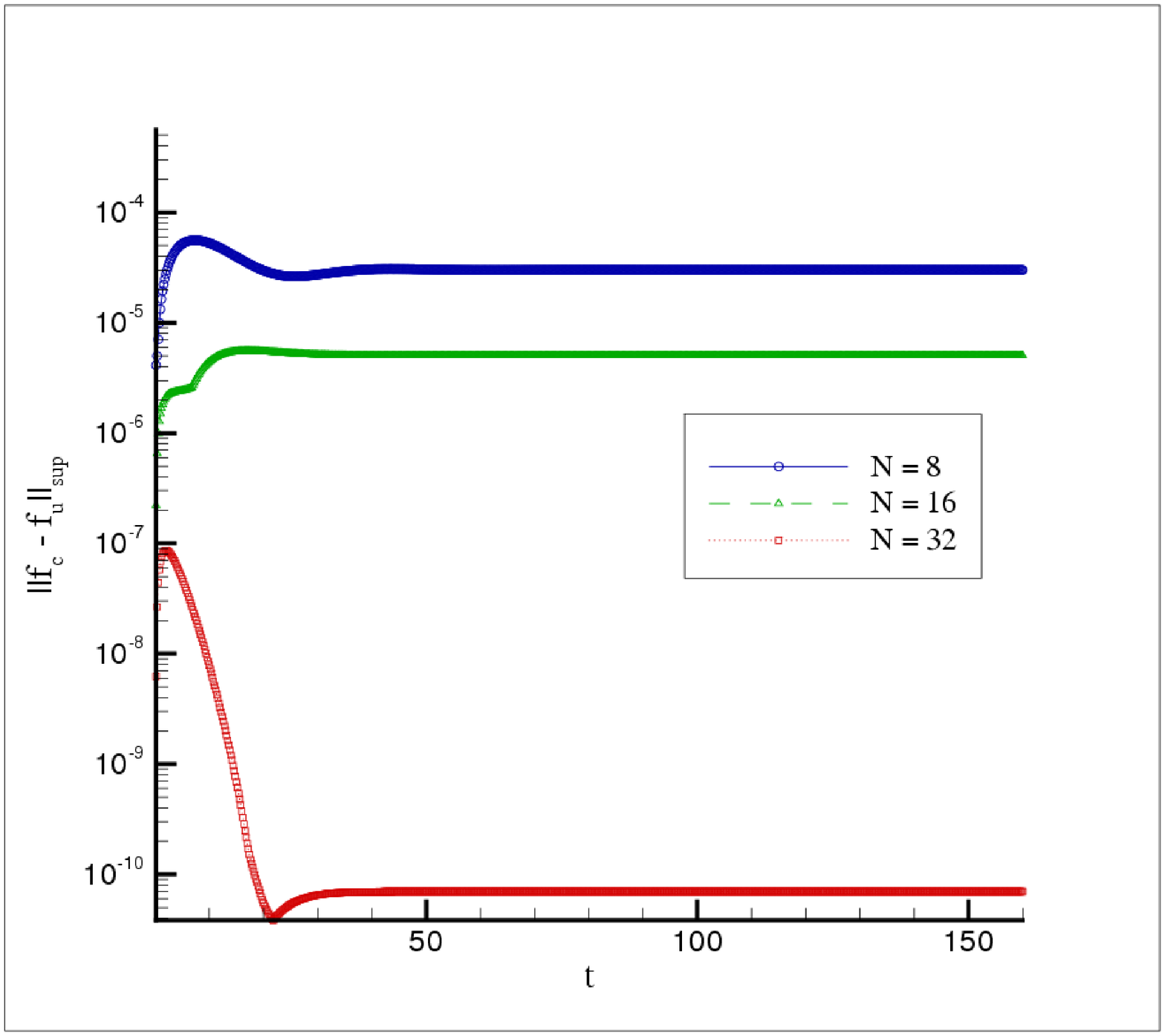}
\caption{Maxwell type Elastic Collisions: Conservation Correction for Elastic Collisions}
\label{fig:consError}
\end{figure}
In order to check the conservation accuracy of the method, let
$f_u$ - unconserved distribution given as input to the conservation
routine and $f_c$ - conserved distribution resulting from the
conservation routine. With a convex combination of two Gaussians as
input, the numerical method is allowed to run and $\| f_c - f_u
\|_{\infty}$ is plotted for all times for different values of $N$ in
figure {\ref{fig:consError}}. As expected, for $t$ approaching the final time,
the largest value of $N$ gives the smallest conservation correction.
\subsection{Maxwell type of Elastic collisions - Bobylev-Krook-Wu (BKW) Solution}
\begin{figure}[h]
\centering
\begin{tabular}{cc}
\epsfig{file=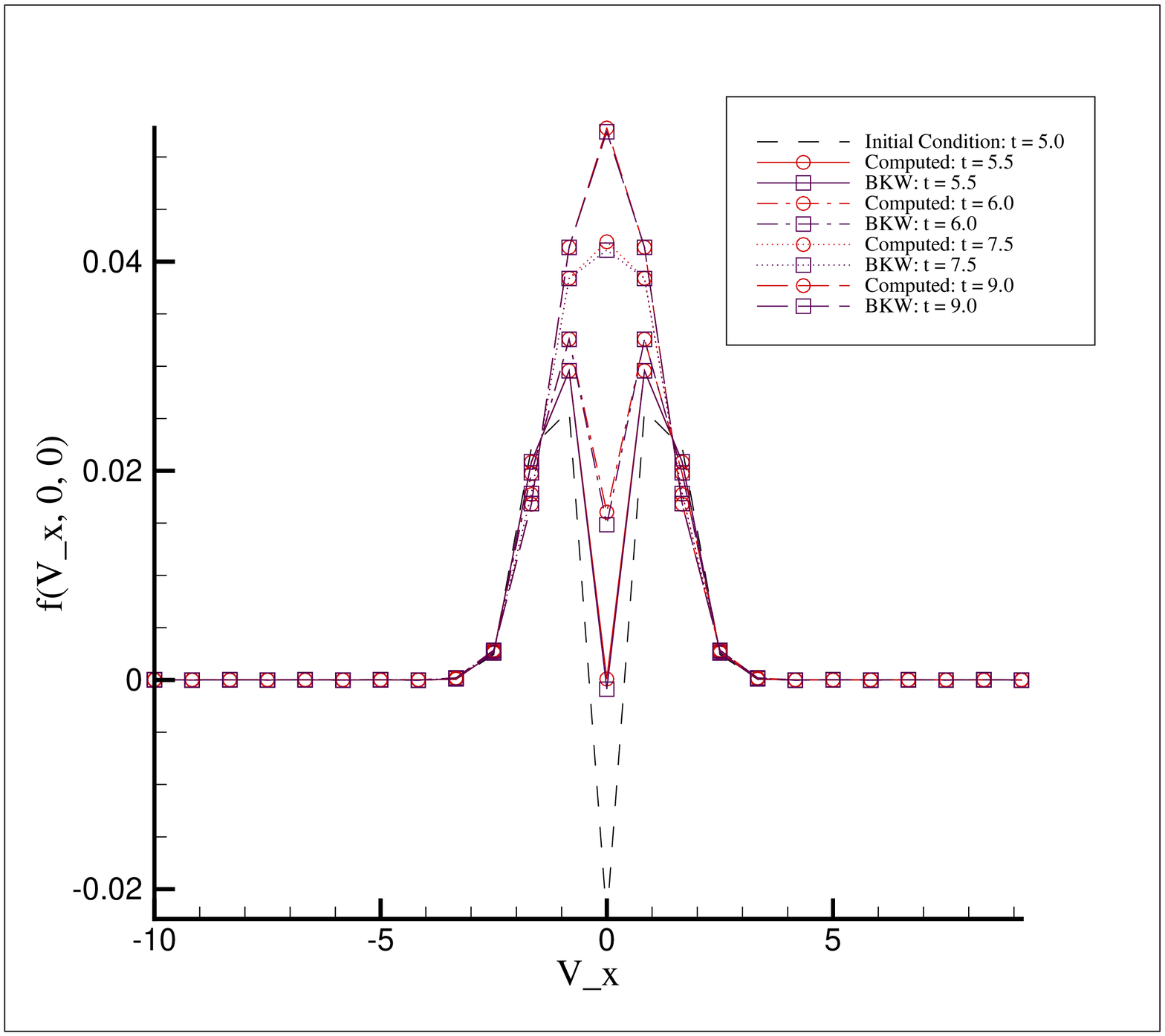,width=2.5in,height=2.5in} &
\epsfig{file=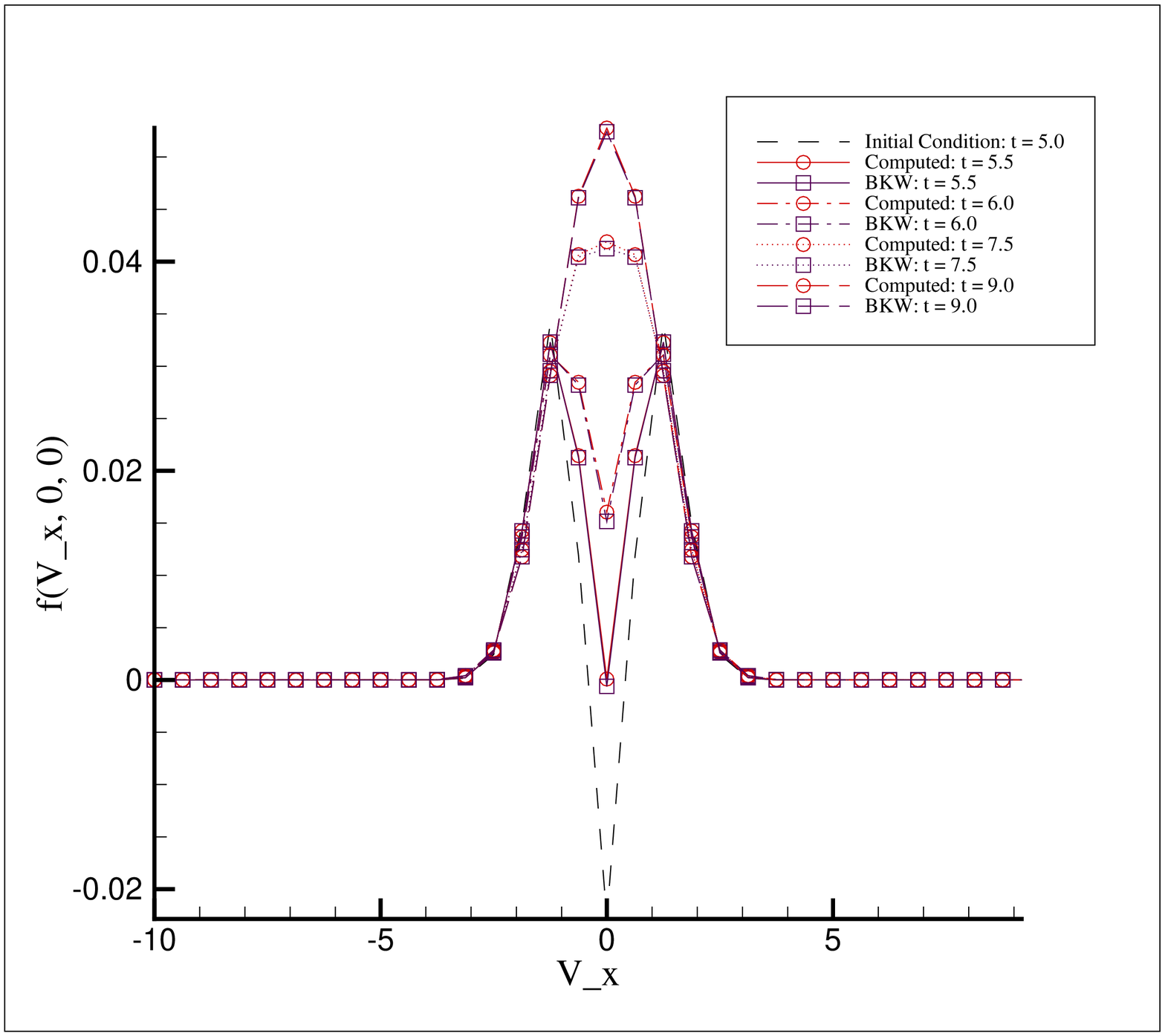,width=2.5in,height=2.5in} \\
($N = 24$) & ($N = 32$)\\
\end{tabular}
\caption{BKW, $\rho, E(t)$ conserved}
\label{fig:bkw}
\end{figure}

An explicit solution to the initial value
problem \eqref{singleEq} for elastic, Maxwell type of interactions
($\beta=1, \lambda=0$)
was derived in~\cite{Bob75} and independently in \cite{KW} for
initial states that have at least $2+\delta$-moments bounded. It is
not of self-similar type, but it can be shown to converge to a
Maxwellian distribution. This solution takes the form
\begin{equation}
f(v, t) = \frac{e^{-|v|^2/(2K\eta^2)}}{2(2\pi K \eta^2)^{3/2}} (\frac{5K - 3}{K} + \frac{1 - K}{K^2} \frac{|v|^2}{\eta^2}) \, ,
\label{bkw}
\end{equation}
where $K = 1 - e^{-t/6}$ and $\eta = $initial distribution
temperature. It is interesting that it is negative for small values
of $t$. So in order to obtained a physically meaning probability
distribution, $f$ must be non-negative. This is indeed the case for
any $K \geqslant \frac{3}{5}$ or $t \geqslant t_0 \equiv 6
ln(\frac{5}{2}) \sim 5.498$. In order to test the accuracy of
our solver, set the initial distribution function to be the BKW
solution, the numerical approximation to the BKW solution and the
exact solution are plotted for different values of $N$ at various time steps
in Figure \ref{fig:bkw}.
\subsection{Hard-Sphere Elastic Collisions}
\begin{figure}
\centering
\begin{tabular}{cc}
\epsfig{file=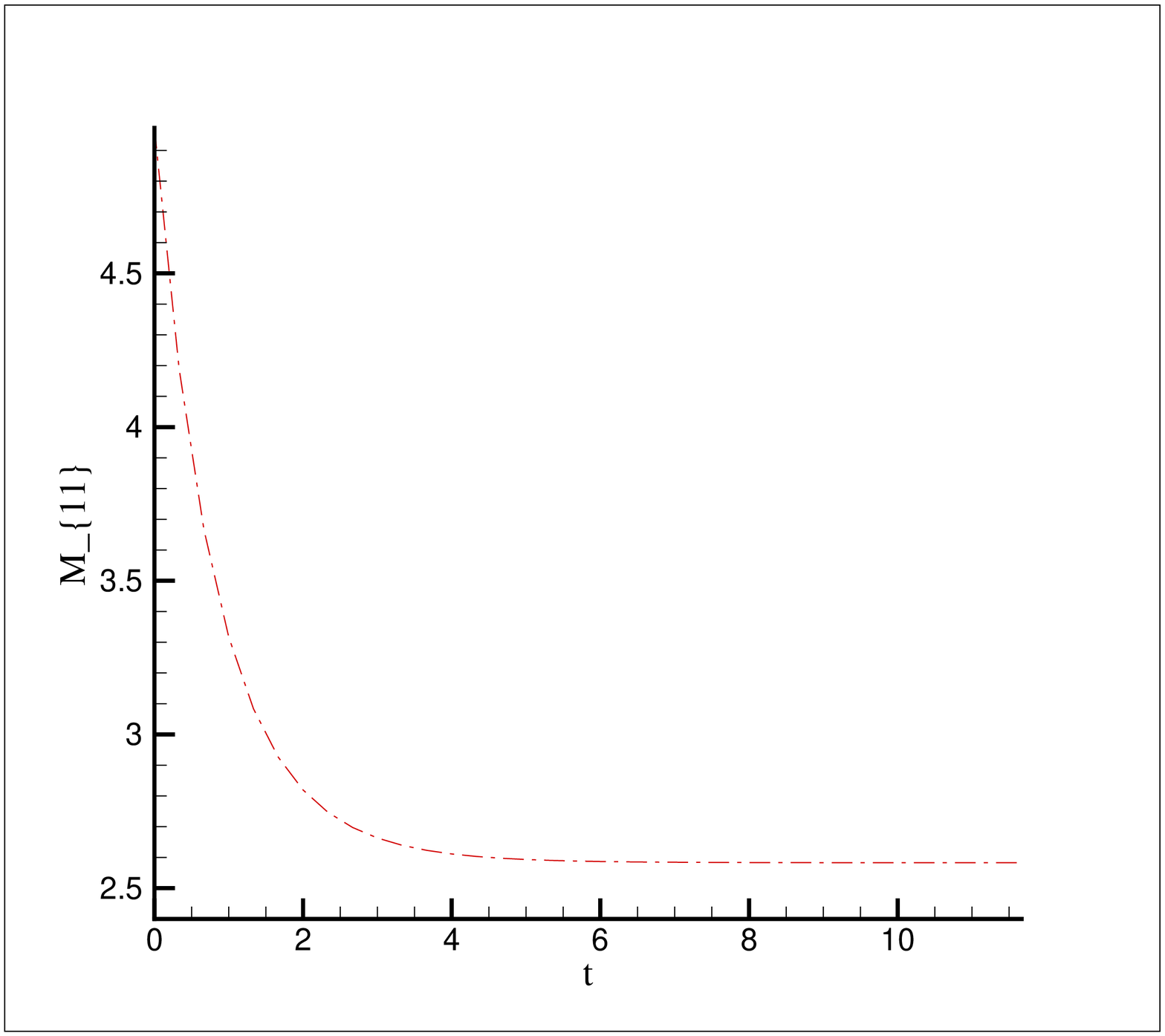,width=2.5in,height=2in} &
\epsfig{file=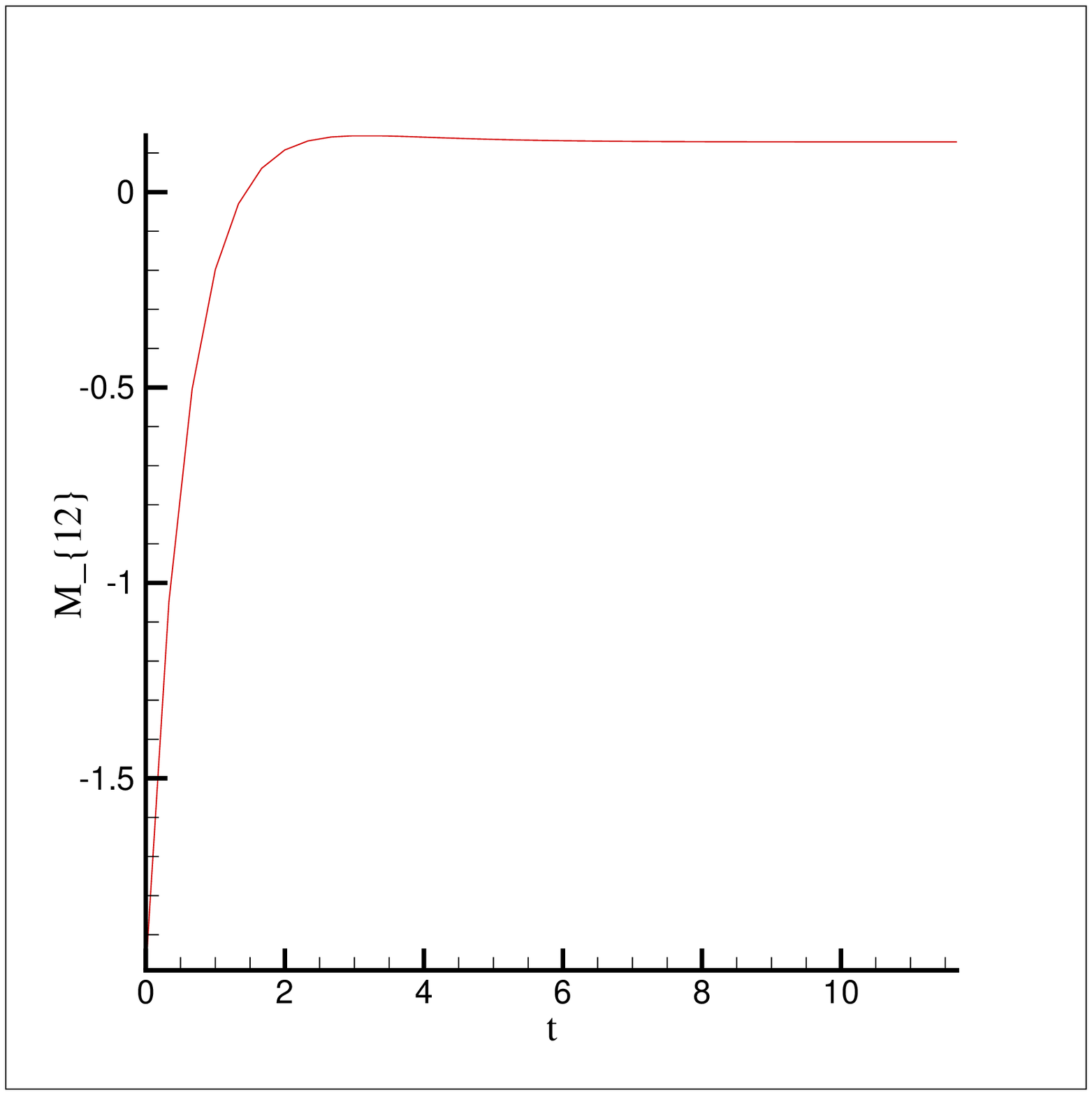,width=2.5in,height=2in} \\
($M_{11}$) & ($M_{12}$)\\
\epsfig{file=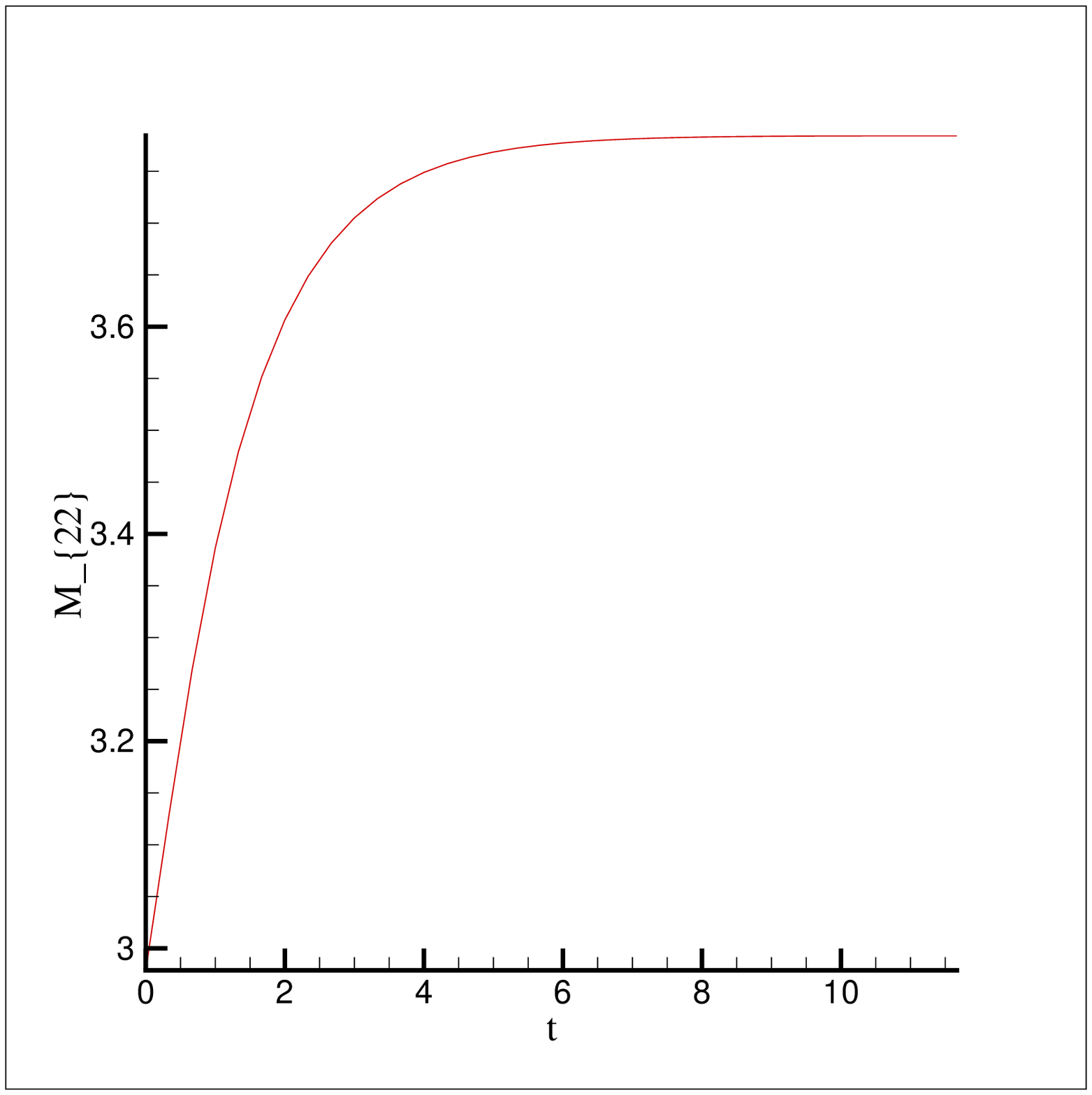,width=2.5in,height=2in} &
\epsfig{file=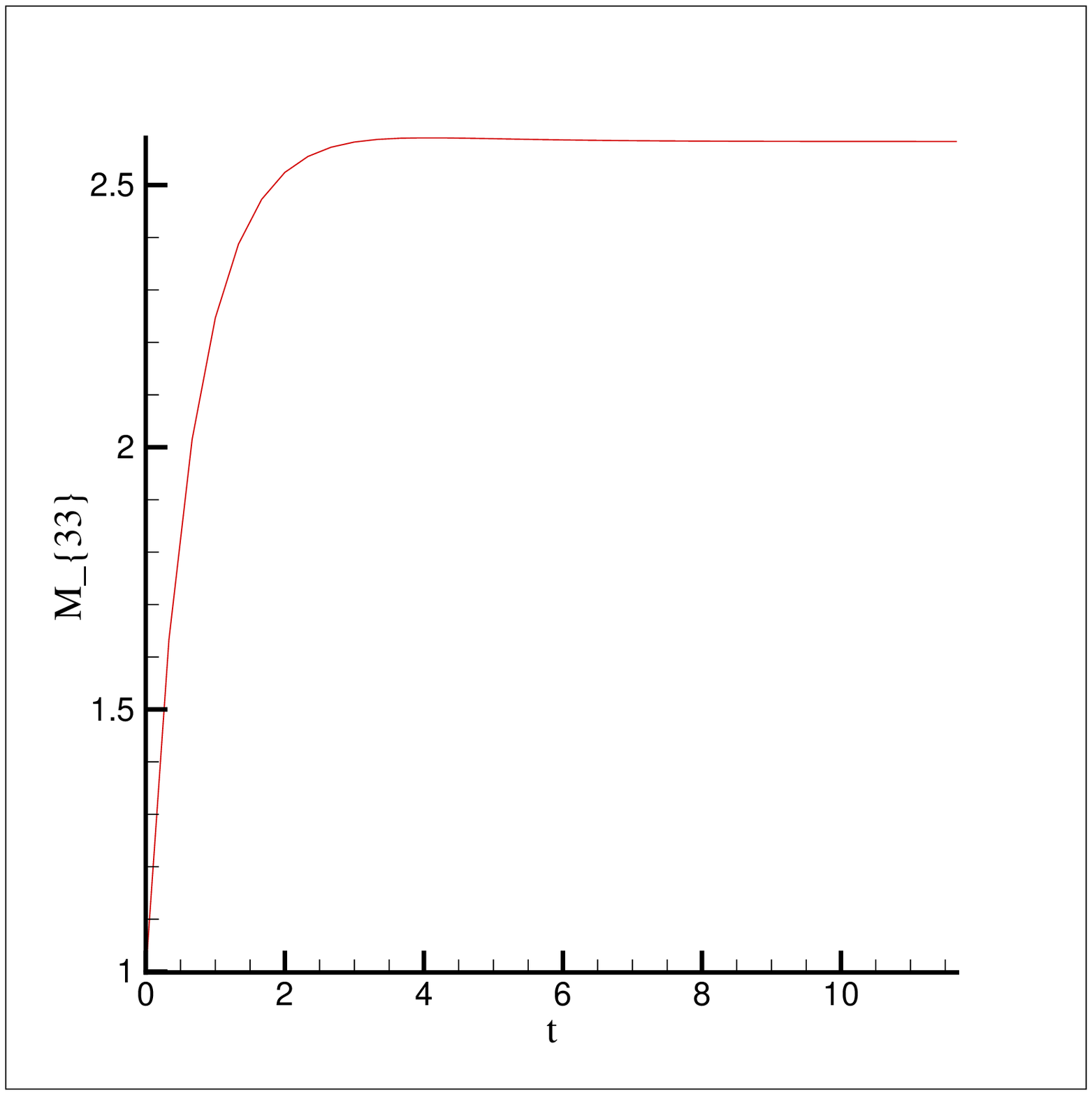,width=2.5in,height=2in} \\
($M_{22}$) & ($M_{33}$)\\
\epsfig{file=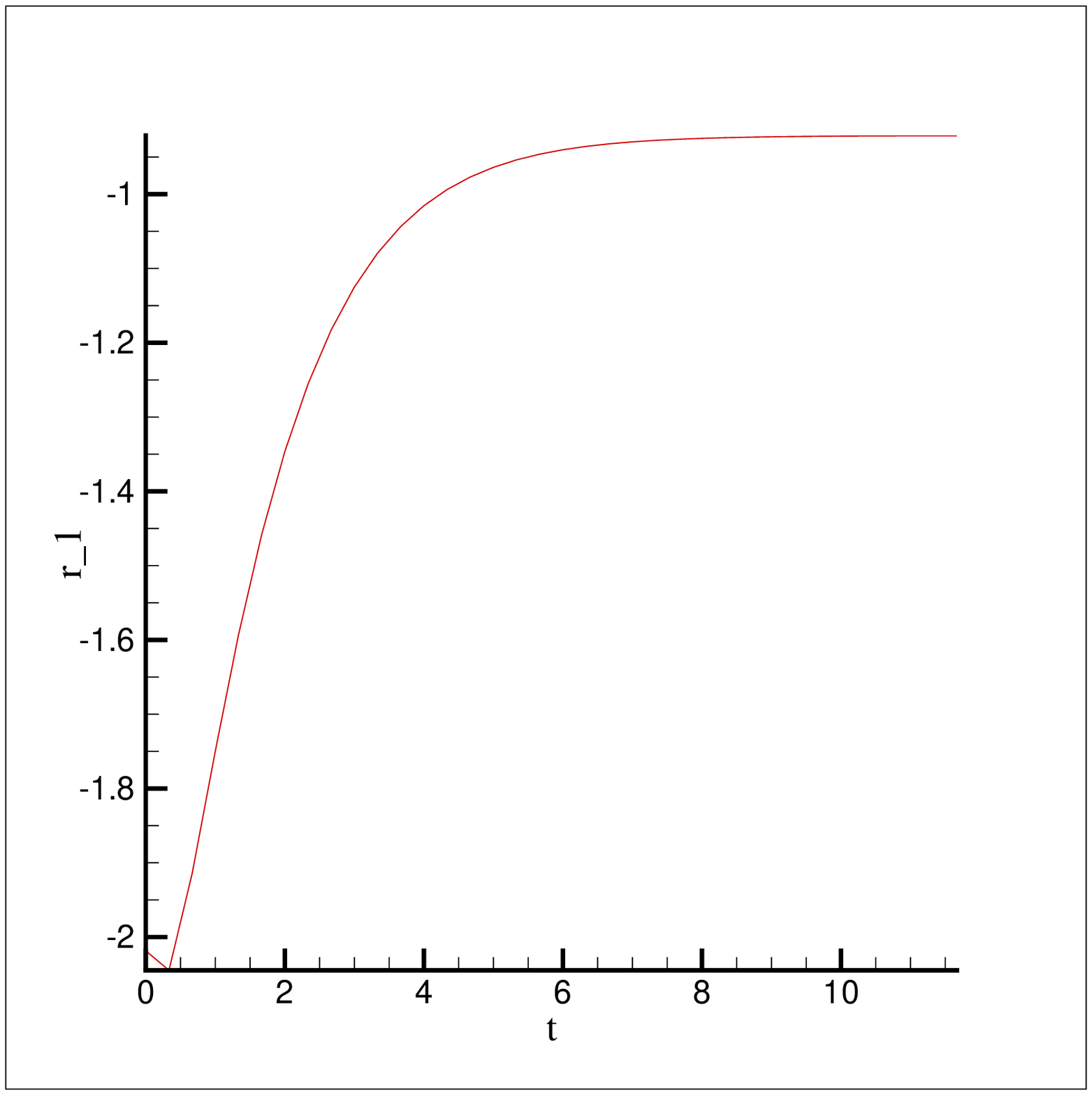,width=2.5in,height=2in} &
\epsfig{file=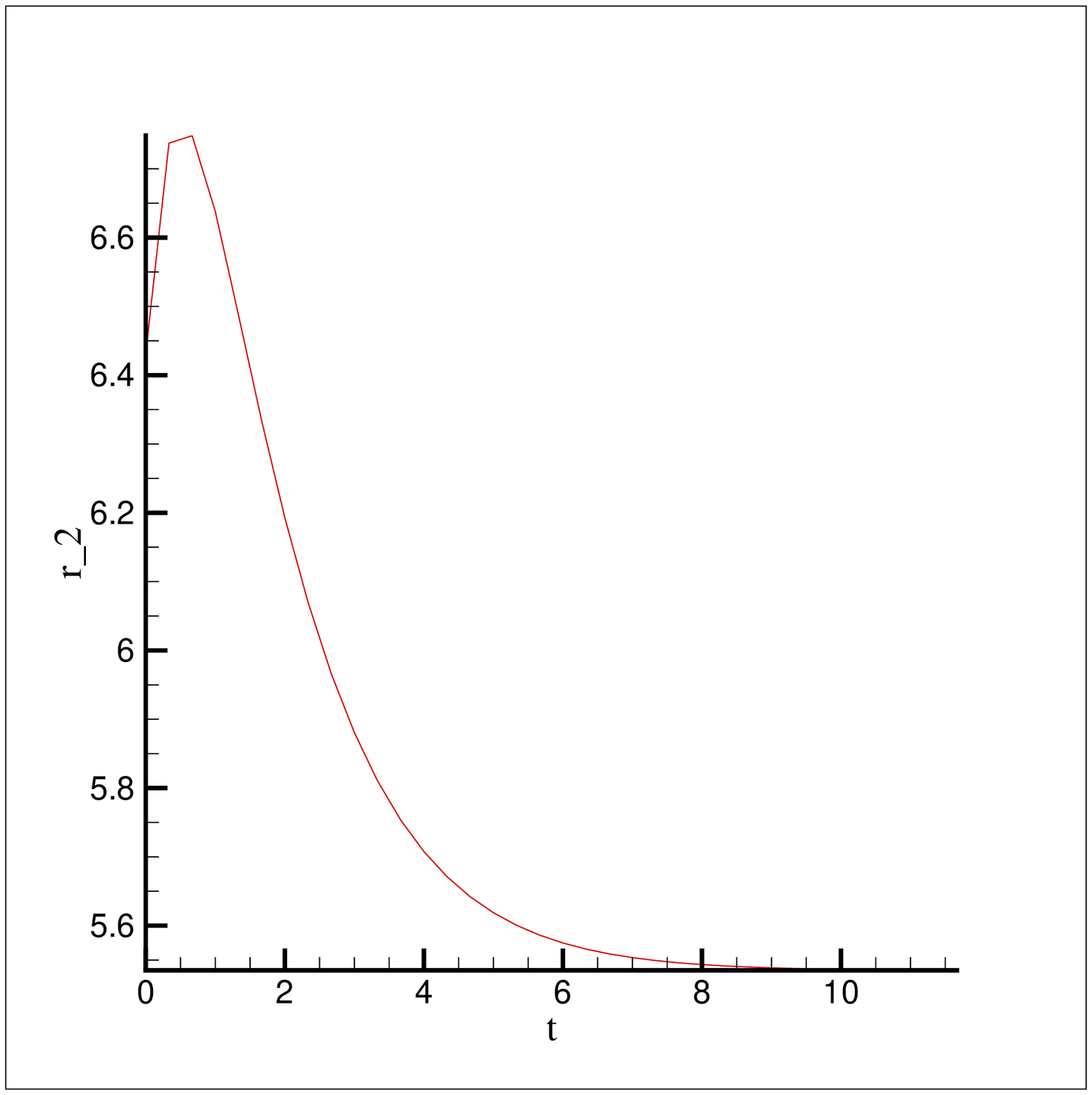,width=2.5in,height=2in} \\
($r_1$) & ($r_2$)\\
\end{tabular}
\caption{Hard Sphere, Elastic: Momentum Flow $M_{11}, M_{12}, M_{22}, M_{33}$, Energy Flow $r_1, r_2$}
\label{fig:HSAllMom}
\end{figure}

\begin{figure}
\centering
\includegraphics[width=2.75in,height=2.75in]{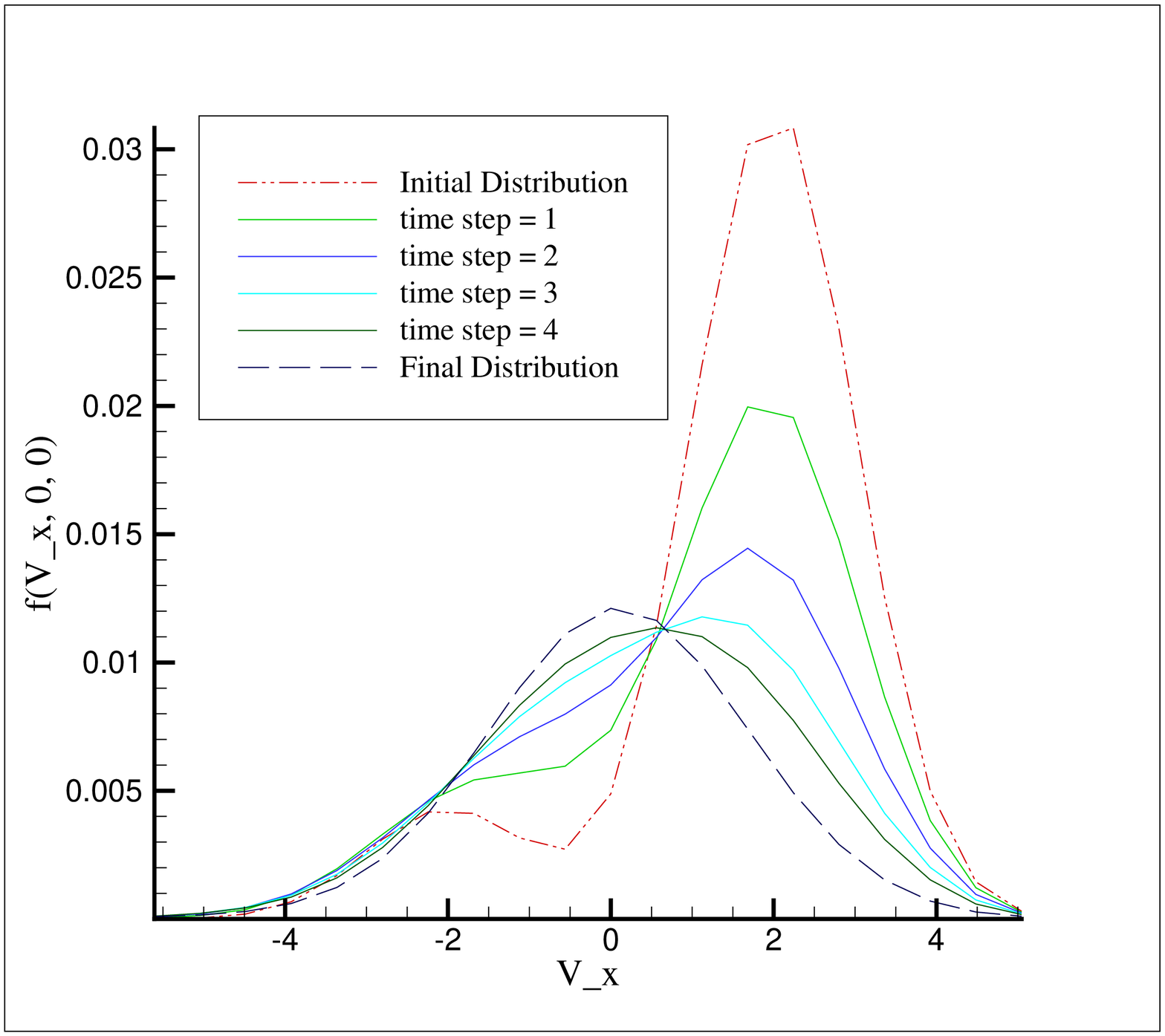}
\caption{Hard-Sphere, Elastic: Evolution of the Distribution function, N = 32}
\label{fig:evolFHS_32}
\end{figure}

In \eqref{singleEq}, \eqref{singleEq2}, we have $\beta = 1, J_{\beta} = 1$ and $\lambda = 1$ with the post-collision velocities defined from \eqref{singleEq}.
Unlike Maxwell type of interactions, there is no explicit expression
for the moment equations and neither is there any explicit solution
expression as in the BKW solution scenario. For Hard Sphere
isotropic collisions, the expected behavior of the moments is similar to that of the Maxwell type of interactions case
except that in this case, the moments somewhat evolve to the
equilibrium a bit faster than in the former case i.e. figure \ref{fig:HSAllMom}.
Also plotted is the time evolution of the distribution function
starting from the convex combination of Maxwellians as described in
a previous subsection in Figure \ref{fig:evolFHS_32}.
\subsection{Inelastic Collisions}
\begin{figure}
\centering
\begin{tabular}{cc}
\epsfig{file=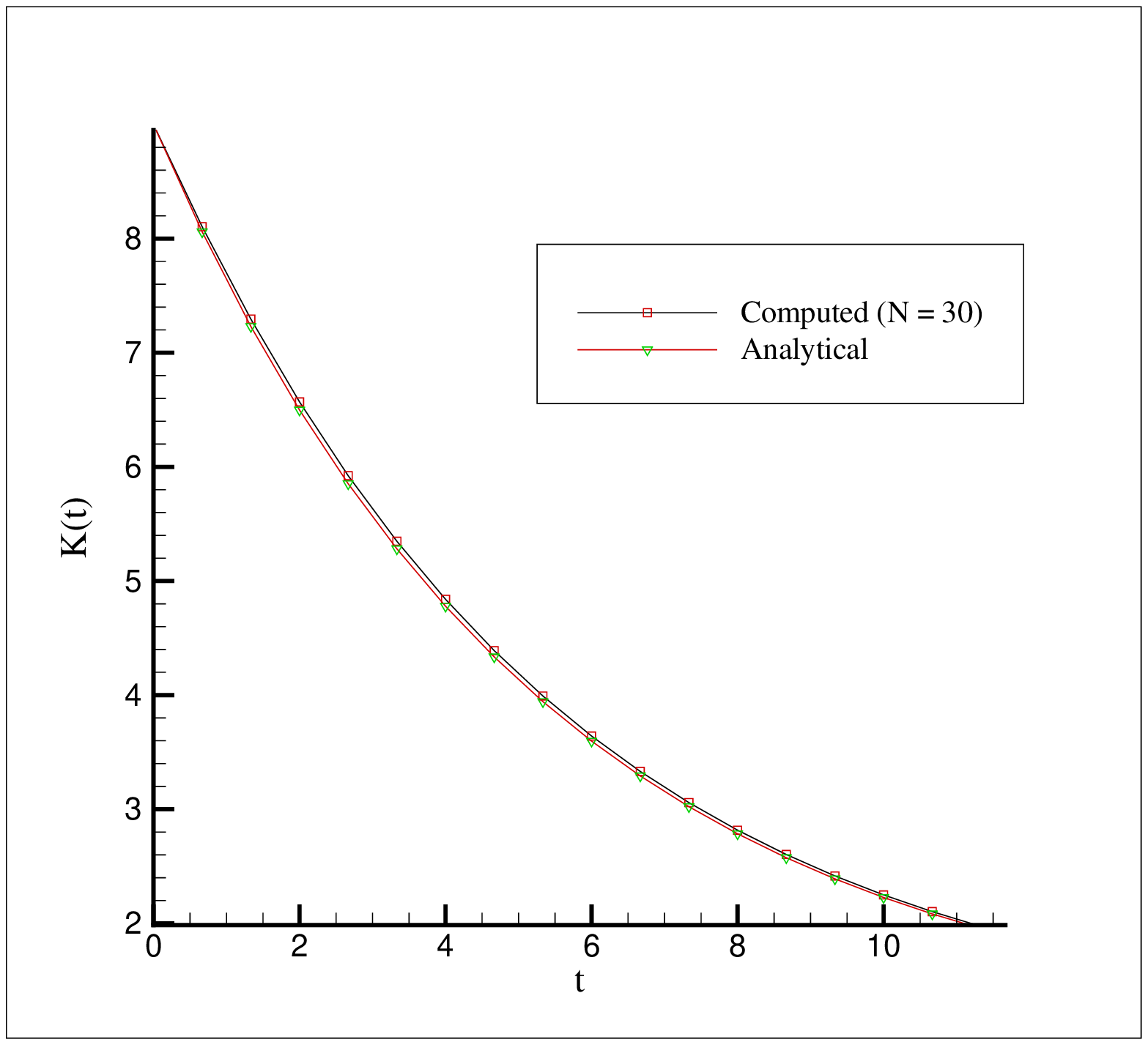,width=2.5in,height=2.75in} &
\epsfig{file=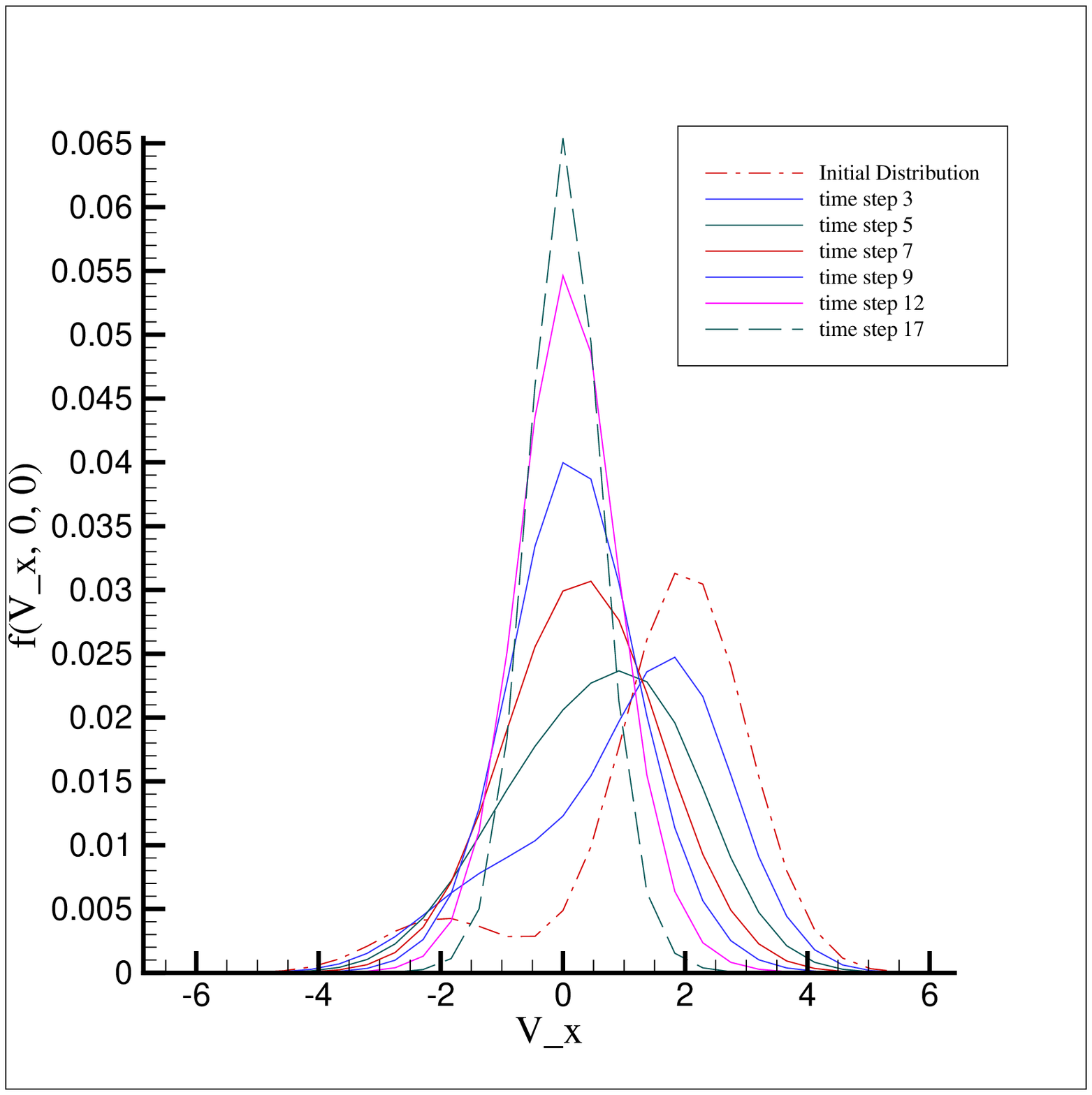,width=2.5in,height=2.75in} \\
\end{tabular}
\caption{Inelastic: Kinetic Energy (left) \& $f(v, t)$ (right)}
\label{fig:kinEnergy_fEvol}
\end{figure}

This is the scenario wherein the utility of the proposed method is
clearly seen. No other deterministic method can compute the
distribution function in the case of inelastic collisions
(isotropic), but the current method computed this $3-D$ evolution without much
complication and with the exactly same number of operations as used
in an elastic collision case. This model works for all sorts of
Variable Hard Potential interactions. Consider the special case of
Maxwell ($\lambda = 0$) type of inelastic ($\beta \neq 1$)
collisions in a space homogeneous Boltzmann Equation in \eqref{singleEq}, \eqref{singleEq2}.
Let $\phi(v) = |v|^2$ be a smooth enough test function. Using the weak form of the Boltzmann equation with such a test function one can obtain the ODE governing the evolution of the kinetic energy $K(t)$
\begin{equation}
K'(t) = \beta (1 - \beta) (\frac{|V|^2}{2} - K(t)) \, ,
\label{kPrime1}
\end{equation}
where $V$ - conserved (constant) bulk velocity of the distribution
function. This gives the following solution for the kinetic energy as computed in \eqref{moments3}
\begin{equation}
K(t) = K(0) e^{-\beta (1 - \beta)t} + \frac{|V|^2}{2} (1 - e^{-\beta (1 - \beta)t}) \, ,
\label{kineticEnergy}
\end{equation}
where $K(0) = $ kinetic energy at time $t = 0$. As we have an
explicit expression for the kinetic energy evolving in time, this
analytical moment can be compared with its numerical approximation
for accuracy and the corresponding graph is given in Figure
\ref{fig:kinEnergy_fEvol}. Also the general evolution of the
distribution in an inelastic collision environment is also shown in
Figure \ref{fig:kinEnergy_fEvol}. In the conservation routine
(constrained Lagrange multiplier method), energy is not used as a
constraint and just density and momentum equations are used for
constraints. Figure \ref{fig:kinEnergy_fEvol} shows the numerical accuracy of
the method even though the energy (plotted quantity) is not being conserved as part of the constrained optimization method.
\subsection{Inelastic Collisions with Diffusion Term}
\begin{figure}
\centering
\begin{tabular}{cc}
\includegraphics[height=2.75in, width=2.75in]{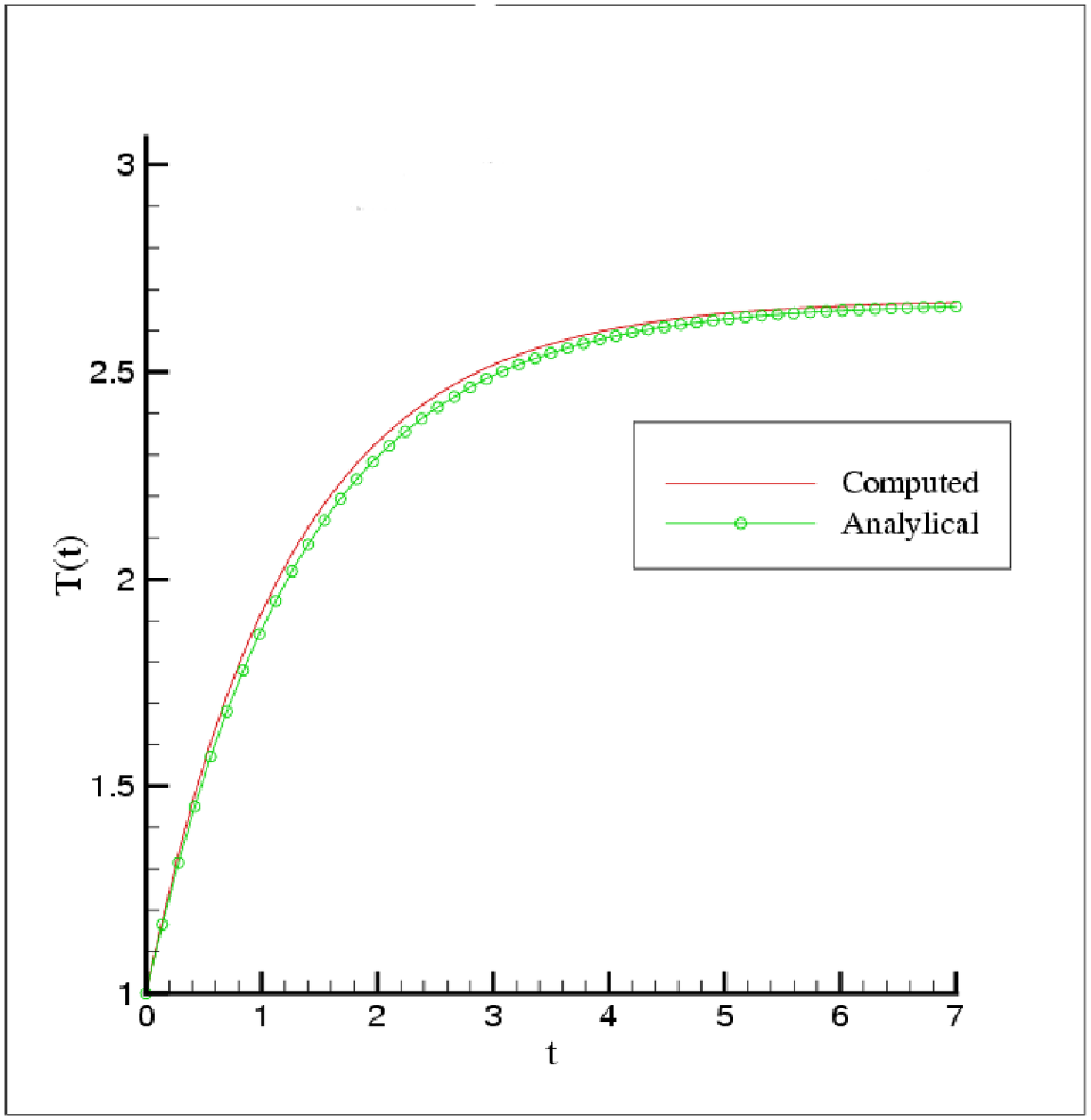} &
\includegraphics[height=2.75in, width=2.75in]{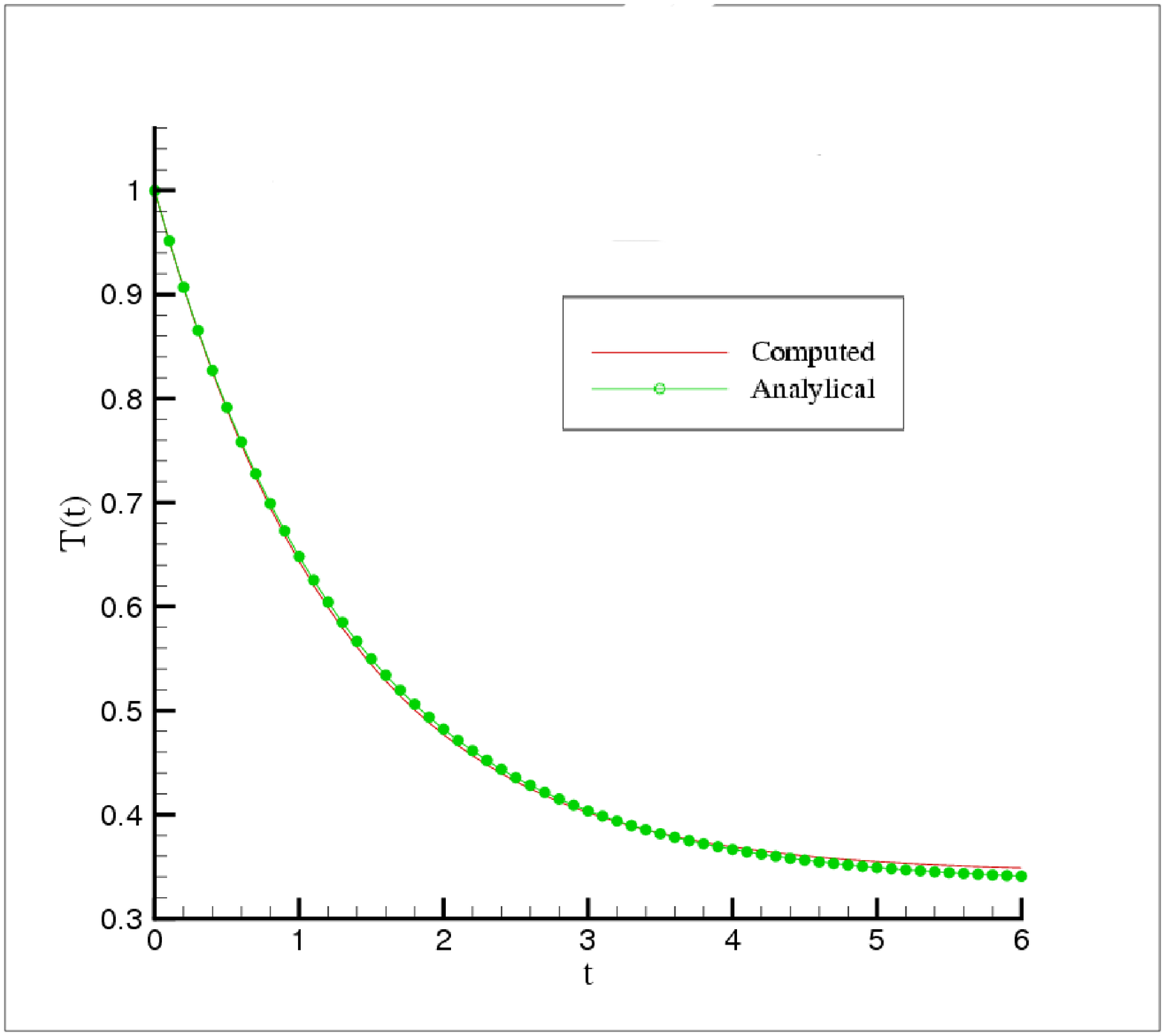} \\
($T^{MM}_{\infty} > T_0$) & ($T^{MM}_{\infty} < T_0$) \\
\end{tabular}
\caption{Maxwell type of Inelastic collisions, Diffusion Term for N = 16}
\label{fig:TMaxwell}
\end{figure}

\begin{figure}
\centering
\includegraphics[height=3in, width=3in]{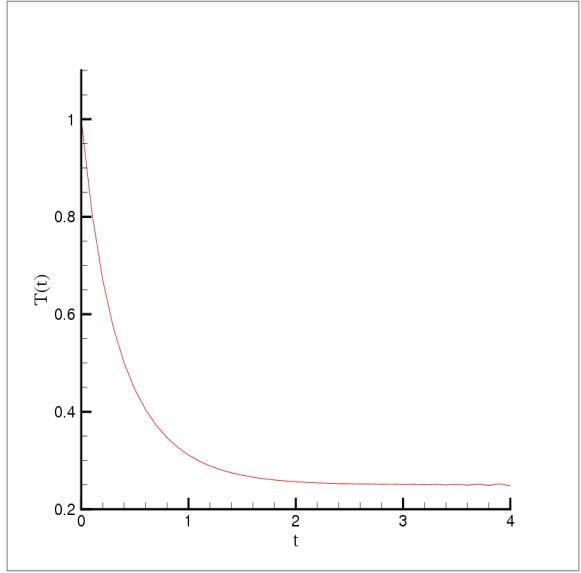}
\caption{Hard-Sphere, Inelastic Collisions, Diffusion Term, $T^{HS}_{\infty} < T_0$ for N = 16}
\label{fig:THSLow}
\end{figure}

Here we simulate, the equations \eqref{bte-source}, \eqref{force:Gauss}. Here we simulate a model corresponding to inelastic interactions in a randomly excited heat bath with constant temperature $\eta$. The evolution equation for kinetic temperature as a function of time is given by:
\begin{equation}
\frac{dT}{dt} = 2\eta - \zeta \frac{1-e^2}{24} \int_{v \in \mathbb{R}^3} \int_{w \in \mathbb{R}^3} \int_{\sigma \in \mathbb{S}^2} (1 - \mu) B(|u|, \mu) |u|^2 f(v) f(w) \, d\sigma dw dv \, ,
\label{tEqDiff}
\end{equation}
which, in the case of inelastic Maxwell type of interactions according to \eqref{moments3}, \eqref{tEqDiff} becomes
\begin{equation}
\frac{dT}{dt} = 2 \eta - \zeta \pi C_0 (1 - e^2) T \, .
\label{inelastic_ODE}
\end{equation}
The above equation gives a closed form expression for the time
evolution of the kinetic temperature and can be expressed as follows:
\begin{equation}
T(t) = T_0 e^{-\zeta \pi C_0 (1- e^2)t} + T^{MM}_{\infty} [1 - e^{-\zeta \pi C_0 (1- e^2)t}] \, ,
\label{inelastic_T}
\end{equation}
where
\begin{equation}
T_0 = \frac{1}{3} \int_{v \in \mathbb{R}^3} |v|^2 f(v) dv \qquad \text{and} \qquad T^{MM}_{\infty} = \frac{2 \eta}{\zeta \pi C_0 (1 - e^2)} \, . \nonumber
\end{equation}
As it can be seen from the expression for T, in the absence of the
diffusion term (i.e. $\eta = 0$) and for $e \neq 1$ (inelastic
collisions), the kinetic temperature of the distribution function decays like an
exponential just like in the previous section. So, the presence of
the diffusion term pushes the temperature to an equilibrium value of
$T^{MM}_{\infty} > 0$ even in the case of inelastic collisions. Also note that if the interactions were elastic and the diffusion coefficient positive then, $T^{MM}_{\infty} = + \infty$, so there would be no equilibrium states with finite kinetic temperature. These properties were shown in \cite{diffGran} and similar time asymptotic behavior is expected in the case of hard-sphere interactions where $T^{HS}_{\infty} > 0$ is shown to exist. However, the time evolution of the kinetic temperature is a non-local integral \eqref{tEqDiff} does not satisfy a close ODE form \eqref{inelastic_ODE}. The
proposed numerical method for the calculation of the collision
integral is tested for these two cases. We compared with the analytical
expression \eqref{inelastic_T} for different initial data, the corresponding computed kinetic temperatures for Maxwell type interactions in Figure \ref{fig:TMaxwell}. The asymptotic behavior is observed in the case of hard-sphere interactions in Figure \ref{fig:THSLow}. The conservation properties for this case of inelastic collisions with a diffusion term are set exactly like in the previous subsection (inelastic collisions without the diffusion term).
\subsection{Maxwell type of Elastic Collisions - Slow down process problem}
\begin{figure}
\centering
\begin{tabular}{cc}
\includegraphics[height=2.75in, width=2.75in]{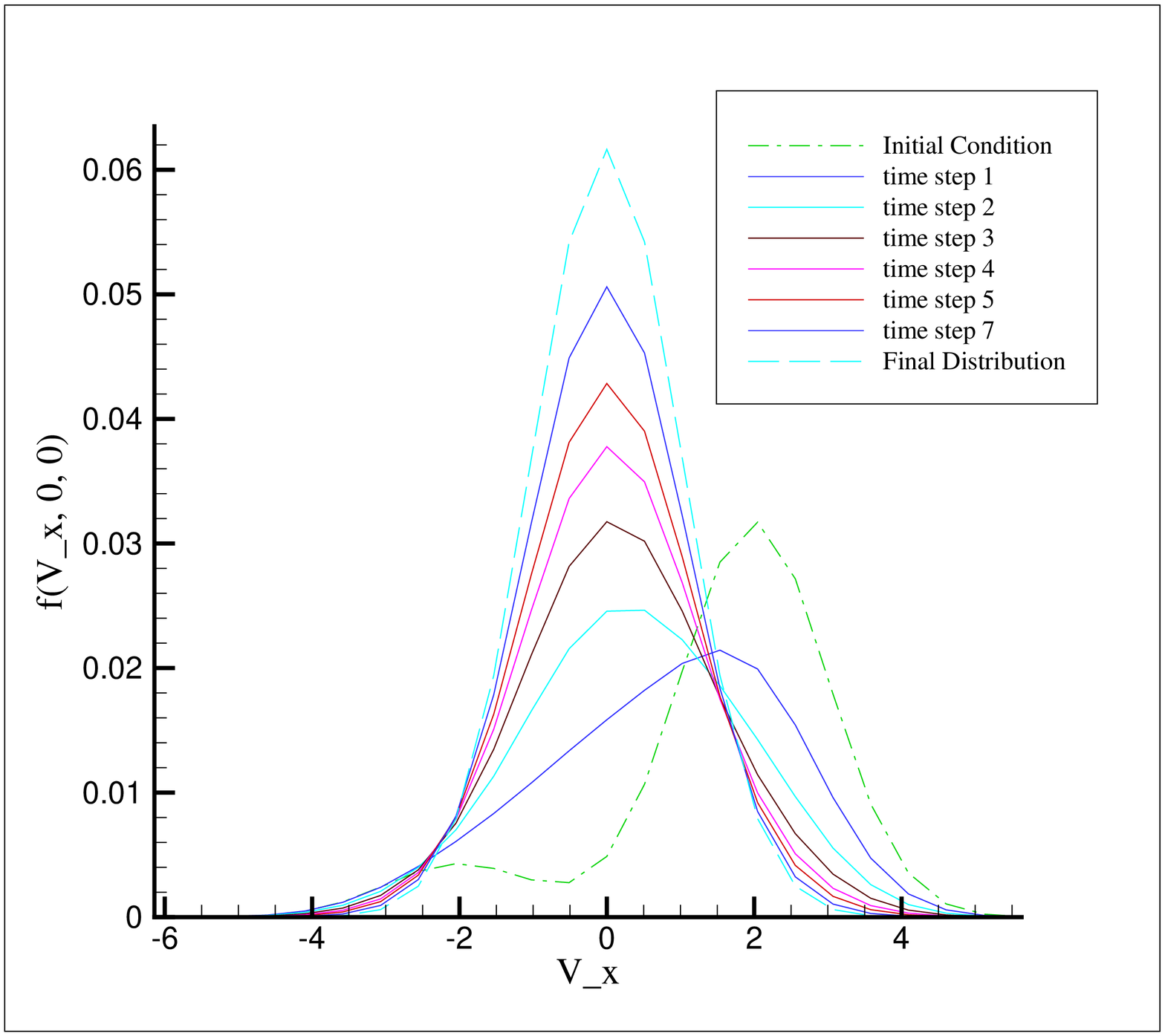} &
\includegraphics[height=2.75in, width=2.75in]{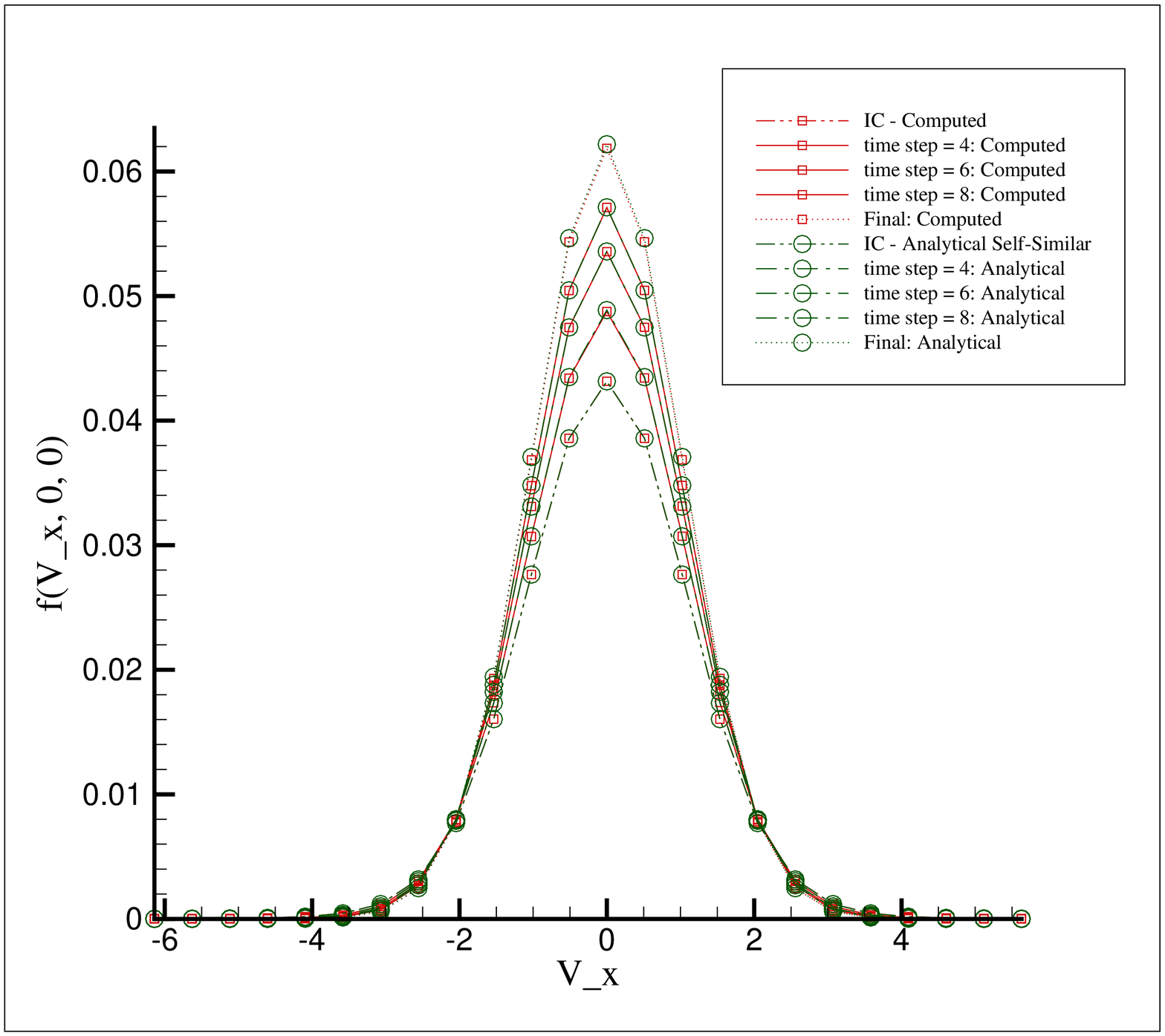} \\
Evolving $f(v, t)$ & Computed Vs. $f^{ss}_{\mathcal{T}}: \mathcal{T}= 1$ \\
\epsfig{file=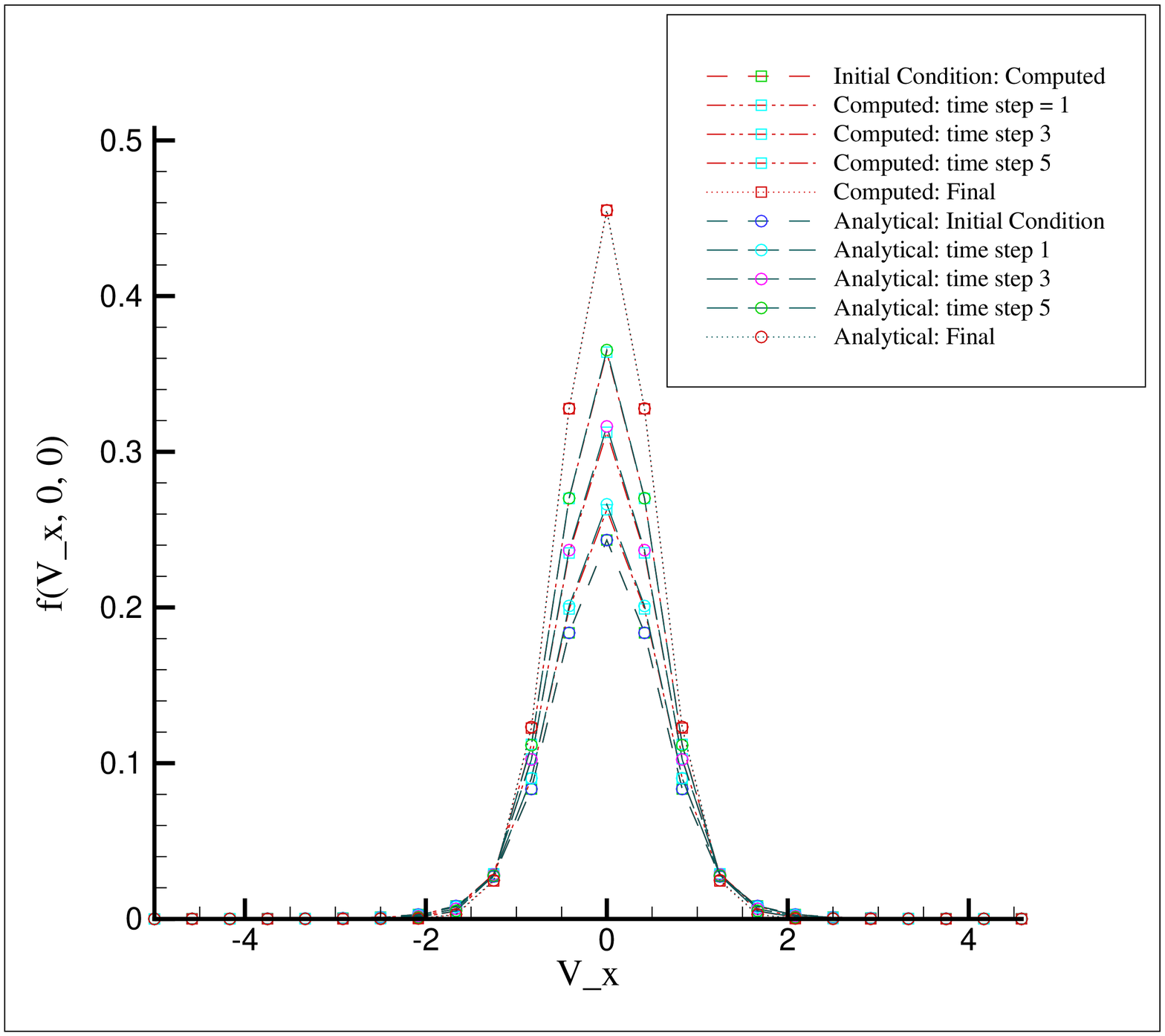,height=2.75in, width=2.75in} &
\epsfig{file=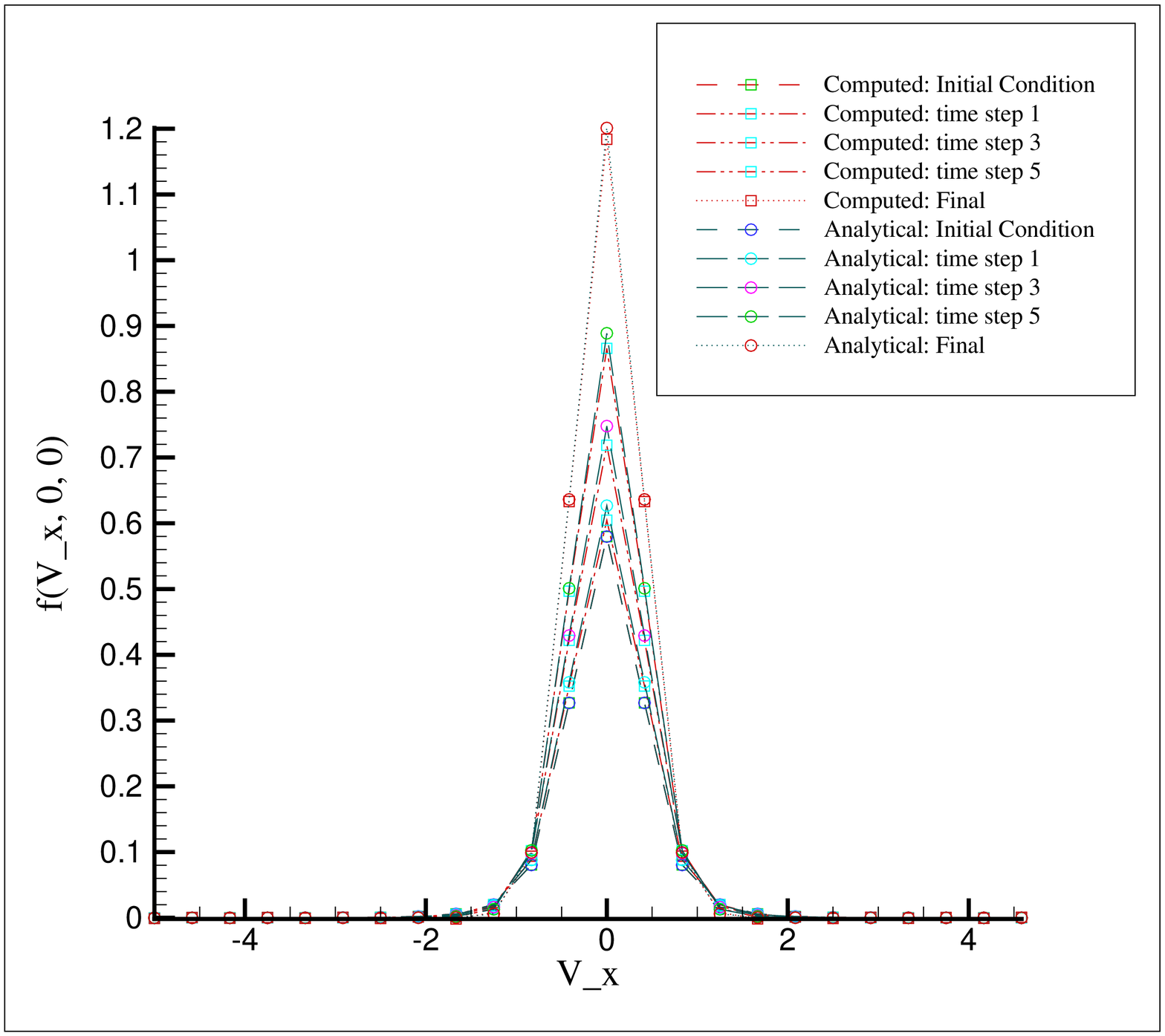,height=2.75in, width=2.75in} \\
Computed Vs. $f^{ss}_{\mathcal{T}}: \mathcal{T}= 0.25$ & Computed Vs. $f^{ss}_{\mathcal{T}}: \mathcal{T}= 0.125$\\
\end{tabular}
\caption{Maxwell type collisions, Slow down process with $\Theta = 4/3, \mu = 2/3, N = 24$}
\label{fig:mixSS}
\end{figure}

Next, consider \eqref{mixEq} with $\beta = 1, J_{\beta} = 1$ and $B(|u|, \mu) = \frac{1}{4\pi}$ i.e. isotropic collisions.
The second term is the linear collision integral which
conserves only  density and the 
the first term is the classical collision integral from \eqref{mixEq}
conserving density, momentum and energy. $M(v)$ in \eqref{mixEq}
refers to the Maxwellian defined by $M_{\mathcal{T}}(v) = e^{\frac{-|v|^2}{(2{\mathcal{T}})}}\frac{1}{(2\pi {\mathcal{T}})^{3/2}}$, with ${\mathcal{T}}$ the constant thermostat temperature. In particular, any initial distribution function 
converges to the background distribution $M_{\mathcal{T}}$. Such behavior is
well captured by the numerical method. Indeed, Fig.~\ref{fig:mixSS} corresponds to an 
initial state of a convex combination of two Maxwellians.
In addition, from \eqref{fTSolution2}:
\begin{equation}
f^{ss}_{\mathcal{T}}(v, t) =\frac{\sqrt(2)}{\pi^{5/2}} \int_0^{\infty} \frac{1}{(1 + s^2)^2} \frac{e^{-|v|^2/2 \bar{{T}}}}{\bar{T}^{\frac{3}{2}}} ds \quad \bar{T} = {\mathcal{T}} + as^2 e^{\frac{-2t}{3}} \, , \nonumber
\end{equation}
which is the finite energy solution for $p = 1, a = 1,
\mu = \frac{2}{3}, \theta = \frac{4}{3}$ in \eqref{equz}, i.e. $p = 1$ in \eqref{self-lim} and \eqref{self-lim2}. As $t \rightarrow \infty$, the time rescaled numerical distribution is compared with the analytical solution $f^{ss}_{\mathcal{T}}$ for a positive background temperature ${\mathcal{T}}$ and it converges to a Maxwellian $M_{\mathcal{T}}$.
From Figure \ref{fig:mixSS}, it can be seen that the numerical method is quite accurate and the computed distribution is in very good agreement with the analytical self-similar distribution $f^{ss}_{\mathcal{T}}$ from \eqref{fTSolution2}.
Similar agreement is observed for different constant values
of ${\mathcal{T}}$ approaching $0$ (Figure \ref{fig:mixSS}). The interesting
asymptotics \eqref{ssAsym} corresponding to power-like tails and infinitely many particles at zero energies occur only when ${\mathcal{T}} = 0$ as shown in \eqref{ssAsym} and \eqref{ssAsym}. 

Since letting $\mathcal{T} = 0$ in the scheme created an instability, we proposed the following new methodology to counter this effect. We let, instead,  $\mathcal{T} = \zeta e^{-\alpha t}$  ensuring
that the thermostat temperature vanishes for large time and set
\begin{eqnarray}
\bar{T}  = \zeta e^{-\alpha t} + as^2 e^{\frac{-2t}{3}} \, ,
\label{tHat}
\end{eqnarray}
where the role of $\alpha$ is very important and a proper choice needs to be made. In our simulations, we take $\zeta = 0.25$ and the values of $\alpha$ need to be chosen exactly as $\alpha = \mu(1) = 2/3$, the energy dissipation rate as described in section 4.2 to recover the asymptotics as in \eqref{ssAsym}. 

{\bf Remark:} \textsl{Due to the exponential time rescaling of Fourier modes,  our procedure to compute self-similar solutions in free space may also be viewed as a non-uniform grid of Fourier modes that are distributed according to the continuum spectrum of the associated problem.  This choice plays the equivalent role to the corresponding  spectral approximation of the free space problem  of the heat kernel, that is, the  Green's function for the heat equation, which happens to be a similarity solution as well, due to the linearity of the problem in this case. In particular, we expect  optimal algorithm complexity using such non-equispaced Fast Fourier Transform,  as obtained by Greengard and Lin \cite{greengardLin} for spectral approximation of the free space heat kernel. This problem will be addressed in a forthcoming paper. 
 }

\begin{figure}
\centering
\begin{tabular}{cc}
\epsfig{file=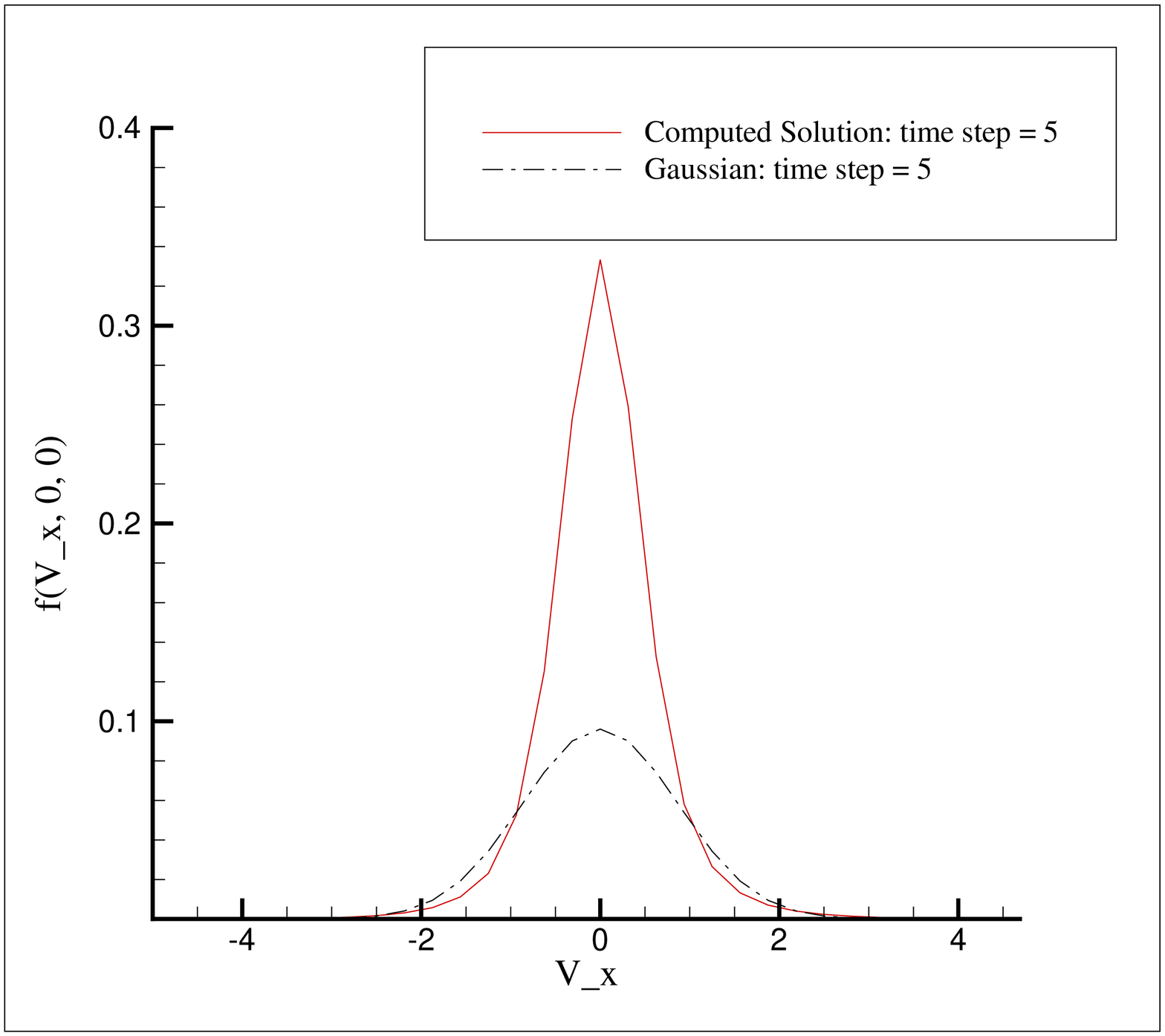,height=2.65in, width=2.65in} &
\epsfig{file=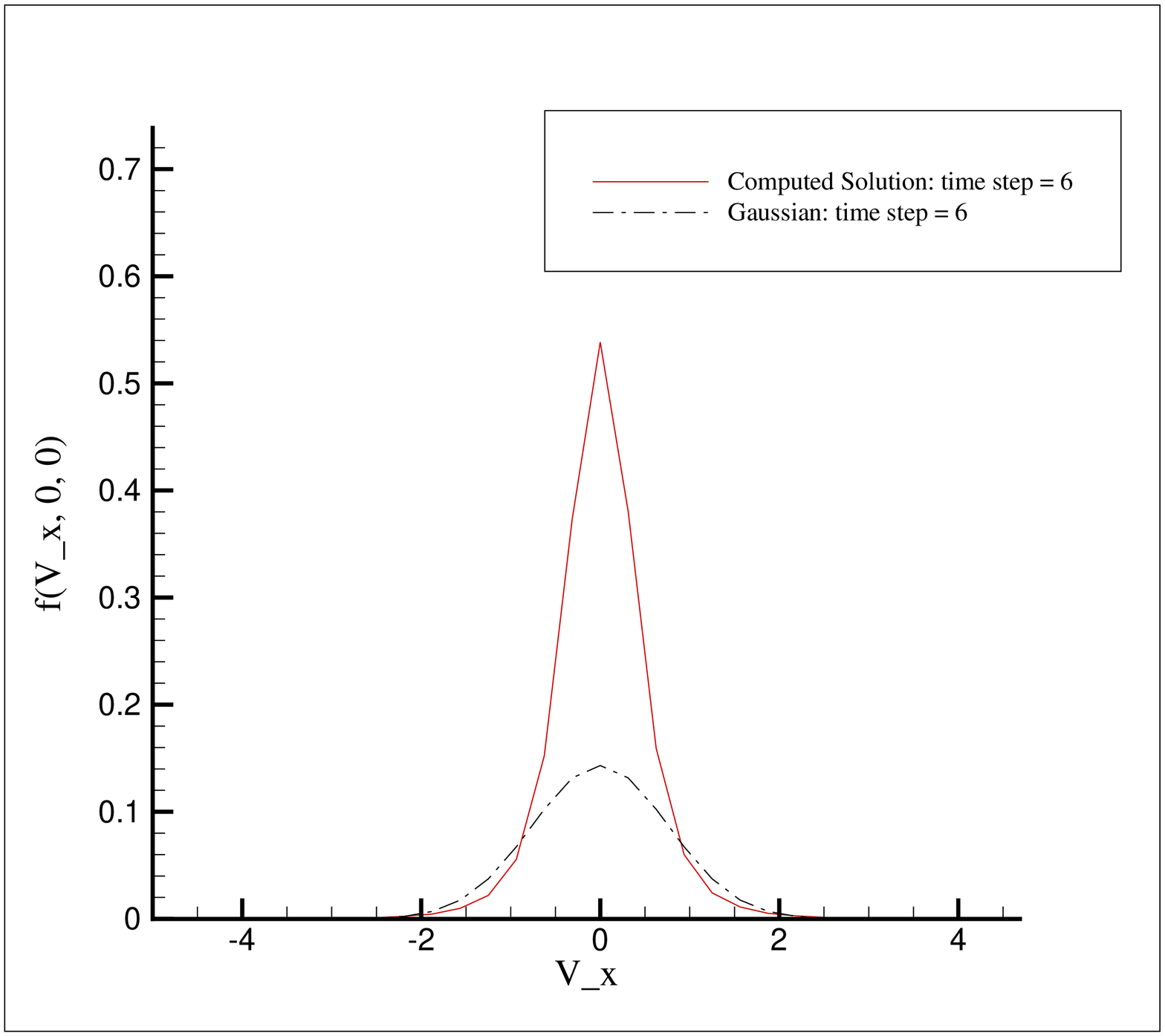,height=2.65in, width=2.65in} \\
(time step = 5) & (time step = 6)\\
\end{tabular}
\caption{Computed distribution Vs. Maxwellian with temperature of the computed distribution}
\label{fig:ssg}
\end{figure}

The following plots elucidate the fact that power-like tails are achieved asymptotically with a decaying $\mathcal{T}$.
\begin{figure}
\centering
\includegraphics[height=3in, width=3in]{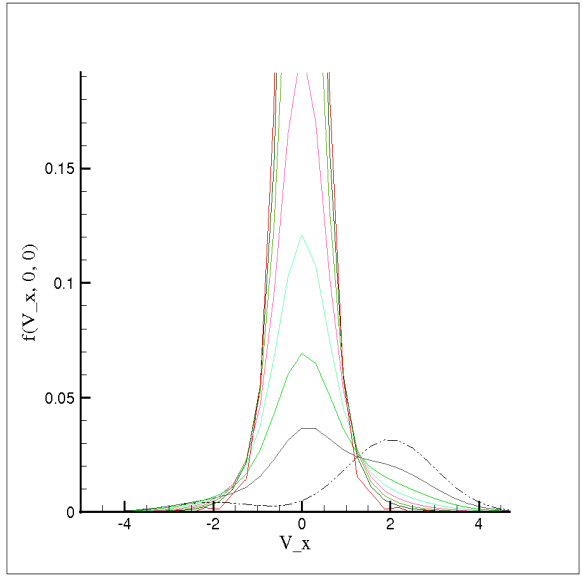}
\caption{Slow down process: N = 32, $\mathcal{T}= \frac{1}{4} e^{-2t/3}$}
\label{fig:ssf_32}
\end{figure}
For a decaying background temperature as in \eqref{tHat}, Figure
\ref{fig:ssf_32} shows evolution of a convex combination of Maxwellians to a self-similar (blow up for zero energies and power-like for high energies) behavior. Figure \ref{fig:ssg} plots the computed distribution along with a Maxwellian with temperature of the computed solution. This illustrates that the computed self-similar solution is largely deviated from a Maxwellian equilibrium.
In order to better capture the power-like effect using this
numerical method, we  set 
 $\mathcal{T} = \zeta e^{-2t/3} = \zeta e^{-\mu t}$, see  \eqref{tHat}, where $\mu$ is related the spectral properties of the Fourier
transformed equation as described in section 4.2 on the slow down process
problem with $\mu = \mu(1)$ the energy dissipation rate.  Thus, as it was computed in~\cite{powerLikeGB} 
and revised in section 4 of this paper, we know that  for initial states with finite energy, $p=1$ and the corresponding energy dissipation rate is $\mu(1) = \mu = 2/3$ is positive. In particular   $p_{*} = 1.5$
is the  conjugate of $p=1$ of the spectral curve $m_q$ in Theorem 4.1  part (i).  In addition 
the rescaled probability will converge to the moments of the self-similar state \eqref{self-lim}, \eqref{self-lim2}, that is  
\begin{equation}
e^{-q t 2/3} \int_{v \in \mathbb{R}^3} f(v) |v|^{2q} dv \rightarrow m_q \, ,
\nonumber
\end{equation}
and we know any moment $m_q$ is unbounded for $q> p_{*} = 1.5$. 

\begin{figure}
\centering
\begin{tabular}{cc}
\epsfig{file=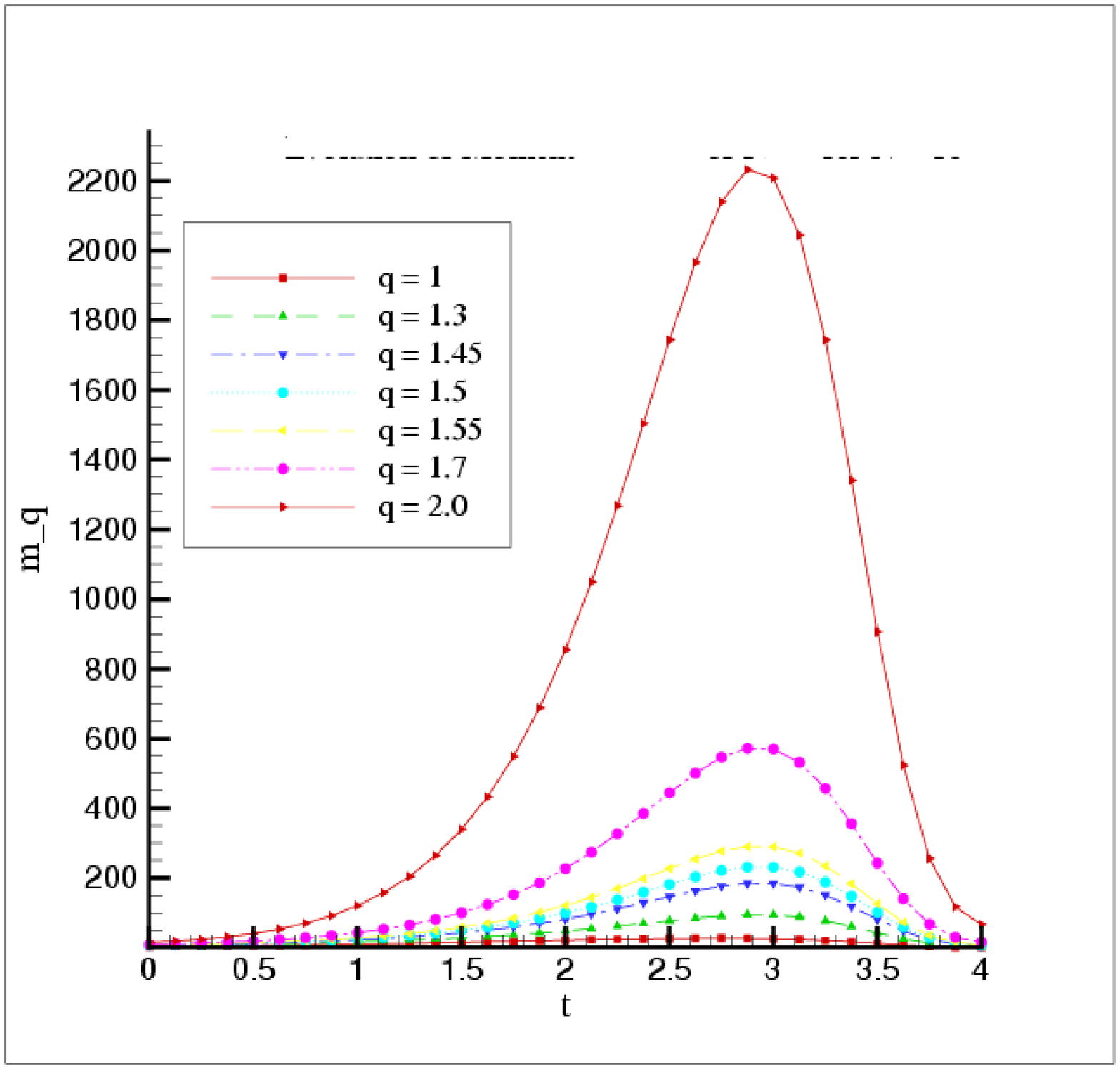,height=2.5in, width=2.5in} &
\epsfig{file=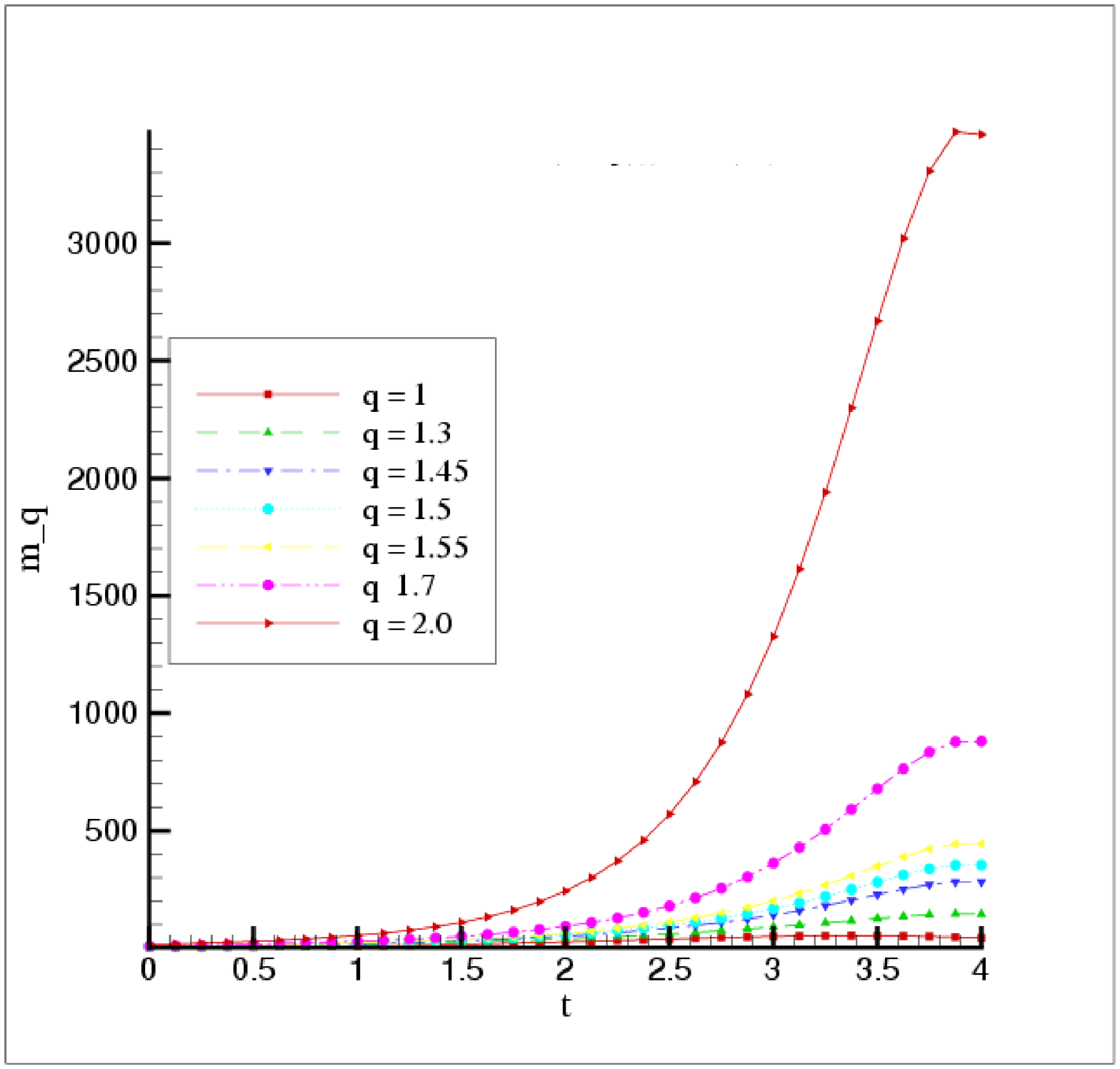,height=2.5in, width=2.5in}\\
$N = 10$ & $N = 14$\\
\epsfig{file=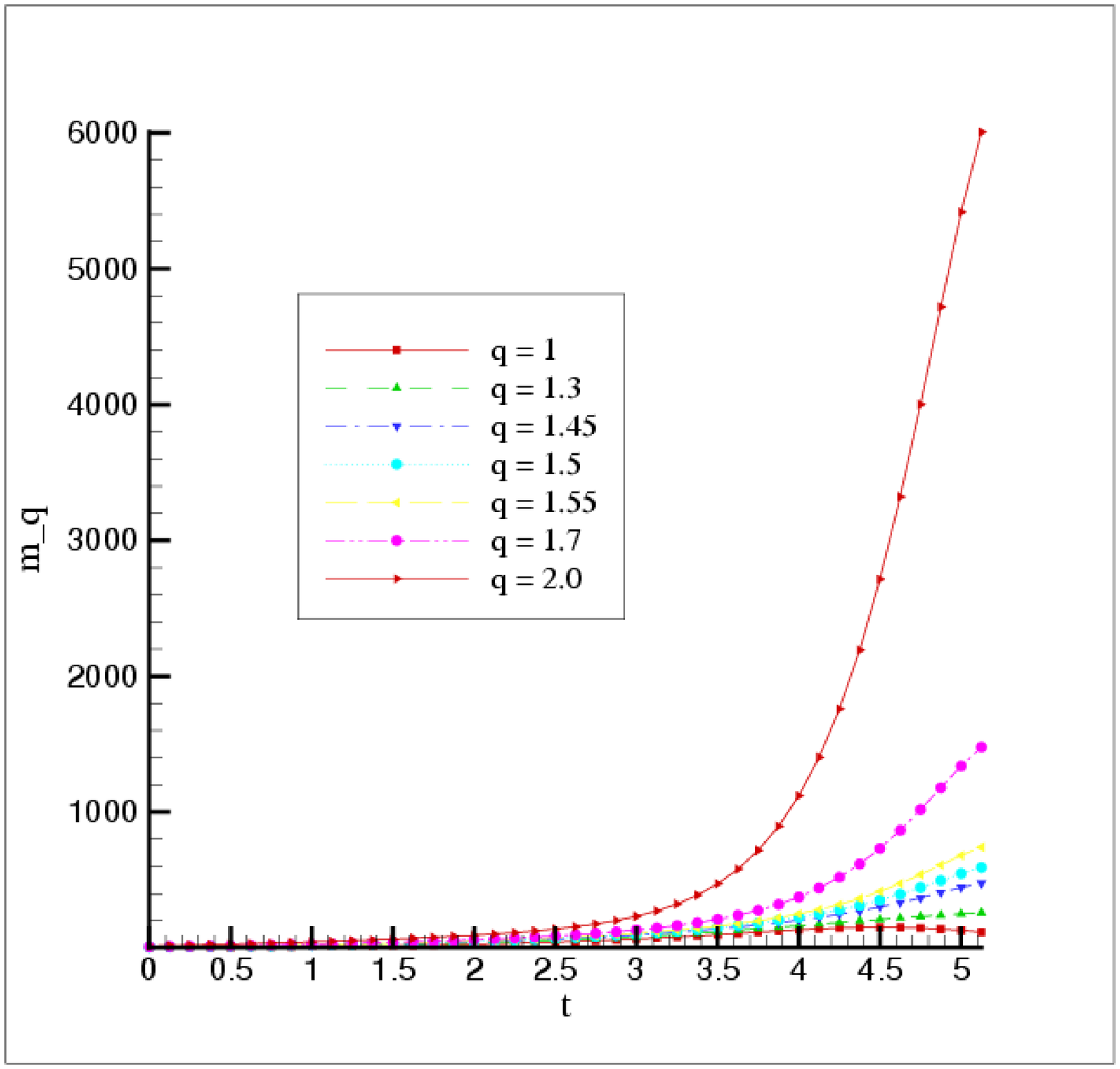,height=2.5in, width=2.5in}&
\epsfig{file=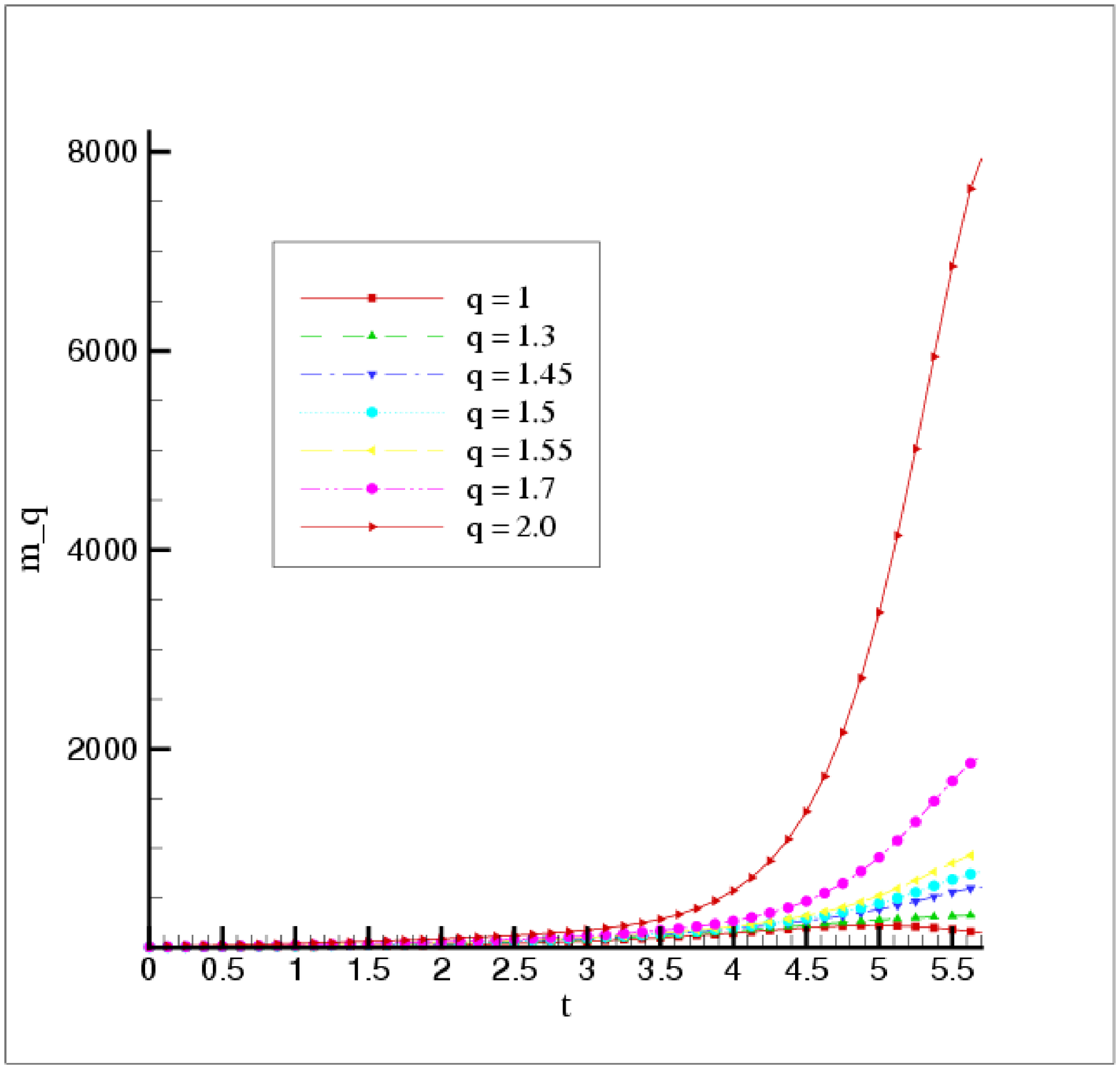,height=2.5in, width=2.5in}\\
$N = 22$ & $N = 26$\\
\end{tabular}
\caption{$m_q (t)$ for $\mathcal{T}=e^{-2t/3}$}
\label{fig: m_q}
\end{figure}

\begin{figure}
\centering
\begin{tabular}{cc}
\epsfig{file=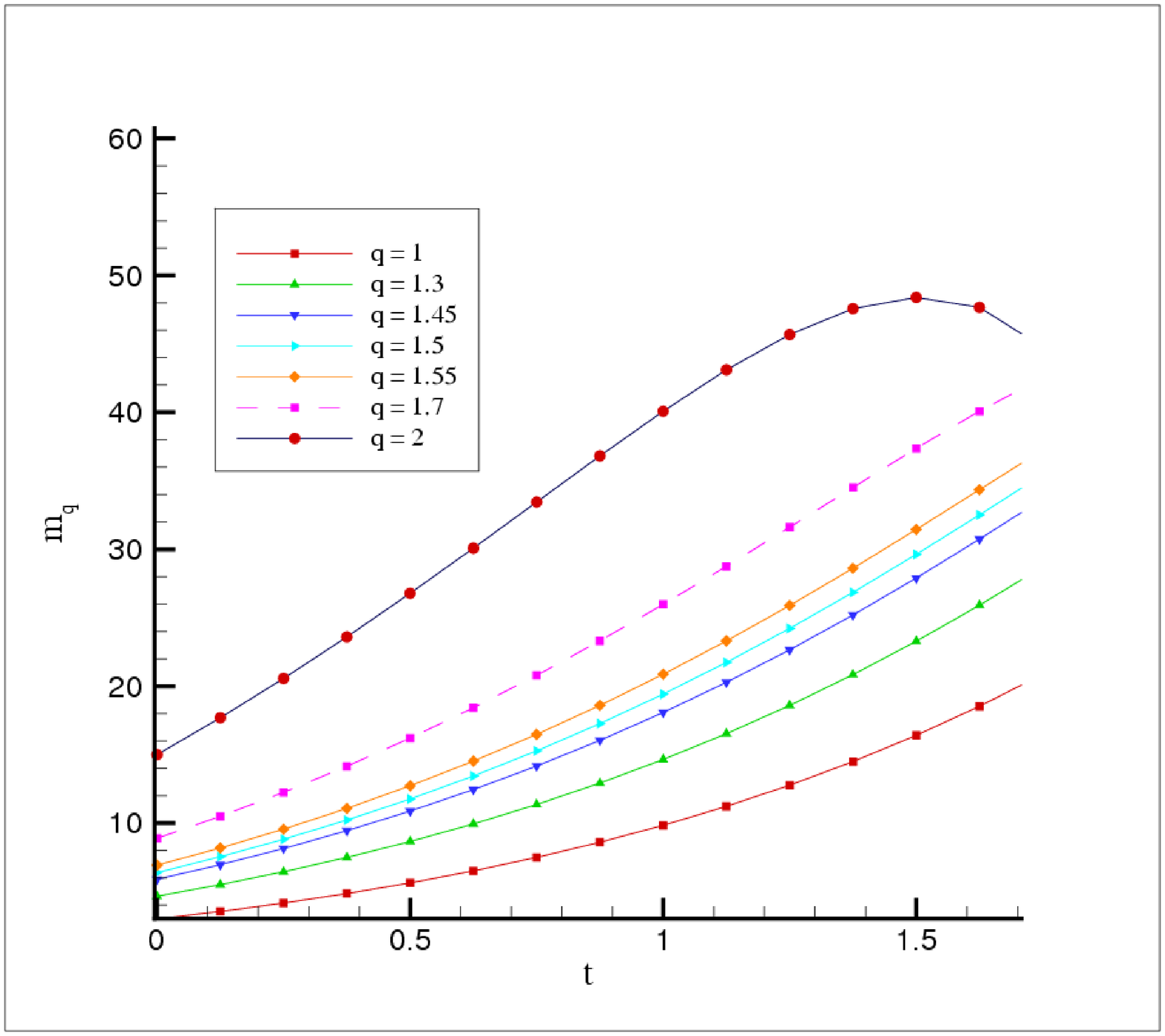,height=2.5in, width=2.5in} &
\epsfig{file=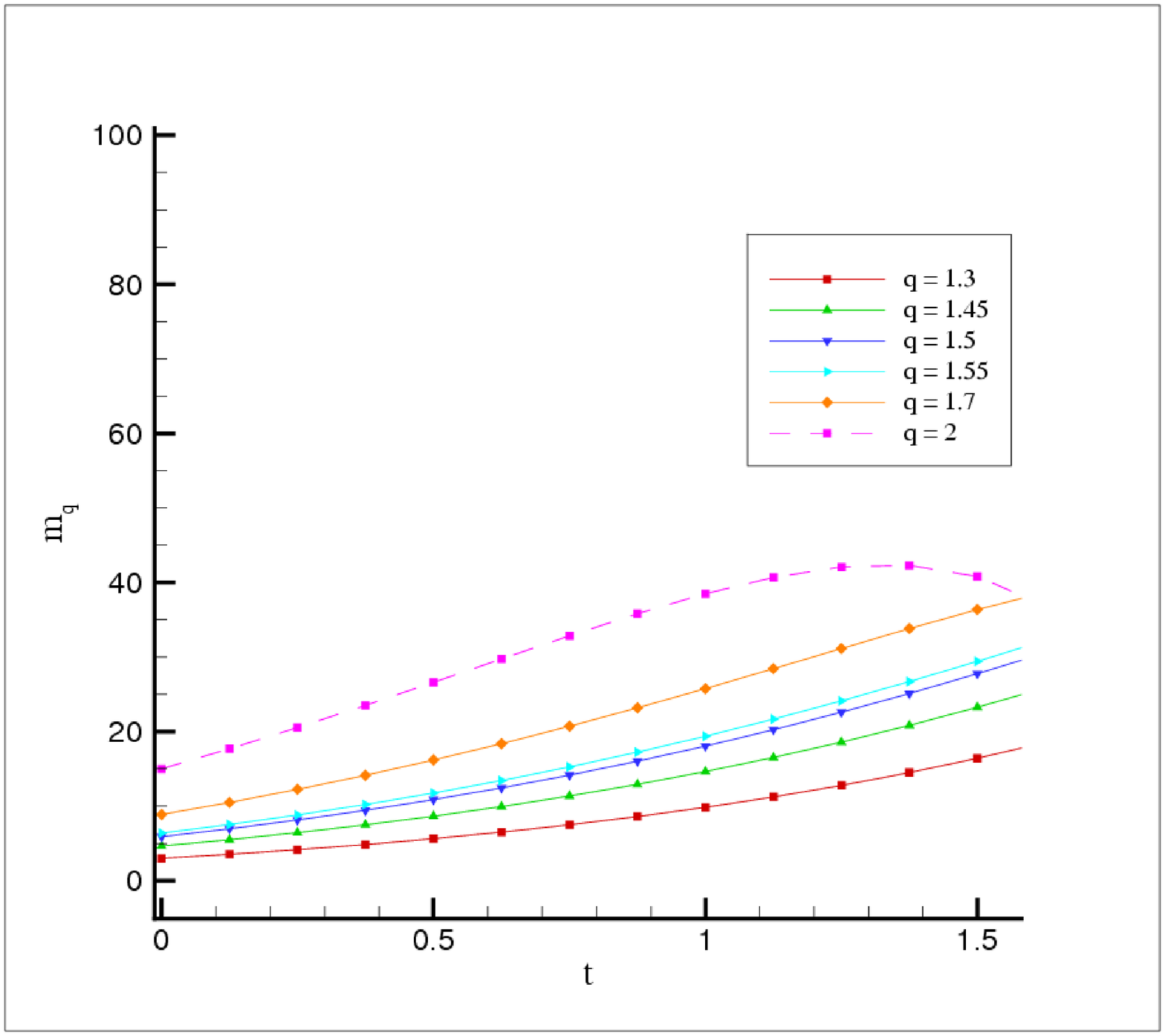,height=2.5in, width=2.5in}\\
$N = 16$ & $N = 20$\\
\end{tabular}
\caption{$m_q (t)$ for $\mathcal{T} = e^{-2t/3}$}
\label{fig: m_qm}
\end{figure}

We have plot  in Fig.~\ref{fig: m_q} the evolution of $e^{-q t 2/3} \int_{v \in \mathbb{R}^3} f(v) |v|^{2q} dv  $   for $q = 1, 1.3, 1.45, 1.5, 1.55, 1.7, 2.0$, 
computed  for different values of $N = 10, 14, 16$, $18,
22, 26$. It can be seen that as time progresses (and as the thermostat temperature $\mathcal{T}$ decreases to $0$), the approximated numerically computed moments to $m_q, q \geq 1.5$ start to blow up as predicted. The value $q = 1.5$ is the threshold value, as any moment $m_{q > 1.5}(t) \rightarrow \infty$.
The expected spectral accuracy, as the value of $N$
increases, improves  the  growth zone of such moments  for larger final
times. 
The reason for such effect is that since the truncation of Fourier modes that results in the
truncation in velocity domain, makes the distribution
function to take small negative values for large velocities contributing to numerical errors 
that may cause
 $m_q$ to peak and then relax back.  In particular, larger order moments of the computed self-similar asymptotics  with 
the negative oscillating parts on large energy tails,  result in the large negative
moment values for the above mentioned values of $N$ crating large negative errors. 
However it is noticed that
the negative oscillation values of $f(t, v)$ coincide with large
velocity values used in getting its $q$-moments  approximating  $m_q$, for $q>1.5$, and that such error is reduced in time for larger number of Fourier modes.
Finally we point out that a FFTW package has been used. We have noticed in  our numerics that for the specific choice of  values  $N \neq 6, 10, 14, 18, 22, 26, . . . , 6 + 4k; k= 0, 1, 2, 3, . . .$,  the approximating  moments to $m_q(t)$ start to take negative values very quickly, as seen in Fig.~\ref{fig: m_qm} for $N=16$ and $20$,  making the numerical solution  inadmissible since analytically $m_q(t) >0, \forall t$. Such effect may be due to the particular choice of the FTTW solver.
\section{Conclusions and Future Work}
In conclusion, the presented numerical method works for elastic and inelastic variable hard potential interactions. This is first of its kind as no additional modification is required to compute for elastic and inelastic collisions. In comparison with the known analytical results (moment equations for elastic BTE, BKW self-similar solution, attracting Bobylev-Cercignani-Gamba self-similar solutions for elastic collisions in a slow down process), the computed ones are found to be very close. The method employs a Fast Fourier Transform for faster evaluation of the collision integral. Even though the method is implemented for a uniform grid in velocity space, it can even be implemented for a non-uniform velocity grid. The only challenge in this case is computing the Fast Fourier Transform on such a non-uniform grid. There are available packages for this purpose, but such a non-uniform FFT can also be implemented using certain high degree polynomial interpolation and this possibility is currently being explored. The integration over the unit sphere is avoided completely and only a simple integration over a regular velocity grid is needed. Even though a trapezoidal rule is used as an integration rule, other integration rules like a Gaussian quadrature can be used to get better accuracy. For time discretization, a simple second-order Runge Kutta scheme is used. The proposed method has a big advantage over other non-deterministic methods as the exact distribution function can actually be computed instead of just the averages.
\\
\\
Implementations of this scheme for the space inhomogeneous case is currently developed by the authors by means of splitting algorithms in advection and collision components. Next step in this direction would be to implement the method for a practical $1$ and $2-D$ space inhomogeneous problems such  shock tube phenomena for specular and diffusive boundary conditions, and   Rayleigh-Benard instability or a Couette flow problem.

\section{Acknowledgements}
The authors would like to thank Sergej Rjasanow for discussions about the conservation properties of the numerical method and for other comments. Both authors are partially supported under the NSF grant DMS-0507038. Support from the Institute of Computational Engineering and Sciences and the University of Texas Austin is also gratefully acknowledged.

%

\begin{thebibliography}{10}
\expandafter\ifx\csname url\endcsname\relax
  \def\url#1{\texttt{#1}}\fi
\expandafter\ifx\csname urlprefix\endcsname\relax\def\urlprefix{URL }\fi

\bibitem{AlaGa07}
R.~J. Alonso, I.~M. Gamba, Propagation of $l^1$ and $l^\infty$ maxwellian
  weighted bounds for derivatives of solutions to the homogeneous elastic
  boltzmann equation, To appear in Journal of Mathematiques Pures et
  Appliqu\'ees.

\bibitem{bird}
G.~A. Bird, Molecular Gas Dynamics, Clarendon Press, Oxford, 1994.

\bibitem{Bob75}
A.~V. Bobylev, Exact solutions of the boltzmann equation, (Russian) Dokl. Akad.
  Nauk SSSR 225 (1975) 1296--1299.

\bibitem{bobylevFT}
A.~V. Bobylev, Exact solutions of the nonlinear boltzmann equation and the
  theory of relaxation of a maxwellian gas, Translated from Teoreticheskaya i
  Mathematicheskaya Fizika 60 (1984) 280 -- 310.

\bibitem{Bob88}
A.~V. Bobylev, The theory of the nonlinear spatially uniform boltzmann equation
  for maxwell molecules, Mathematical physics reviews; Soviet Sci. Rev. Sect. C
  Math. Phys. Rev. 7 (1988) 111--233.

\bibitem{BoCaGa00}
A.~V. Bobylev, J.~A. Carrillo, I.~M. Gamba, On some properties of kinetic and
  hydrodynamic equations for inelastic interactions, Journal of Statistical
  Physics 98 (2000) 743--773.

\bibitem{bobyCerci99}
A.~V. Bobylev, C.~Cercignani, Discrete velocity models without nonphysical
  invariants, Journal of Statistical Physics 97 (1999) 677--686.

\bibitem{bcSS1}
A.~V. Bobylev, C.~Cercignani, Exact eternal solutions of the boltzmann
  equation, Journal of Statistical Physics 106 (2002) 1019--1038.

\bibitem{BCLT}
A.~V. Bobylev, C.~Cercignani, The inverse laplace transform of some analytic
  functions with an application to the eternal solutions of the boltzmann
  equation, Applied Mathematics Letters 15 (2002) 807--813(7).

\bibitem{BoCe1}
A.~V. Bobylev, C.~Cercignani, Moment equations for a granular material in a
  thermal bath, Journal of Statistical Physics 106 (2002) 547--567(21).

\bibitem{bcSS2}
A.~V. Bobylev, C.~Cercignani, Self-similar asymptotics for the boltzmann
  equation with inleastic and elastic interactions, Journal of Statistical
  Physics 110 (2003) 333--375.

\bibitem{BCG}
A.~V. Bobylev, C.~Cercignani, I.~M. Gamba, On the self-similar asymptotics for
  generalized non-linear kinetic maxwell models, to appear in Communication in
  Mathematical Physics.
\newline\urlprefix\url{http://arxiv.org/abs/math-ph/0608035}

\bibitem{asympGranular}
A.~V. Bobylev, C.~Cercignani, G.~Toscani, Proof of an asymptotic property of
  self-similar solutions of the boltzmann equation for granular materials,
  Journal of Statistical Physics 111 (2003) 403--417.

\bibitem{powerLikeGB}
A.~V. Bobylev, I.~M. Gamba, Boltzmann equations for mixtures of maxwell gases:
  Exact solutions and power like tails, Journal of Statistical Physics 124
  (2006) 497--516.

\bibitem{highEnergyTails}
A.~V. Bobylev, I.~M. Gamba, V.~Panferov, Moment inequalities and high-energy
  tails for boltzmann equations with inelastic interactions, Journal of
  Statistical Physics 116 (2004) 1651--1682.

\bibitem{BoGrSp}
A.~V. Bobylev, M.~Groppi, G.~Spiga, Approximate solutions to the problem of
  stationary shear flow of smooth granular materials, Eur. J. Mech. B Fluids 21
  (2002) 91--103.

\bibitem{bobylevRjasanow0}
A.~V. Bobylev, S.~Rjasanow, Difference scheme for the boltzmann equation based
  on the fast fourier transform, European journal of mechanics. B, Fluids 16:22
  (1997) 293--306.

\bibitem{bobylevRjasanow3}
A.~V. Bobylev, S.~Rjasanow, Fast deterministic method of solving the boltzmann
  equation for hard spheres, European journal of mechanics. B, Fluids 18:55
  (1999) 869--887.

\bibitem{bobylevRjasanow2}
A.~V. Bobylev, S.~Rjasanow, Numerical solution of the boltzmann equation using
  fully conservative difference scheme based on the fast fourier transform,
  Transport Theory Statist. Phys. 29 (2000) 289--310.

\bibitem{broadwell}
J.~E. Broadwell, Study of rarefied shear flow by the discrete velocity method,
  J. Fluid Mech. 19 (1964) 401--414.

\bibitem{cabannes}
H.~Cabannes, Global solution of the initial value problem for the discrete
  boltzmann equation, Arch. Mech. (Arch. Mech. Stos.) 30 (1978) 359--366.

\bibitem{carcer95}
C.~Cercignani, Recent developments in the mechanics of granular materials,
  Fisica matematica e ingegneria delle strutture, Bologna: Pitagora Editrice
  (1995) 119--132.

\bibitem{Ce1}
C.~Cercignani, Shear flow of a granular material, Journal of Statistical
  Physics 102 (2001) 1407--1415.

\bibitem{cerci-corn00}
C.~Cercignani, H.~Cornille, Shock waves for a discrete velocity gas mixture,
  Journal of Statistical Physics 99 (2000) 115--140.

\bibitem{ErBr2}
M.~H. Ernst, R.~Brito, Driven inelastic maxwell models with high energy tails,
  Phys. Rev. E 65~(4) (2002) 040301.

\bibitem{ernst-brito}
M.~H. Ernst, R.~Brito, Scaling solutions of inelastic boltzmann equations with
  over-populated high energy tails, Journal of Statistical Physics 109 (2002)
  407--432.

\bibitem{filbetRusso1}
F.~Filbet, C.~Mouhot, L.~Pareschi, Solving the boltzmann equation in nlogn,
  SIAM J. Sci. Comput. 28 (2006) 1029--1053.

\bibitem{filbetRusso}
F.~Filbet, G.~Russo, High order numerical methods for the space non homogeneous
  boltzmann equation, Journal of Computational Physics 186 (2003) 457--480.
\newline\urlprefix\url{citeseer.ist.psu.edu/filbet03high.html}

\bibitem{filbet03high}
F.~Filbet, G.~Russo, High order numerical methods for the space non homogeneous
  boltzmann equation, Journal of Computational Physics 186 (2003) 457 -- 480.

\bibitem{elInCoa}
N.~Fournier, S.~Mischler, A boltzmann equation for elastic, inelastic and
  coalescing collisions, Journal de mathematiques pures et appliques 84 (2005)
  1173--1234.

\bibitem{fftw}
M.~Frigo, S.~G. Johnson, Fast fourier transform of the west.
\newline\urlprefix\url{www.fftw.org}

\bibitem{Gab-Par-Tos}
E.~Gabetta, L.~Pareschi, G.~Toscani, Relaxation schemes for nonlinear kinetic
  equations, SIAM J. Numer. Anal. 34 (1997) 2168--2194.

\bibitem{Ga-Pa-Vi07}
I.~M. Gamba, V.~Panferov, C.~Villani, Upper maxwellian bounds for the spatially
  homogeneous boltzmann equation, To appear in Arch.Rat.Mec.Anal.
\newline\urlprefix\url{http://arxiv.org/abs/math/0701081}

\bibitem{diffGran}
I.~M. Gamba, V.~Panferov, C.~Villani, On the boltzmann equation for diffusively
  excited granular media, Communications in Mathematical Physics 246 (2004)
  503--541(39).

\bibitem{GRW04}
I.~M. Gamba, S.~Rjasanow, W.~Wagner, Direct simulation of the uniformly heated
  granular boltzmann equation, Mathematical and Computer Modelling 42 (2005)
  683--700.

\bibitem{GTConv08}
I.~M. Gamba, S.~H. Tharkabhushanam, Convergence and error analysis of
  spectral-lagrange boltzmann solver, In preparation.

\bibitem{GreLe05}
R.~L. Greenblatt, J.~L. Lebowitz, Product measure steady states of generalized
  zero range processes, J. Phys. A 39 (2006) 1565--1573.

\bibitem{greengardLin}
L.~Greengard, P.~Lin, Spectral approximation of the free-space heat kernel,
  Appl. Comput. Harmon. Anal. 9~(1) (2000) 83--97.

\bibitem{HerParSea}
M.~Herty, L.~Pareschi, M.~Seaid, Discrete-velocity models and relaxation
  schemes for traffic flows, SIAM J. Sci. Comput. 28 (2006) 1582--1596.

\bibitem{ibragRjasanow}
I.~Ibragimov, S.~Rjasanow, Numerical solution of the boltzmann equation on the
  uniform grid, Computing 69 (2002) 163--186.

\bibitem{illner78}
R.~Illner, On the derivation of the time-dependent equations of motion for an
  ideal gas with discrete velocity distribution, J. de Mecanique 17 (1978)
  781--796.

\bibitem{kawa81}
S.~Kawashima, Global solution of the initial value problem for a discrete
  velocity model of the boltzmann equation, Proc. Japan Acad. Ser. A Math. Sci.
  57 (1981) 19--24.

\bibitem{KW}
K.~Max, W.~T. Tsun, Formation of maxwellian tails, Physical Review Letters 36
  (1976) 1107--1109.

\bibitem{mieuss00}
L.~Mieussens, Discrete-velocity models and numerical schemes for the
  boltzmann-bgk equation in plane and axisymmetric geometries, Journal of
  Computational Physics 162 (2000) 429--466.

\bibitem{MoSa}
J.~M. Montanero, A.~Santos, Computer simulation of uniformly heated granular
  fluids, Gran. Matt. 2 (2000) 53--64.

\bibitem{MSS}
S.~J. Moon, M.~D. Shattuck, J.~Swift, Velocity distributions and correlations
  in homogeneously heated granular media, Physical Review E 64 (2001) 031303.

\bibitem{mouhotPareschi}
C.~Mouhot, L.~Pareschi, Fast algorithms for computing the boltzmann collision
  operator, Math. Comp. 75 (2006) 1833--1852.

\bibitem{nanbu}
K.~Nanbu, Direct simulation scheme derived from the boltzmann equation
  i.monocomponent gases, J. Phys. Soc. Japan 52 (1983) 2042 -- 2049.

\bibitem{vanNoi-Ernst}
T.~V. Noije, M.~Ernst, Velocity distributions in homogeneously cooling and
  heated granular fluids, Gran. Matt. 1:57(1998).

\bibitem{panferovRjasanow}
V.~Panferov, S.~Rjasanow, Deterministic approximation of the inelastic
  boltzmann equation, Unpublished manuscript.

\bibitem{pareschiPerthame}
L.~Pareschi, B.~Perthame, A fourier spectral method for homogenous boltzmann
  equations, Transport Theory Statist. Phys. 25 (2002) 369--382.

\bibitem{pareschiRusso}
L.~Pareschi, G.~Russo, Numerical solution of the boltzmann equation. i.
  spectrally accurate approximation of the collision operator, SIAM J.
  Numerical Anal. (Online) 37 (2000) 1217--1245.

\bibitem{Par-Tos}
L.~Pareschi, G.~Toscani, Self-similarity and power-like tails in
  nonconservative kinetic models, J. Stat. Phys. 124 (2006) 747--779.

\bibitem{RjaWa96}
S.~Rjasanow, W.~Wagner, A stochastic weighted particle method for the boltzmann
  equation, Journal of Computational Physics (1996) 243--253.

\bibitem{RjaWa05}
S.~Rjasanow, W.~Wagner, Stochastic Numerics for the Boltzmann Equation,
  Springer, Berlin, 2005.

\bibitem{villani}
C.~Villani, Handbook of Fluid dynamics, chap. A Review of Mathematical Topics
  in Collisional Kinetic Theory, Elsevier, 2003, pp. 71--306.
\newline\urlprefix\url{http://www.umpa.ens-lyon.fr/~cvillani}

\bibitem{Wagner92}
W.~Wagner, A convergence proof for bird's direct simulation monte carlo method
  for the boltzmann equation, Journal of Statistical Physics (1992) 1011--1044.

\bibitem{WiMa}
D.~R.~M. Williams, F.~MacKintosh, Driven granular media in one dimension:
  Correlations and equation of state, Phys. Rev. E 54 (1996) 9--12.

\bibitem{zheng-struchtrup}
Y.~Zheng, H.~Struchtrup, A linearization of mieussens's discrete velocity model
  for kinetic equations, Eur. J. Mech. B Fluids 26 (2007) 182--192.

\end{thebibliography}

\end{document}